\documentclass[prd,aps,nofootinbib,floatfix,10pt]{revtex4}
\usepackage{graphicx,epsfig,mathtools}
\usepackage[usenames]{color}
\usepackage{slashed}
\usepackage{epstopdf}

\begin{document}
\title{A comprehensive analysis of weak transition form factors for  doubly heavy baryons in the light front approach}

\author{Xiao-Hui Hu$^1$~\footnote{Email:huxiaohui@sjtu.edu.cn}, Run-Hui Li$^2$~\footnote{Email:lirh@imu.edu.cn},   Zhi-Peng Xing$^1$~\footnote{Email:zpxing@sjtu.edu.cn}}
\affiliation{
$^{1}$ INPAC,SKLPPC,
MOE KLPPC,
School of Physics and Astronomy, Shanghai Jiao-Tong University, Shanghai 200240, P.R. China.\\
$^{2}$ School of Physical Science and Technology, Inner Mongolia University, Hohhot 010021, P.R. China.}
\begin{abstract}
The transition form factors for doubly heavy baryons into a spin-$1/2$ or spin-$3/2$ ground-state baryon induced by both the charged current and the flavor changing neutral current are systematically  studied  within the light-front quark model. In the transition the two spectator quarks have two spin configurations and both are considered in this calculation.  We use an updated vertex functions, and inspired by the flavor SU(3) symmetry, we also provide a new approach to derive the flavor-spin  factors.   With the obtained transition form factors, we perform a phenomenological study of the corresponding semi-leptonic decays of doubly heavy baryons induced by the $c\to d/s \ell^+\nu$, $b\to c/u\ell^-\bar \nu$ and $b\to d/s\ell^+\ell^-$. Results for   partial decay widths, branching ratios and  the polarization ratios $\Gamma_{L}/\Gamma_{T}$s are given.
We find that most  branching ratios for the semi-leptonic decays  induced by the $c\to d,s$ transitions
are at the order of $10^{-3}\sim10^{-2}$, which might be useful for the search of other doubly-heavy baryons.
Uncertainties in form factors, the  flavor SU(3) symmetry and sources of symmetry breaking effects are discussed.
We find that the SU(3) symmetry breaking effects could be  sizable in   charmed
baryon decays while in the bottomed case, the SU(3) symmetry breaking effects are less significant. Our results can be examined at the experimental facilities in the future.
\end{abstract}

\maketitle

\section{Introduction}
In hadron physics, quark model has become a well-established tool for the classification of various hadronic states.  Most predictions of the quark model have already been experimentally confirmed, but the quest for  doubly heavy baryons, baryonic states made of two heavy charm/bottom quarks,  has  been conducted  for a long time. These baryonic states  had never been observed  in experiments until 2017, when the LHCb collaboration announced the discovery of $\Xi_{cc}^{++}$ via $\Xi_{cc}^{++} \to \Lambda_c^{+} K^- \pi^+ \pi^+$~\cite{Aaij:2017ueg} with the decay mode  suggested in Ref.~\cite{Yu:2017zst}. This discovery is subsequently confirmed  in 2018 in the $\Xi_{cc}^{++} \to \Xi_c^+ \pi^+$ decay~\cite{Aaij:2018gfl}, and  meanwhile triggered  a series of  experimental investigations~\cite{Aaij:2018wzf,Aaij:2019jfq,Aaij:2019dsx,Aaij:2019uaz}.  Now  studies of doubly-heavy baryons now open a  window to study the hadron spectroscopy and strong interactions  in a baryonic system in the presence of two heavy constituent quarks. 

Among various properties  on doubly heavy baryons,     weak decays are of special importance. In the experimental searches for new type of particles the firstly-discovered ones are usually the ground states, which can only be reconstructed via weak decay final states.  Thus theoretical analysis of  their weak decays can greatly help  to optimize the experimental resources. Meanwhile, there exists rich dynamics in weak decay processes and currently only few theoretical approaches are available, which makes them a wonderland full of challenges and opportunities.

On the theoretical side an ingredient in the weak decay is the transition matrix element of the parent particle to a daughter particle, which can be parameterized as form factors. Fortunately,  there are various available   methods for this part of transition on the market. Thus the  decays of a doubly heavy baryon to a singly heavy baryon transition are studied intensively~\cite{Yu:2017zst,Wang:2017mqp,Hu:2017dzi,Gutsche:2017hux,Sharma:2017txj,Zhao:2018zcb,Zhao:2018mrg,Yu:2019lxw,Gutsche:2019iac,Onishchenko:2000yp,Ebert:2004ck,Ebert:2005ip,Albertus:2006wb,Albertus:2012jt,Dhir:2018twm,Xing:2018lre,Zhang:2018llc,Jiang:2018oak,Ke:2019lcf,Shi:2019hbf,Shi:2019fph,Hu:2019bqj,Gerasimov:2019jwp}. In particular, some of  the form factors are studied  under the light front quark model~\cite{Wang:2017mqp,Zhao:2018mrg,Xing:2018lre,Ke:2019lcf}, QCD sum rules~\cite{Shi:2019hbf} and light cone sum rules~\cite{Shi:2019fph,Hu:2019bqj}.

The light-front quark model is originally developed in meson decays~ \cite{Jaus:1999zv,Jaus:1989au,Jaus:1991cy,Cheng:1996if,Cheng:2003sm,Cheng:2004yj,Ke:2009ed,Ke:2009mn,Cheng:2009ms,Lu:2007sg,Wang:2007sxa,Wang:2008xt,Wang:2008ci,Wang:2009mi,Chen:2009qk,Li:2010bb,Verma:2011yw,Shi:2016gqt,Chang:2019mmh,Chang:2019obq}. During the last decade it was applied to baryon decays with the help of quark-diquark picture~\cite{Ke:2007tg,Wei:2009np,Ke:2012wa,Zhu:2018jet,Ke:2017eqo}, and it   is also interesting to notice that Ref.~\cite{Ke:2019lcf} has adopted the three-quark transition for the form factors.  Under the quark-diquark  picture,  the two spectator quarks play the role of the antiquark in a mesonic system and are treated as a system of spin-$0$ or $1$. In  the calculation the vertex functions are associated with the couplings of a baryon to its quark and diquark constituents.  Following a recent work~\cite{Chua:2018lfa} we revise the vertex function concerned with the spin-$1$ diquark system in this paper, and consequently we update  the form factors of a doubly heavy baryon to a singly heavy baryon transitions under the light front quark model. As argued in a previous work \cite{Zhao:2018mrg}, both spin-$1/2$ to spin-$1/2$ and spin-$1/2$ to spin-$3/2$ transitions are important to the potential discovery channels, and thus both transitions are studied  in this work. Meanwhile, we investigate the charged current induced transitions as well as the flavor changing neutral current (FCNC) induced ones. To be more specific, we will explore  the following transitions:
\begin{enumerate}
	\item  the spin-${1}/{2}$ to spin-${1}/{2}$ transition  with the charged current~\footnote{In the following, we will abbreviate  the spin-$S_1$ to spin-$S_2$ transition as the $S_1\to S_2$ transition. If there is no special note, spin-$1/2$ and spin-$3/2$ are all with positive parity. We will omit the positive sign of positive parity in the following.},
	\begin{itemize}
	\item $c\to d,s$ process,
	\begin{align*}
	&\left.
            \begin{array}{lcl}
		\Xi_{cc}^{++}(ccu) & \to&\Lambda_{c}^{+}(dcu)/\Sigma_{c}^{+}(dcu)/\Xi_{c}^{(\prime)+}(scu),\\
		\Xi_{cc}^{+}(ccd) & \to&\Sigma_{c}^{0}(dcd)/\Xi_{c}^{0}(scd)/\Xi_{c}^{\prime0}(scd),\\
		\Omega_{cc}^{+}(ccs) & \to&\Xi_{c}^{0}(dcs)/\Xi_{c}^{\prime0}(dcs)/\Omega_{c}^{0}(scs),
		\end{array}
       \right.
       \left.
            \begin{array}{lcl}
            \Xi_{bc}^{+}/\Xi_{bc}^{\prime+}(cbu) & \to&\Lambda_{b}^{0}(dbu)/\Sigma_{b}^{0}(dbu)/\Xi_{b}^{(\prime)0}(sbu),\\
	\Xi_{bc}^{0}/\Xi_{bc}^{\prime0}(cbd) & \to&\Sigma_{b}^{-}(dbd)/\Xi_{b}^{-}(sbd)/\Xi_{b}^{\prime-}(sbd),\\
	\Omega_{bc}^{0}/\Omega_{bc}^{\prime0}(cbs) & \to&\Xi_{b}^{-}(dbs)/\Xi_{b}^{\prime-}(dbs)/\Omega_{b}^{-}(sbs);\end{array}
       \right.
	\end{align*}
	\item $b\to u,c$ process,
\begin{align*}
&\left.
            \begin{array}{lcl}
	\Xi_{bb}^{0}(bbu) & \to&\Sigma_{b}^{+}(ubu)/\Xi_{bc}^{+}(cbu)/\Xi_{bc}^{\prime+}(cbu),\\
	\Xi_{bb}^{-}(bbd) & \to&\Lambda_{b}^{0}(ubd)/\Sigma_{b}^{0}(ubd)/\Xi_{bc}^{0}(cbd)/\Xi_{bc}^{\prime0}(cbd),\\
	\Omega_{bb}^{-}(bbs) & \to&\Xi_{b}^{0}(ubs)/\Xi_{b}^{\prime0}(ubs)/\Omega_{bc}^{0}(cbs)/\Omega_{bc}^{\prime0}(cbs),
\end{array}
       \right.\quad
       \left.
            \begin{array}{lcl}
	\Xi_{bc}^{+}/\Xi_{bc}^{\prime+}(bcu) & \to&\Sigma_{c}^{++}(ucu)/\Xi_{cc}^{++}(ccu),\\
	\Xi_{bc}^{0}/\Xi_{bc}^{\prime0}(bcd) & \to&\Lambda_{c}^{+}(ucd)/\Sigma_{c}^{+}(ucd)/\Xi_{cc}^{+}(ccd),\\
	\Omega_{bc}^{0}/\Omega_{bc}^{\prime0}(bcs) & \to&\Xi_{c}^{+}(ucs)/\Xi_{c}^{\prime+}(ucs)/\Omega_{cc}^{+}(ccs);
	\end{array}
       \right.
\end{align*}
\end{itemize}
	\item the   ${1}/{2}\to {1}/{2}$ transition   with FCNC,
\begin{itemize}
		\item $c\to u$ process,
		\begin{align*}
       \left.
            \begin{array}{lcl}
            \Xi_{cc}^{++}(ccu) & \to&\Sigma_{c}^{++}(ucu),\\
		\Xi_{cc}^{+}(ccd) & \to&\Lambda_{c}^{+}(ucd)/\Sigma_{c}^{+}(ucd),\\
		\Omega_{cc}^{+}(ccs) & \to&\Xi_{c}^{+}(ucs)/\Xi_{c}^{\prime+}(ucs),
            \end{array}
       \right.\quad\quad
       \left.
            \begin{array}{lcl}
           \Xi_{cb}^{+}/\Xi_{cb}^{\prime+}(cbu) & \to& \Sigma_{b}^{+}(ubu),\\
		\Xi_{cb}^{0}/\Xi_{cb}^{\prime0}(cbd) & \to& \Lambda_{b}^{0}(ubd)/\Sigma_{b}^{0}(ubd),\\
		\Omega_{cb}^{0}/\Omega_{cb}^{\prime0}(cbs) & \to&\Xi_{b}^{0}(ubs)/\Xi_{b}^{\prime0}(ubs);
            \end{array}
       \right.
        \end {align*}
\end{itemize}
\begin{itemize}
		\item $b\to d,s$ process,
		\begin{align*}
       \left.
            \begin{array}{lcl}
            \Xi_{bb}^{0}(bbu) & \to&\Lambda_{b}^{0}(dbu)/\Sigma_{b}^{0}(dbu)/\Xi_{b}^{(\prime)0}(sbu),\\
		\Xi_{bb}^{-}(bbd) & \to&\Sigma_{b}^{-}(dbd)/\Xi_{b}^{-}(sbd)/\Xi_{b}^{\prime-}(sbd),\\
		\Omega_{bb}^{-}(bbs) & \to&\Xi_{b}^{-}(dbs)/\Xi_{b}^{\prime-}(dbs)/\Omega_{b}^{-}(sbs),
            \end{array}
       \right.\quad\quad
       \left.
            \begin{array}{lcl}
           \Xi_{bc}^{+}/\Xi_{bc}^{\prime+}(bcu) & \to&\Lambda_{c}^{+}(dcu)/\Sigma_{c}^{+}(dcu)/\Xi_{c}^{(\prime)+}(scu),\\
		\Xi_{bc}^{0}/\Xi_{bc}^{\prime0}(bcd) & \to&\Sigma_{c}^{0}(dcd)/\Xi_{c}^{0}(scd)/\Xi_{c}^{\prime0}(scd),\\
		\Omega_{bc}^{0}/\Omega_{bc}^{\prime0}(bcs) & \to&\Xi_{c}^{0}(dcs)/\Xi_{c}^{\prime0}(dcs)/\Omega_{c}^{0}(scs);
            \end{array}
       \right.
        \end {align*}
\end{itemize}
\item the  ${1}/{2}\to {3}/{2}$ transition   induced by the charged current,
\begin{itemize}
	\item $c\to d,s$ process,
	\begin{align*}
	\left.
\begin{array}{lcl}
		\Xi_{cc}^{++}(ccu) & \to&\Sigma_{c}^{*+}(dcu)/\Xi_{c}^{\prime*+}(scu),\\
		\Xi_{cc}^{+}(ccd) & \to&\Sigma_{c}^{*0}(dcd)/\Xi_{c}^{\prime*0}(scd),\\
		\Omega_{cc}^{+}(ccs) & \to&\Xi_{c}^{\prime*0}(dcs)/\Omega_{c}^{*0}(scs),
		\end{array}
       \right.\quad\quad
       \left.
            \begin{array}{lcl}
            \Xi_{bc}^{+}/\Xi_{bc}^{\prime+}(cbu) & \to&\Sigma_{b}^{*0}(dbu)/\Xi_{b}^{\prime*0}(sbu),\\
		\Xi_{bc}^{0}/\Xi_{bc}^{\prime0}(cbd) & \to&\Sigma_{b}^{*-}(dbd)/\Xi_{b}^{\prime*-}(sbd),\\
		\Omega_{bc}^{0}/\Omega_{bc}^{\prime0}(cbs) & \to&\Xi_{b}^{\prime*-}(dbs)/\Omega_{b}^{*-}(sbs).
		\end{array}
       \right.
	\end{align*}
    \item  $b\to u,c$ process,
	\begin{align*}
	\left.
\begin{array}{lcl}
		\Xi_{bb}^{0}(bbu) & \to&\Sigma_{b}^{*+}(ubu)/\Xi_{bc}^{*+}(cbu),\\
		\Xi_{bb}^{-}(bbd) & \to&\Sigma_{b}^{*0}(ubd)/\Xi_{bc}^{*0}(cbd),\\
		\Omega_{bb}^{-}(bbs) & \to&\Xi_{b}^{\prime*0}(ubs)/\Omega_{bc}^{*0}(cbs),
		\end{array}
       \right.\quad\quad
       \left.
            \begin{array}{lcl}
            \Xi_{bc}^{+}/\Xi_{bc}^{\prime+}(bcu) & \to&\Sigma_{c}^{*++}(ucu)/\Xi_{cc}^{*++}(ccu),\\
		\Xi_{bc}^{0}/\Xi_{bc}^{\prime0}(bcd) & \to&\Sigma_{c}^{*+}(ucd)/\Xi_{cc}^{*+}(ccd),\\
		\Omega_{bc}^{0}/\Omega_{bc}^{\prime0}(bcs) & \to&\Xi_{c}^{\prime*+}(ucs)/\Omega_{cc}^{*+}(ccs);
		\end{array}
       \right.
	\end{align*}
\end{itemize}
\item the  ${1}/{2}\to {3}/{2}$ transition   with FCNC,
\begin{itemize}
		\item $c\to u$ process,
		\begin{align*}
       \left.
            \begin{array}{lcl}
            \Xi_{cc}^{++}(ccu) & \to&\Sigma_{c}^{*++}(ucu),\\
		\Xi_{cc}^{+}(ccd) & \to&\Sigma_{c}^{*+}(ucd),\\
		\Omega_{cc}^{+}(ccs) & \to&\Xi_{c}^{\prime*+}(ucs),
            \end{array}
       \right.\quad\quad
       \left.
            \begin{array}{lcl}
           \Xi_{cb}^{+}/\Xi_{cb}^{\prime+}(cbu) & \to& \Sigma_{b}^{*+}(ubu),\\
		\Xi_{cb}^{0}/\Xi_{cb}^{\prime0}(cbd) & \to& \Sigma_{b}^{*0}(ubd),\\
		\Omega_{cb}^{0}/\Omega_{cb}^{\prime0}(cbs) & \to&\Xi_{b}^{\prime*0}(ubs);
            \end{array}
       \right.
        \end {align*}
\end{itemize}
\begin{itemize}
\item $b\to d,s$ process,
\begin{align*}
\left.
\begin{array}{lcl}
\Xi_{bb}^{0}(bbu) & \to\Sigma_{b}^{*0}(dbu)/\Xi_{b}^{\prime*0}(sbu),\\
\Xi_{bb}^{-}(bbd) & \to\Sigma_{b}^{*-}(dbd)/\Xi_{b}^{\prime*-}(sbd),\\
\Omega_{bb}^{-}(bbs) & \to\Xi_{b}^{\prime*-}(dbs)/\Omega_{b}^{*-}(sbs),
\end{array}
       \right.\quad\quad
       \left.
            \begin{array}{lcl}
\Xi_{bc}^{+}/\Xi_{bc}^{\prime+}(bcu) & \to\Sigma_{c}^{*+}(dcu)/\Xi_{c}^{\prime*+}(scu),\\
\Xi_{bc}^{0}/\Xi_{bc}^{\prime0}(bcd) & \to\Sigma_{c}^{*0}(dcd)/\Xi_{c}^{\prime*0}(scd),\\
\Omega_{bc}^{0}/\Omega_{bc}^{\prime0}(bcs) & \to\Xi_{c}^{\prime*0}(dcs)/\Omega_{c}^{*0}(scs);
\end{array}
       \right.
\end{align*}
\end{itemize}
\end{enumerate}
In the above, the quark components have been explicitly given  in the brackets,
in which the first quarks denote the quarks participating in the  weak decays.
The initial baryons are all doubly heavy baryons.
The spin-parity $J^{P}$ quantum numbers of the doubly heavy baryons has been listed in Tab.~\ref{tab:JPC}.

\begin{table}[!htb]
\footnotesize
\caption{The spin-parity $J^{P}$ quantum numbers and quark composition for doubly heavy baryons. The symbol $S_{h}^{\pi}$ indicates the spin-parity of the system consisting of two heavy quarks. The light quark $q$ represents $u,d$ quark.}\label{tab:JPC}
\begin{center}
\begin{tabular}{cccc|cccccc} \hline \hline
Baryon      & Quark Content  &  $S_h^\pi$  &$J^P$   & Baryon & Quark Content &   $S_h^\pi$  &$J^P$   \\ \hline
$\Xi_{cc}$ & $\{cc\}q$  & $1^+$ & $1/2^+$ &   $\Xi_{bb}$ & $\{bb\}q$  & $1^+$ & $1/2^+$ & \\
$\Xi_{cc}^*$ & $\{cc\}q$  & $1^+$ & $3/2^+$ &   $\Xi_{bb}^*$ & $\{bb\}q$  & $1^+$ & $3/2^+$ & \\ \hline
$\Omega_{cc}$ & $\{cc\}s$  & $1^+$ & $1/2^+$ &   $\Omega_{bb}$ & $\{bb\}s$  & $1^+$ & $1/2^+$ & \\
$\Omega_{cc}^*$ & $\{cc\}s$  & $1^+$ & $3/2^+$ &   $\Omega_{bb}^*$ & $\{bb\}s$  & $1^+$ & $3/2^+$ &  \\ \hline
$\Xi_{bc}'$ & $[bc]q$  & $0^+$ & $1/2^+$ &   $\Omega_{bc}'$ & $[bc]s$  & $0^+$ & $1/2^+$ & \\
$\Xi_{bc}$ & $\{bc\}q$  & $1^+$ & $1/2^+$ &   $\Omega_{bc}$ & $\{bc\}s$  & $1^+$ & $1/2^+$ & \\
$\Xi_{bc}^*$ & $\{bc\}q$  & $1^+$ & $3/2^+$ &   $\Omega_{bc}^*$ & $\{bc\}s$  & $1^+$ & $3/2^+$ &
 \\ \hline \hline
\end{tabular}
\end{center}
\end{table}

\begin{figure}[!]
\includegraphics[width=0.3\columnwidth]{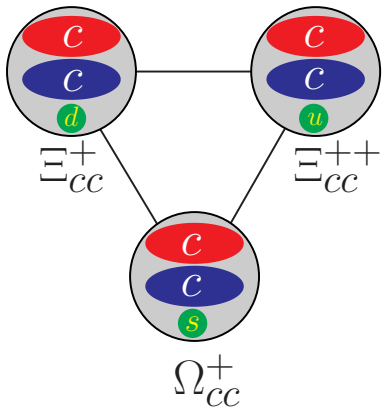}
\caption{Spin-$1/2$ doubly charmed baryons. It is similar for the doubly bottom baryons and the bottom-charm baryons.}
\label{fig:doubly_heavy_baron}
\end{figure}

\begin{figure}[!]
\includegraphics[width=0.6\columnwidth]{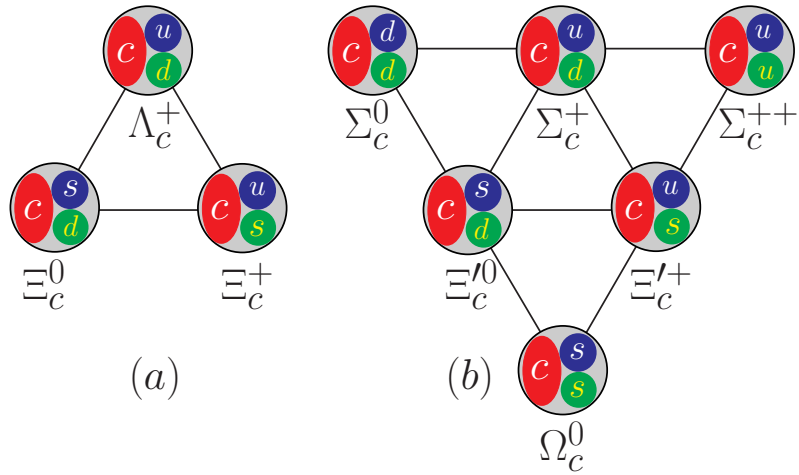}
\caption{Spin-$1/2$ singly charmed baryons. Here (a) represents SU(3) anti-triplets $\boldsymbol{\bar{3}}$ and (b) represents SU(3) sextets $\boldsymbol{6}$. The spin-3/2 singly charmed baryons only have SU(3) sextets $\boldsymbol{6}$ as shown by panel (b) just with the replacement ``${\cal B}_{c}\to {\cal B}_{c}^{*}$".
For spin-$1/2$ and spin-$3/2$ singly bottomed baryons, a replacement $c\to b$ is needed.}
\label{fig:singly_heavy}
\end{figure}

The lowest-lying  doubly heavy baryons with $J^{P}=1/2^{+}$ for example the doubly charm SU(3) triplets $\Xi_{cc}^{++}$ ($ccu$), $\Xi_{cc}^{+}$ ($ccd$), and $\Omega^{+}_{cc}$($ccs$) shown in Fig.~\ref{fig:doubly_heavy_baron},  can only weak decay.
Three doubly bottom baryons $\Xi_{bb}^{0}$ ($bbu$), $\Xi_{bb}^{-}$ ($bbd$), and $\Omega^{-}_{bb}$($bbs$) can also constitute one  SU(3) triplet  similar to Fig.~\ref{fig:doubly_heavy_baron} with the replacement $c\to b$. While the bottom-charm baryons could form two sets of SU(3) triplets, ($\Xi_{bc},\Omega_{bc}$) and ($\Xi_{bc}^{\prime},\Omega_{bc}^{\prime}$). The difference between the two sets is the different total spin of $bc$ system as shown in Tab.~\ref{tab:JPC}, In fact there could be   mixing between them. However, the detailed mixing scheme between the two triplets is still unclear, the initial baryons include two triplets, ($\Xi_{bc},\Omega_{bc}$) and ($\Xi_{bc}^{\prime},\Omega_{bc}^{\prime}$) in this work. The doubly heavy baryons with $J^{P}=3/2^{+}$ can decay into the lowest-lying ones radiatively if the mass splitting is small, or decay into the lowest-lying ones with the emission of a light pion when they are   heavy enough. The final baryons include doubly heavy baryons and singly heavy baryons. The singly heavy baryons can compose one SU(3) anti-triplets $\boldsymbol{\bar{3}}$ and one SU(3) sextet $\boldsymbol{6}$ as shown in Fig.~\ref{fig:singly_heavy}.
Taking the transition ${\cal B}_{bc}\to {\cal B}_{c}$ with $b\to s$ as an example, the final baryons $\Xi_{c}^{\prime+}$, $\Xi_{c}^{\prime0}$ and $\Omega_{c}^{0}$ belong to the presentation of $\boldsymbol{6}$, while $\Xi_{c}^{+}$ and $\Xi_{c}^{0}$ are included in the $\boldsymbol{\bar{3}}$, as can be seen from Fig.~\ref{fig:singly_heavy}.

This paper is organized as follows. In Sec.~II, we will present the framework of the light-front approach under the diquark picture,  and then the flavor-spin wave functions will be discussed. In the appendix, we will provide a new approach to derive the flavor-spin factors.  Numerical results of various transition  form factors are shown in this section. In Sec.~III,   phenomenological applications of the doubly heavy baryon decays will be carried out, including numerical results of the decay widths, branching ratios and $\Gamma_{L}/\Gamma_{T}$s of the semileptonic weak decays of doubly heavy baryons. The SU(3) symmetry breaking effect and error estimations will be also discussed in Sec.~III. A brief summary is given in the last section. The appendix  also contains some brief description of the flavor-spin wave functions, and  helicity amplitudes.

\section{Theoretical framework}

The theoretical framework for the charged current and FCNC induced baryonic   transitions    will be briefly introduced in this section, including the definitions of the states for spin-$1/2$ and $3/2$ baryons, and the extraction of the transition form factors. More details can be found in Refs.~\cite{Ke:2017eqo,Ke:2007tg}.
Flavor-spin wave functions will be given in the second subsection, while a new derivation is given in the appendices.

\subsection{Light-front quark model}\label{subsec_lightfrontquarkmodel}


\begin{figure}[htp]
\includegraphics[width=0.4\columnwidth]{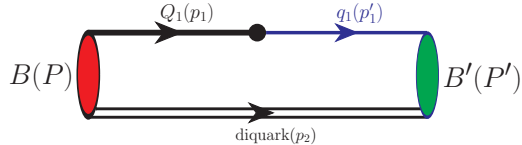} \caption{Feynman diagram for doubly heavy baryons $B$ into a spin-1/2 and spin-3/2 ground-state baryons $B^{\prime}$ with two spectator quarks as a diquark. Here $P$ and $P^{\prime}$ are the momentum of the initial and final baryons, respectively. In quark level, the transition is one heavy quark $Q_{1}$ with momentum $p_{1}$ decays into a lighter quark $q_{1}$ with momentum $p_{1}^{\prime}$, and the diquark with momentum $p_{2}$. The black ball means the weak interaction vertex.}
\label{fig:decay}
\end{figure}

For the $J^{P}=1/2^{+}$ baryon states, their wave functions in light-front quark model can be written as
\begin{eqnarray}
	|{\cal B}(P,S,S_{z})\rangle & = & \int\{d^{3}p_{1}\}\{d^{3}p_{2}\}2(2\pi)^{3}\delta^{3}(\tilde{P}-\tilde{p}_{1}-\tilde{p}_{2})\nonumber \\
	&  & \times\sum_{\lambda_{1},\lambda_{2}}\Psi^{SS_{z}}(\tilde{p}_{1},\tilde{p}_{2},\lambda_{1},\lambda_{2})|Q_{1}(p_{1},\lambda_{1})({\rm{di}})(p_{2},\lambda_{2})\rangle,\label{eq:state_vector}
	\end{eqnarray}
here $Q_{1}=b,c$ is initial heavy quark, and ``$({\rm{di}})$" presents the diquark shown in Fig.~\ref{fig:decay}. $\Psi^{SS_{z}}$ is the momentum-space wave function and can be shown with the following equation,
	\begin{equation}
	\Psi^{SS_{z}}(\tilde{p}_{1},\tilde{p}_{2},\lambda_{1},\lambda_{2})=\frac{1}{\sqrt{2(p_{1}\cdot\bar{P}+m_{1}M_{0})}}\bar{u}(p_{1},\lambda_{1})\Gamma_{S(A)} u(\bar{P},S_{z})\phi(x,k_{\perp}),\label{eq:momentum_wave_function_1/2}
	\end{equation}
here $\Gamma$ is the coupling vertex of the decay quark $Q_{1}$ and the diquark in the baryon state, and when the diquark is a scalar diquark, the coupling vertex is defined as $\Gamma_{S}=1$.
In Ref.~\cite{Chua:2018lfa}, when an axial-vector diquark is involved,
the vertex should be
\begin{align}
	\Gamma_{A} & =\frac{\gamma_{5}}{\sqrt{3}}\left(\slashed\epsilon^{*}(p_{2},\lambda_{2})-\frac{M_0+m_1+m_2}{\bar{P}\cdot p_2+m_2M_0}\epsilon^{*}(p_{2},\lambda_{2})\cdot\bar{P}\right).\label{eq:momentum_wave_function_1/2gamma}
	\end{align}
In Eq.~(\ref{eq:momentum_wave_function_1/2}), $\phi$ is a Gaussian-type function:
	\begin{equation}
	\phi=4\left(\frac{\pi}{\beta^{2}}\right)^{3/4}\sqrt{\frac{e_{1}e_{2}}{x_{1}x_{2}M_{0}}}\exp\left(\frac{-\vec{k}^{2}}{2\beta^{2}}\right).\label{eq:Gauss}
	\end{equation}

In analogy to the $1/2^{+}$ baryon case, a  $3/2^{+}$
baryon state has a similar expression to Eq. (\ref{eq:state_vector}) except a different coupling vertex:
\begin{equation}
\Psi^{SS_{z}}(\tilde{p}_{1},\tilde{p}_{2},\lambda_{1},\lambda_{2})=
\frac{1}{\sqrt{2(p_{1}\cdot\bar{P}+m_{1}M_{0})}}\bar{u}(p_{1},\lambda_{1})\Gamma^{\alpha}_{A}(p_{2},\lambda_{2})u_{\alpha}(\bar{P},S_{z})\phi(x,k_{\perp}),\label{eq:momentum_wave_fuction_3/2}
\end{equation}
where
\begin{equation}
	\Gamma_{A}^{\alpha}=-\left(\epsilon^{*\alpha}(p_{2},\lambda_{2})-\frac{p_2^{\alpha}}{\bar{P}\cdot p_2+m_2M_0}\epsilon^{*}(p_{2},\lambda_{2})\cdot\bar{P}\right).
\end{equation}
With the help of Eqs. (\ref{eq:state_vector}) and (\ref{eq:momentum_wave_function_1/2}),  the spin-$1/2$ to spin-$1/2$ transition matrix element with $\rm{(V-A)}$ and tensor current can be derived as
\begin{eqnarray}
&  & \langle{\cal B}^{\prime}_{f}(P^{\prime},S^{\prime}=\frac{1}{2}, S_{z}^{\prime})|\bar{q}_{1}\gamma_{\mu}(1-\gamma_{5})Q_{1}|{\cal B}_{i}(P, S=\frac{1}{2}, S_{z})\rangle\nonumber \\
& = & \int\{d^{3}p_{2}\}\frac{\phi^{\prime}(x^{\prime},k_{\perp}^{\prime})\phi(x,k_{\perp})}{2\sqrt{p_{1}^{+}p_{1}^{\prime+}(p_{1}\cdot\bar{P}+m_{1}M_{0})(p_{1}^{\prime}\cdot\bar{P}^{\prime}+m_{1}^{\prime}M_{0}^{\prime})}}\nonumber \\
&  & \times\sum_{\lambda_{2}}\bar{u}(\bar{P}^{\prime},S_{z}^{\prime})\bar{\Gamma}^{\prime}_{S(A)}(\slashed p_{1}^{\prime}+m_{1}^{\prime})\gamma_{\mu}(1-\gamma_{5})(\slashed p_{1}+m_{1})\Gamma_{S(A)} u(\bar{P},S_{z}),\label{eq:matrix_element_half}\\
&  & \langle{\cal B}^{\prime}_{f}(P^{\prime},S^{\prime}=\frac{1}{2}, S_{z}^{\prime})|\bar{q}_{1}i\sigma_{\mu\nu}\frac{q^{\nu}}{M}(1+\gamma_{5})Q_{1}|{\cal B}_{i}(P, S=\frac{1}{2}, S_{z})\rangle\nonumber \\
& = & \int\{d^{3}p_{2}\}\frac{\phi^{\prime}(x^{\prime},k_{\perp}^{\prime})\phi(x,k_{\perp})}{2\sqrt{p_{1}^{+}p_{1}^{\prime+}(p_{1}\cdot\bar{P}+m_{1}M_{0})(p_{1}^{\prime}\cdot\bar{P}^{\prime}+m_{1}^{\prime}M_{0}^{\prime})}}\nonumber \\
&  & \times\sum_{\lambda_{2}}\bar{u}(\bar{P}^{\prime},S_{z}^{\prime})\bar{\Gamma}^{\prime}_{S(A)}(\slashed p_{1}^{\prime}+m_{1}^{\prime})i\sigma_{\mu\nu}\frac{q^{\nu}}{M}(1+\gamma_{5})(\slashed p_{1}+m_{1})\Gamma_{S(A)} u(\bar{P},S_{z}).\label{eq:matrix_element_half_tensor}
\end{eqnarray}
With the help of Eqs. (\ref{eq:state_vector}), (\ref{eq:momentum_wave_function_1/2})
and (\ref{eq:momentum_wave_fuction_3/2}), the spin-1/2 to spin-$3/2$ transition matrix element with $\rm{(V-A)}$ and tensor current can be derived as
\begin{eqnarray}
&  & \langle{\cal B}^{\prime*}_{f}(P^{\prime},S^{\prime}=\frac{3}{2},S_{z}^{\prime})|\bar{q}_{1}\gamma^{\mu}(1-\gamma_{5})Q_{1}|{\cal B}_{i}(P,S=\frac{1}{2},S_{z})\rangle\nonumber \\
& = & \int\{d^{3}p_{2}\}\frac{\phi^{\prime}(x^{\prime},k_{\perp}^{\prime})\phi(x,k_{\perp})}{2\sqrt{p_{1}^{+}p_{1}^{\prime+}(p_{1}\cdot\bar{P}+m_{1}M_{0})(p_{1}^{\prime}\cdot\bar{P}^{\prime}+m_{1}^{\prime}M_{0}^{\prime})}}\nonumber \\
&  & \times\sum_{\lambda_{2}}\bar{u}_{\alpha}(\bar{P}^{\prime},S_{z}^{\prime})\left[\bar{\Gamma}^{\prime\alpha}_{A}(\slashed p_{1}^{\prime}+m_{1}^{\prime})\gamma^{\mu}(1-\gamma_{5})(\slashed p_{1}+m_{1})\Gamma_{A}\right]u(\bar{P},S_{z}),\label{eq:matrix_element_onehalf}\\
&  & \langle{\cal B}^{\prime*}_{f}(P^{\prime},S^{\prime}=\frac{3}{2},S_{z}^{\prime})|\bar{q}_{1}i\sigma_{\mu\nu}\frac{q^{\nu}}{M}(1+\gamma_{5})Q_{1}|{\cal B}_{i}(P,S=\frac{1}{2},S_{z})\rangle\nonumber \\
& = & \int\{d^{3}p_{2}\}\frac{\phi^{\prime}(x^{\prime},k_{\perp}^{\prime})\phi(x,k_{\perp})}{2\sqrt{p_{1}^{+}p_{1}^{\prime+}(p_{1}\cdot\bar{P}+m_{1}M_{0})(p_{1}^{\prime}\cdot\bar{P}^{\prime}+m_{1}^{\prime}M_{0}^{\prime})}}\nonumber \\
&  & \times\sum_{\lambda_{2}}\bar{u}_{\alpha}(\bar{P}^{\prime},S_{z}^{\prime})\left[\bar{\Gamma}^{\prime\alpha}_{A}(\slashed p_{1}^{\prime}+m_{1}^{\prime})i\sigma_{\mu\nu}\frac{q^{\nu}}{M}(1+\gamma_{5})(\slashed p_{1}+m_{1})\Gamma_{A}\right]u(\bar{P},S_{z}).\label{eq:matrix_element_onehalf_tensor}
\end{eqnarray}
In Eqs.~(\ref{eq:momentum_wave_function_1/2})-(\ref{eq:momentum_wave_function_1/2gamma})
and (\ref{eq:momentum_wave_fuction_3/2})-(\ref{eq:matrix_element_onehalf_tensor}),
\begin{equation}
m_{1}=m_{Q},\quad m_{1}^{\prime}=m_{q},\quad m_{2}=m_{(di)},
\end{equation}
and $p_{1}$ and $p_{1}^{\prime}$ are the four-momentum of the initial and final quark, respectively. $P$ and $P^{\prime}$ are
the four-momentum of the initial baryons ${\cal B}$ and final baryon states ${\cal B}^{\prime}$, respectively.
$q_{1}=u,d,s,c$ means the lighter quark in the final states shown in Fig.~\ref{fig:decay}. When the diquark is a scalar diquark, the coupling vertex is defined as,
\begin{equation}
\Gamma_{S}=\bar{\Gamma}^{\prime}_{S}=1, \label{scalar diquark}
\end{equation}
and when an axial-vector diquark is involved, the vertex should be
\begin{align}
\bar{\Gamma}^{\prime}_{A} & =\frac{1}{\sqrt{3}}\left(-\slashed\epsilon(p_{2},\lambda_{2})+\frac{M_0^{\prime}+m_1^{\prime}+m_2}{\bar{P}^{\prime}\cdot p_2+m_2M_0^{\prime}}\epsilon(p_{2},\lambda_{2})\cdot\bar{P}^{\prime}\right)\gamma_{5},\label{axial-vector diquarkprime}
\end{align}
and
\begin{equation}
	\bar{\Gamma}_{A}^{\prime\alpha}=-\left(\epsilon^{\alpha}(p_{2},\lambda_{2})-\frac{p_2^{\alpha}}{\bar{P}^{\prime}\cdot p_2+m_2M_0^{\prime}}\epsilon(p_{2},\lambda_{2})\cdot\bar{P}^{\prime}\right).
\end{equation}
The $1/2\to1/2$ transition matrix elements can be parameterized as
\begin{eqnarray}
&&\langle{\cal B}^{\prime}_{f}(P^{\prime}, S^{\prime}=\frac{1}{2},S_{z}^{\prime})|\bar{q}_{1}\gamma_{\mu}(1-\gamma_{5})Q_{1}|{\cal B}_{i}(P,S=\frac{1}{2},S_{z})\rangle\nonumber \\
& &=  \bar{u}(P^{\prime},S_{z}^{\prime})
\Big[\gamma_{\mu}f_{1,S(A)}(q^{2})
+i\sigma_{\mu\nu}\frac{q^{\nu}}{M}f_{2,S(A)}(q^{2})
+\frac{q_{\mu}}{M}f_{3,S(A)}(q^{2})\Big]u(P,S_{z})\nonumber \\
 && \quad-  \bar{u}(P^{\prime},S_{z}^{\prime})
\Big[\gamma_{\mu}g_{1,S(A)}(q^{2})
+i\sigma_{\mu\nu}\frac{q^{\nu}}{M}g_{2,S(A)}(q^{2})
+\frac{q_{\mu}}{M}g_{3,S(A)}(q^{2})\Big]\gamma_{5}u(P,S_{z}),\label{eq:matrix_element_2}
\end{eqnarray}
\begin{eqnarray}
&&\langle{\cal B}^{\prime}_{f}(P^{\prime},S^{\prime}=\frac{1}{2},S_{z}^{\prime})
|\bar{q}_{1}i\sigma_{\mu\nu}\frac{q^{\nu}}{M}(1+\gamma_{5})Q_{1}|{\cal B}_{i}(P,S=\frac{1}{2},S_{z})\rangle \nonumber \\
&&=\bar{u}(P^{\prime},S_{z}^{\prime})
\Big[\frac{f_{1,S(A)}^{T}(q^2)}{M(M^{\prime}-M)}(q^2\gamma_{\mu}-{\slashed q} q_{\mu})
+i\sigma_{\mu\nu}\frac{q^{\nu}}{M}f_{2,S(A)}^{T}(q^{2})\Big]u(P,S_{z})\nonumber \\
& &\quad+ \bar{u}(P^{\prime},S_{z}^{\prime})
\Big[\frac{g_{1,S(A)}^{T}(q^2)}{M(M+M^{\prime})}(q^2\gamma_{\mu}-{\slashed q} q_{\mu})+i\sigma_{\mu\nu}\frac{q^{\nu}}{M}g_{2,S(A)}^{T}(q^{2})
\Big]\gamma_{5}u(P,S_{z}).\label{eq:matrix_element_2p}
\end{eqnarray}
	
Then the extraction of these form factors $f_{1,2,3,S(A)}$ can be performed as Ref.~\cite{Ke:2017eqo}.
Multiply $\bar{u}(\bar{P},S_{z})(\bar{\Gamma}^{\mu})_{i}u(\bar{P}^{\prime},S_{z}^{\prime})$
and $\bar{u}(P,S_{z})(\Gamma^{\mu})_{i}u(P^{\prime},S_{z}^{\prime})$
on the $\langle{\cal B}^{\prime}_{f}(P^{\prime},S^{\prime}=\frac{1}{2},S_{z}^{\prime})
|\bar{q}_{1}\gamma_{\mu}Q_{1}|{\cal B}_{i}(P,S=\frac{1}{2},S_{z})\rangle$
part of Eq.~(\ref{eq:matrix_element_half}) and Eq.~(\ref{eq:matrix_element_2}), respectively.
At the same time, the approximation $P^{(\prime)}\to\bar{P}^{(\prime)}$ need to be taken for the integral.
After summing the polarizations of the initial and final baryon states up, we can get the three linear equations as follows,
\begin{eqnarray}
&  & {\rm Tr}\Big\{(\Gamma^{\mu})_{i}(\slashed P^{\prime}+M^{\prime})\Big[\gamma_{\mu}f_{1,S(A)}(q^2)
+i\sigma_{\mu\nu}\frac{q^{\nu}}{M}f_{2,S(A)}(q^2)
+\frac{q_{\mu}}{M}f_{3,S(A)}(q^2)\Big](\slashed P+M)\Big\}\nonumber \\
& = & \int\{d^{3}p_{2}\}
\frac{\phi^{\prime}(x^{\prime},k_{\perp}^{\prime})\phi(x,k_{\perp})}
{2\sqrt{p_{1}^{+}p_{1}^{\prime+}
(p_{1}\cdot\bar{P}+m_{1}M_{0})(p_{1}^{\prime}\cdot\bar{P}^{\prime}
+m_{1}^{\prime}M_{0}^{\prime})}}\nonumber \\
&  & \times\sum_{\lambda_{2}}{\rm Tr}\left[(\bar{\Gamma}^{\mu})_{i}(\bar{\slashed P}^{\prime}+M_{0}^{\prime})\bar{\Gamma}^{\prime}_{S(A)}(\slashed p_{1}^{\prime}+m_{1}^{\prime})\gamma_{\mu}(\slashed p_{1}+m_{1})\Gamma_{S(A)}(\bar{\slashed P}+M_{0})\right],
\label{eq:solve_fi}
\end{eqnarray}
with $(\bar{\Gamma}^{\mu})_{i}=\{\gamma^{\mu},\bar{P}^{\mu},\bar{P}^{\prime\mu}\}$ and $(\Gamma^{\mu})_{i}=\{\gamma^{\mu},P^{\mu},P^{\prime\mu}\}$.
Using the above Eq.~(\ref{eq:solve_fi}), we can get the specific expression of the form factors $f_{1,2,3,S(A)}$ as follows:
\begin{eqnarray}
f_1&=&\frac{q^2 [B_1 (M+M')^2-2 B_2 (2 M+M')-2 B_3 (M+2
   M')]-B_1 (q^2)^2-2 (B_2-B_3) (M-M') (M+M')^2}{4
 [(M-M')^2-q^2] [(M+M')^2-q^2]^2},\notag\\
f_2&=&\frac{M (M+M') [B_1 (M+M')^2+2 B_2 (M'-2 M)+2 B_3 (M-2 M')]-M
   q^2 [B_1 (M+M')+2 B_2+2 B_3]}{4
 [(M-M')^2-q^2] [(M+M')^2-q^2]^2},\notag\\
f_3&=&\frac{M \{B_1 (M-M') [q^2-(M+M')^2]+2 B_2 \left(4 M^2-M
   M'+M'^2-q^2\right)+2 B_3 \left(-M^2+M M'-4 M'^2+q^2\right)\}}{4
 [(M-M')^2-q^2] [(M+M')^2-q^2]^2},\notag
\end{eqnarray}
with
\begin{eqnarray}
B_i&=&\int\{d^{3}p_{2}\}
\frac{\phi^{\prime}(x^{\prime},k_{\perp}^{\prime})\phi(x,k_{\perp})}
{2\sqrt{p_{1}^{+}p_{1}^{\prime+}
(p_{1}\cdot\bar{P}+m_{1}M_{0})(p_{1}^{\prime}\cdot\bar{P}^{\prime}
+m_{1}^{\prime}M_{0}^{\prime})}}\nonumber \\
&  &\times\sum_{\lambda_{2}}{\rm Tr}\left[(\bar{\Gamma}^{\mu})_{i}(\bar{\slashed P}^{\prime}+M_{0}^{\prime})\bar{\Gamma}^{\prime}_{S(A)}(\slashed p_{1}^{\prime}+m_{1}^{\prime})\gamma_{\mu}(\slashed p_{1}+m_{1})\Gamma_{S(A)}(\bar{\slashed P}+M_{0})\right].
\end{eqnarray}
The form factors $g_{1,2,3,S(A)}$ can be calculated using the similar process,
\begin{eqnarray}
g_1&=&\frac{q^2 [R_1 (M-M')^2+2 R_2 (M'-2 M)-2 R_3 (M-2
   M')]-R_1 (q^2)^2-2 (R_2-R_3) (M+M') (M-M')^2}{4
   [(M-M')^2-q^2]^2 [(M+M')^2-q^2]},\notag\\
g_2&=&\frac{M q^2 [R_1 (M-M')+2 R_2+2 R_3]-M (M-M') [R_1 (M-M')^2-2
   R_2 (2 M+M')+2 R_3 (M+2 M')]}{4
   [(M-M')^2-q^2]^2 [(M+M')^2-q^2]},\notag\\
g_3&=&\frac{M \{R_1 (M+M') [(M-M')^2-q^2]-2 R_2 \left(4 M^2+M
   M'+M'^2-q^2\right)+2 R_3 \left(M^2+M M'+4 M'^2-q^2\right)\}}{4
   [(M-M')^2-q^2]^2 [(M+M')^2-q^2]},\notag
 \end{eqnarray}
with
\begin{eqnarray}
R_i&=&\int\{d^{3}p_{2}\}
\frac{\phi^{\prime}(x^{\prime},k_{\perp}^{\prime})\phi(x,k_{\perp})}
{2\sqrt{p_{1}^{+}p_{1}^{\prime+}
(p_{1}\cdot\bar{P}+m_{1}M_{0})(p_{1}^{\prime}\cdot\bar{P}^{\prime}
+m_{1}^{\prime}M_{0}^{\prime})}}\nonumber \\
& &\times\sum_{\lambda_{2}}{\rm Tr}\left[(\bar{\Gamma}^{\mu})_{i}(\bar{\slashed P}^{\prime}+M_{0}^{\prime})\bar{\Gamma}^{\prime}_{S(A)}(\slashed p_{1}^{\prime}+m_{1}^{\prime})\gamma_{\mu}\gamma_{5}(\slashed p_{1}+m_{1})\Gamma_{S(A)}(\bar{\slashed P}+M_{0})\right].
\end{eqnarray}
Then tensor form factors $f_{1,2,S(A)}^{T}$ or $g_{1,2,S(A)}^{T}$ defined by Eq.~(\ref{eq:matrix_element_2p}) can also be extracted in the similar way with the form factors $f_{1,2,3,S(A)}$ and $g_{1,2,3,S(A)}$, the differences are only $(\Gamma^{\mu})_{i}=\{\gamma^{\mu},P^{\mu}\}$ and $(\bar{\Gamma}^{\mu})_{i}=\{\gamma^{\mu},\bar{P}^{\mu}\}$,
\begin{eqnarray}
f_1^T&=&\frac{M (M-M') \{B^T_1 [q^2-(M+M')^2]+6 B^T_2 (M+M')\}}{4
 [(M-M')^2-q^2] [(M+M')^2-q^2]^2},\notag\\
f_2^T&=&\frac{M (M+M')^2 [B^T_1 (M+M')-2 B^T_2]-M q^2 [B^T_1 (M+M')+4 B^T_2]}{4
 [(M-M')^2-q^2] [(M+M')^2-q^2]^2},\notag\\
g_1^T&=&\frac{M (M+M') \{R^T_1 [(M-M')^2-q^2]+6 R^T_2 (M'-M)\}}{4
   [(M-M')^2-q^2]^2 [(M+M')^2-q^2]},\notag\\
g_2^T&=&\frac{M q^2 [R^T_1 (M-M')+4 R^T_2]+M (M-M')^2 [R^T_1 (M'-M)+2 R^T_2]}{4
   [(M-M')^2-q^2]^2 [(M+M')^2-q^2]},
\end{eqnarray}
with
\begin{eqnarray}
B_i^T&=&\int\{d^{3}p_{2}\}
\frac{\phi^{\prime}(x^{\prime},k_{\perp}^{\prime})\phi(x,k_{\perp})}
{2\sqrt{p_{1}^{+}p_{1}^{\prime+}
(p_{1}\cdot\bar{P}+m_{1}M_{0})(p_{1}^{\prime}\cdot\bar{P}^{\prime}
+m_{1}^{\prime}M_{0}^{\prime})}}\nonumber \\
&  &\times\sum_{\lambda_{2}}{\rm Tr}\left[(\bar{\Gamma}^{\mu})_{i}(\bar{\slashed P}^{\prime}+M_{0}^{\prime})\bar{\Gamma}^{\prime}_{S(A)}(\slashed p_{1}^{\prime}+m_{1}^{\prime})i\sigma_{\mu\nu}\frac{q^{\nu}}{M}(\slashed p_{1}+m_{1})\Gamma_{S(A)}(\bar{\slashed P}+M_{0})\right],\notag\\
R_i^T&=&\int\{d^{3}p_{2}\}
\frac{\phi^{\prime}(x^{\prime},k_{\perp}^{\prime})\phi(x,k_{\perp})}
{2\sqrt{p_{1}^{+}p_{1}^{\prime+}
(p_{1}\cdot\bar{P}+m_{1}M_{0})(p_{1}^{\prime}\cdot\bar{P}^{\prime}
+m_{1}^{\prime}M_{0}^{\prime})}}\nonumber \\
& & \times\sum_{\lambda_{2}}{\rm Tr}\left[(\bar{\Gamma}^{\mu})_{i}(\bar{\slashed P}^{\prime}+M_{0}^{\prime})\bar{\Gamma}^{\prime}_{S(A)}(\slashed p_{1}^{\prime}+m_{1}^{\prime})i\sigma_{\mu\nu}\frac{q^{\nu}}{M}\gamma_{5}(\slashed p_{1}+m_{1})\Gamma_{S(A)}(\bar{\slashed P}+M_{0})\right].
\end{eqnarray}

The $1/2\to3/2$ transition matrix elements can be parameterized in a similar form as follows.
\begin{eqnarray}
&&	\langle{\cal B}^{\prime*}_{f}(P^{\prime},S^{\prime}=\frac{3}{2},S_{z}^{\prime})|\bar{q}_{1}\gamma^{\mu}(1-\gamma_{5})Q_{1}|{\cal B}_{i}(P,S=\frac{1}{2},S_{z})\rangle \nonumber\\
&& =  \bar{u}_{\alpha}(P^{\prime},S_{z}^{\prime})\Big[\mathtt{f}_{1}(q^2)\frac{P^{\alpha}}{M}(\gamma^{\mu}-\frac{\slashed q}{q^2}q^{\mu})+\mathtt{f}_{2}(q^2)\frac{P^{\alpha}}{M^2}(\frac{M^2-M^{\prime 2}}{q^2}q^{\mu}-{\cal P}^{\mu})\nonumber\\
&&\quad\quad\quad\quad\quad\quad
+\mathtt{f}_{3}(q^2)\frac{P^{\alpha}}{M^2}\frac{M^2-M^{\prime 2}}{q^2}q^{\mu}+\mathtt{f}_{4}(q^{2})(g^{\alpha\mu}-\frac{q^{\alpha}q^{\mu}}{q^2})\Big]\gamma_{5}u(P,S_{z})\nonumber\\
& &\quad- \bar{u}_{\alpha}(P^{\prime},S_{z}^{\prime})\Big[\mathtt{g}_{1}(q^2)P^{\alpha}(\gamma^{\mu}-\frac{\slashed q}{q^2}q^{\mu})+\mathtt{g}_{2}(q^2)\frac{P^{\alpha}}{M^2}(\frac{M^2-M^{\prime 2}}{q^2}q^{\mu}-{\cal P}^{\mu})\nonumber\\
&&\quad\quad\quad\quad\quad\quad
+\mathtt{g}_{3}(q^2)\frac{P^{\alpha}}{M^2}\frac{M^2-M^{\prime 2}}{q^2}q^{\mu}+\mathtt{g}_{4}(q^{2})(g^{\alpha\mu}-\frac{q^{\alpha}q^{\mu}}{q^2})\Big]u(P,S_{z}),\label{eq:matrix_element_32nVA}
\end{eqnarray}
\begin{eqnarray}
&&	\langle{\cal B}_{f}^{\prime*}(P^{\prime},S^{\prime}=\frac{3}{2},S_{z}^{\prime})|\bar{q}_{1}i\sigma_{\mu\nu}\frac{q^{\nu}}{M}(1+\gamma_{5})Q_{1}|{\cal B}_{i}(P,S=\frac{1}{2},S_{z})\rangle\nonumber\\
&& =  \bar{u}_{\alpha}(P^{\prime},S_{z}^{\prime})
\Big[\mathtt{f}_{1}^{T}(q^2)\frac{P^{\alpha}}{M}(\gamma^{\mu}-\frac{\slashed q}{q^2}q^{\mu})+\mathtt{f}_{2}^{T}(q^2)\frac{P^{\alpha}}{M^2}(\frac{M^2-M^{\prime 2}}{q^2}q^{\mu}-{\cal P}^{\mu}) \nonumber\\
&&\quad\quad\quad\quad\quad\quad +\mathtt{f}_{3}^{T}(q^2)\frac{P^{\alpha}}{M^2}\frac{M^2-M^{\prime 2}}{q^2}q^{\mu}+\mathtt{f}_{4}^{T}(q^{2})(g^{\alpha\mu}-\frac{q^{\alpha}q^{\mu}}{q^2})\Big]\gamma_{5}u(P,S_{z})\nonumber\\
& &\quad + \bar{u}_{\alpha}(P^{\prime},S_{z}^{\prime})\Big[\mathtt{g}_{1}^{T}(q^2)\frac{P^{\alpha}}{M}(\gamma^{\mu}-\frac{\slashed q}{q^2}q^{\mu})+\mathtt{g}_{2}^{T}(q^2)\frac{P^{\alpha}}{M^2}(\frac{M^2-M^{\prime 2}}{q^2}q^{\mu}-{\cal P}^{\mu})\nonumber\\
&&\quad\quad\quad\quad\quad\quad
+\mathtt{g}_{3}^{T}(q^2)\frac{P^{\alpha}}{M^2}\frac{M^2-M^{\prime 2}}{q^2}q^{\mu}+\mathtt{g}_{4}^{T}(q^{2})(g^{\alpha\mu}-\frac{q^{\alpha}q^{\mu}}{q^2})\Big]u(P,S_{z}).\label{eq:matrix_element_32nT}
\end{eqnarray}
Here $q^{\mu}=P^{\mu}-P^{\prime\mu}$ and ${\cal P}^{\mu}=P^{\mu}+P^{\prime\mu}$.
In the previous work~\cite{Zhao:2018mrg}, the $1/2\to3/2$ transition matrix elements have been parameterized with form factors $\mathtt{f}_{1,2,3,4}^{\prime(T)}$ and $\mathtt{g}_{1,2,3,4}^{\prime(T)}$ in following form,
\begin{eqnarray}
&&	\langle{\cal B}_{f}^{\prime*}(P^{\prime},S^{\prime}=\frac{3}{2},S_{z}^{\prime})|\bar{q}_{1}\gamma^{\mu}(1-\gamma_{5})Q_{1}|{\cal B}_{i}(P,S=\frac{1}{2},S_{z})\rangle \nonumber\\
&& =  \bar{u}_{\alpha}(P^{\prime},S_{z}^{\prime})\Big[\gamma^{\mu}P^{\alpha}\frac{\mathtt{f}_{1}^{\prime}(q^{2})}{M}+\frac{\mathtt{f}_{2}^{\prime}(q^{2})}{M^{2}}P^{\alpha}P^{\mu} +\frac{\mathtt{f}_{3}^{\prime}(q^{2})}{MM^{\prime}}P^{\alpha}P^{\prime\mu}+\mathtt{f}_{4}^{\prime}(q^{2})g^{\alpha\mu}\Big]\gamma_{5}u(P,S_{z})\nonumber\\
& &\quad-\bar{u}_{\alpha}(P^{\prime},S_{z}^{\prime})\Big[\gamma^{\mu}P^{\alpha}\frac{\mathtt{g}_{1}^{\prime}(q^{2})}{M}+\frac{\mathtt{g}_{2}^{\prime}(q^{2})}{M^{2}}P^{\alpha}P^{\mu} +\frac{\mathtt{g}_{3}^{\prime}(q^{2})}{MM^{\prime}}P^{\alpha}P^{\prime\mu}+\mathtt{g}_{4}^{\prime}(q^{2})g^{\alpha\mu}\Big]u(P,S_{z}),\label{eq:matrix_element_32VA} \\
&&\langle{\cal B}^{\prime*}_{f}(P^{\prime},S^{\prime}=\frac{3}{2},S_{z}^{\prime})|\bar{q}_{1}i\sigma_{\mu\nu}\frac{q^{\nu}}{M}(1+\gamma_{5})Q_{1}|{\cal B}_{i}(P,S=\frac{1}{2},S_{z})\rangle \nonumber\\& &=  \bar{u}_{\alpha}(P^{\prime},S_{z}^{\prime})\Big[\gamma^{\mu}P^{\alpha}\frac{\mathtt{f}_{1}^{\prime T}(q^{2})}{M}+\frac{\mathtt{f}_{2}^{\prime T}(q^{2})}{M^{2}}P^{\alpha}P^{\mu} +\frac{\mathtt{f}_{3}^{\prime T}(q^{2})}{MM^{\prime}}P^{\alpha}P^{\prime\mu}+\mathtt{f}_{4}^{\prime T}(q^{2})g^{\alpha\mu}\Big]\gamma_{5}u(P,S_{z})\nonumber\\
& &\quad + \bar{u}_{\alpha}(P^{\prime},S_{z}^{\prime})\Big[\gamma^{\mu}P^{\alpha}\frac{\mathtt{g}_{1}^{\prime T}(q^{2})}{M}+\frac{\mathtt{g}_{2}^{\prime T}(q^{2})}{M^{2}}P^{\alpha}P^{\mu} +\frac{\mathtt{g}_{3}^{\prime T}(q^{2})}{MM^{\prime}}P^{\alpha}P^{\prime\mu}+\mathtt{g}_{4}^{\prime T}(q^{2})g^{\alpha\mu}\Big]u(P,S_{z}).\label{eq:matrix_element_32t}
\end{eqnarray}
Then the form factors $\mathtt{f}_{1,2,3,4}^{(T)}$ and $\mathtt{g}_{1,2,3,4}^{(T)}$ defined by Eqs.~(\ref{eq:matrix_element_32nVA})-(\ref{eq:matrix_element_32nT}) can be related  with $\mathtt{f}_{1,2,3,4}^{\prime(T)}$ and $\mathtt{g}_{1,2,3,4}^{\prime(T)}$ defined by Eqs.~(\ref{eq:matrix_element_32VA})-(\ref{eq:matrix_element_32t}) by the following formulas:
\begin{align}
&\mathtt{f}_{1}^{(T)}(q^2)=\mathtt{f}_{1}^{\prime(T)}(q^2),\quad
\mathtt{f}_{2}^{(T)}(q^2)=-\frac{1}{2}\Big[\mathtt{f}_{2}^{\prime(T)}(q^2)
+\frac{M}{M^{\prime}}\mathtt{f}_{3}^{\prime(T)}(q^2)\Big],\quad
\mathtt{f}_{4}^{(T)}(q^2)=\mathtt{f}_{4}^{\prime(T)}(q^2),\label{eq:ff23f124t}\\
&\mathtt{f}_{3}^{(T)}(q^2)=\frac{M^{2}}{M^2-M^{\prime2}}
\Big[\mathtt{f}_{1}^{\prime(T)}(q^2)\frac{-M-M^{\prime}}{M}+\mathtt{f}_{4}^{\prime(T)}(q^2)\Big]
+\frac{1}{2}\Big[\mathtt{f}_{2}^{\prime(T)}(q^2)
+\frac{M}{M^{\prime}}\mathtt{f}_{3}^{\prime(T)}(q^2)\Big]\nonumber\\
&\qquad\qquad\quad+\frac{1}{2}\frac{q^{2}}{M^2-M^{\prime2}}
\Big[\mathtt{f}_{2}^{\prime(T)}(q^2)
-\frac{M}{M^{\prime}}\mathtt{f}_{3}^{\prime(T)}(q^2)\Big],\label{eq:ff23f3t}\\
&\mathtt{g}_{1}^{(T)}(q^2)=\mathtt{g}_{1}^{\prime(T)}(q^2),\quad
\mathtt{g}_{2}^{(T)}(q^2)=-\frac{1}{2}
\Big[\mathtt{g}_{2}^{\prime(T)}(q^2)
+\frac{M}{M^{\prime}}\mathtt{g}_{3}^{\prime(T)}(q^2)\Big],\quad
\mathtt{g}_{4}(q^2)=\mathtt{g}_4^{\prime(T)}(q^2),\label{eq:ff23g124t}\\
&\mathtt{g}_{3}^{(T)}(q^2)=
\frac{M^{2}}{M^2-M^{\prime2}}
\Big[\mathtt{g}_{1}^{\prime(T)}(q^2)\frac{M-M^{\prime}}{M}+\mathtt{g}_{4}^{\prime(T)}(q^2)\Big]
+\frac{1}{2}\Big[\mathtt{g}_{2}^{\prime(T)}(q^2)
+\frac{M}{M^{\prime}}\mathtt{g}_{3}^{\prime(T)}(q^2)\Big]\nonumber\\
&\qquad\qquad\quad+\frac{1}{2}\frac{q^{2}}{M^2-M^{\prime2}}
\Big[\mathtt{g}_{2}^{\prime(T)}(q^2)
-\frac{M}{M^{\prime}}\mathtt{g}_{3}^{\prime(T)}(q^2)\Big].\label{eq:ff23g3t}
\end{align}
Multiplying Eq.~(\ref{eq:matrix_element_32nT}) by $q^{\mu}$ will  yield
\begin{eqnarray}
\bar{u}_{\alpha}(P^{\prime},S_{z}^{\prime})
\Big[\mathtt{f}_{3}^{ T}(q^2)P^{\alpha}\frac{M^2-M^{\prime 2}}{M^2}\Big]\gamma_{5}u(P,S_{z})&=& 0,\nonumber \\
\bar{u}_{\alpha}(P^{\prime},S_{z}^{\prime})
\Big[\mathtt{g}_{3}^{ T}(q^2)P^{\alpha}\frac{M^2-M^{\prime 2}}{M^2}\Big]u(P,S_{z})&=& 0,\label{eq:matrix_element_32f3g3}
\end{eqnarray}
and one obtains $\mathtt{f}_{3}^{T}(q^2)=\mathtt{g}_{3}^{T}(q^2)=0$.
Then using Eqs.~(\ref{eq:ff23f124t} )- (\ref{eq:ff23g3t}), one could get
\begin{align}
&\mathtt{f}_{1}^{T}(q^2)=\frac{M}{M^{\prime}+M}
\Big\{\mathtt{f}_{4}^{\prime T}(q^2)
+\frac{M^2-M^{\prime2}}{M^{2}}
\Big[\frac{1}{2}\Big(\mathtt{f}_{2}^{\prime T}(q^2)
+\frac{M}{M^{\prime}}\mathtt{f}_{3}^{\prime T}(q^2)\Big)
+\frac{1}{2}\frac{q^{2}}{M^2-M^{\prime2}}
\Big(\mathtt{f}_{2}^{\prime T}(q^2)
-\frac{M}{M^{\prime}}\mathtt{f}_{3}^{\prime T}(q^2)\Big)\Big]\Big\},\\
&\mathtt{g}_{1}^{T}(q^2)=\frac{M}{M^{\prime}-M}
\Big\{\mathtt{g}_{4}^{\prime T}(q^2)
+\frac{M^2-M^{\prime2}}{M^{2}}
\Big[\frac{1}{2}\Big(\mathtt{g}_{2}^{\prime T}(q^2)
+\frac{M}{M^{\prime}}\mathtt{g}_{3}^{\prime T}(q^2)\Big)
+\frac{1}{2}\frac{q^{2}}{M^2-M^{\prime2}}
\Big(\mathtt{g}_{2}^{\prime T}(q^2)-\frac{M}{M^{\prime}}\mathtt{g}_{3}^{\prime T}(q^2)
\Big)\Big]\Big\}.
\end{align}

These form factors $\mathtt{f}_{1,2,3,4}^{\prime}$ and $\mathtt{g}_{1,2,3,4}^{\prime}$ can be extracted in the following way~\cite{Ke:2017eqo}.
Multiply $\bar{u}(P,S_{z})(\bar{\Gamma}_{5}^{\mu\beta})_{i}u_{\beta}(P^{\prime},S_{z}^{\prime})$
and $\bar{u}(P,S_{z})(\Gamma_{5}^{\mu\beta})_{i}u_{\beta}(P^{\prime},S_{z}^{\prime})$
on the ``$\langle{\cal B}^{\prime*}_{f}(P^{\prime},S^{\prime}=\frac{3}{2},S_{z}^{\prime})|\bar{q}_{1}\gamma^{\mu}Q_{1}|{\cal B}_{i}(P,S=\frac{1}{2},S_{z})\rangle$"
part of Eq.~(\ref{eq:matrix_element_onehalf}) and Eq.~(\ref{eq:matrix_element_32VA}), respectively.
At the same time, the approximation $P^{(\prime)}\to\bar{P}^{(\prime)}$ need to be taken for the integral.
After summing the polarizations of the initial and final baryon states up, we can get the four equations as follows,
\begin{eqnarray}
&&{\rm Tr}\Big\{ u_{\beta}(P^{\prime},S_{z}^{\prime})\bar{u}_{\alpha}(P^{\prime},S_{z}^{\prime})
\Big[\gamma^{\mu}P^{\alpha}\frac{\mathtt{f}_{1}^{\prime}(q^{2})}{M}
+\frac{\mathtt{f}_{2}^{\prime}(q^{2})}{M^{2}}P^{\alpha}P^{\mu}
+\frac{\mathtt{f}_{3}^{\prime}(q^{2})}{MM^{\prime}}P^{\alpha}P^{\prime\mu}
+\mathtt{f}_{4}^{\prime}(q^{2})g^{\alpha\mu}\Big]\gamma_{5} (\slashed P+M)(\Gamma_{5}^{\mu\beta})_{i}\Big\}\nonumber\\
 &&= \int\{d^{3}p_{2}\}\frac{\phi^{\prime}(x^{\prime},k_{\perp}^{\prime})\phi(x,k_{\perp})}{2\sqrt{p_{1}^{+}p_{1}^{\prime+}(p_{1}\cdot\bar{P}+m_{1}M_{0})(p_{1}^{\prime}\cdot\bar{P}^{\prime}+m_{1}^{\prime}M_{0}^{\prime})}}\nonumber \\
&  &\quad \times\sum_{S_{z}^{\prime}\lambda_{2}}{\rm Tr}\Big\{ u_{\beta}(\bar{P}^{\prime},S_{z}^{\prime})\bar{u}_{\alpha}(\bar{P}^{\prime},S_{z}^{\prime})\bar{\Gamma}^{\prime\alpha}_{A}(\slashed p_{1}^{\prime}+m_{1}^{\prime})\gamma_{\mu}(\slashed p_{1}+m_{1})\Gamma_{A}(\bar{\slashed P}+M_{0})(\bar{\Gamma}_{5}^{\mu\beta})_{i}\Big\},\label{eq:Fisolving}
\end{eqnarray}
with $(\Gamma_{5}^{\mu\beta})_{i}=\{\gamma^{\mu}P^{\beta},P^{\prime\mu}P^{\beta},P^{\mu}P^{\beta},g^{\mu\beta}\}\gamma_{5}$ and $(\bar{\Gamma}_{5}^{\mu\beta})_{i}=\{\gamma^{\mu}\bar{P}^{\beta},\bar{P}^{\prime\mu}\bar{P}^{\beta},\bar{P}^{\mu}\bar{P}^{\beta},g^{\mu\beta}\}\gamma_{5}$.

The analytic expression of form factors $\mathtt{f}_{1,2,3,4}^{\prime}$ can be got by solving the above four equations,
\begin{eqnarray}
{\mathtt{f}_{1}^{\prime}}(q^2)&=&\frac{{M}{M^{\prime}}}{2\big[{M}^4-2{M}^2
({M^{\prime 2}}+{q^2})+({M^{\prime 2}}-{q^2})^2
\big]^2}
\big\{-4{M^{\prime}}\big[{H_1}
\big(({M}-{M^{\prime}})^2-{q^2}\big)+{H_3}
{M^{\prime}}\big]\nonumber\\
&&-2{H_2}\big({M}^2-4{M}
{M^{\prime}}+{M^{\prime 2}}-{q^2}\big)+{H_4}\big[{M}^4-2
{M}^2
\big({M^{\prime 2}}+{q^2}\big)+\big({M^{\prime 2}}-{q^2}\big)^2
\big]\big\},
\\
{\mathtt{f}_{2}^{\prime}}(q^2)&=&\frac{{M}^2{M^{\prime 2}}}{\big[({M}-{M^{\prime}})^2-{q^2}\big]^3
\big[({M}+{M^{\prime}})^2-{q^2}\big]^2}
\big\{2{M^{\prime}}
\big[{H_1}\big(({M}-{M^{\prime}})^2-{q^2}\big)+10
{H_3}{M^{\prime}}\big]\nonumber\\
&&-4{H_2}\big(2{M}^2+{M}
{M^{\prime}}+2{M^{\prime 2}}-2{q^2}\big)
+{H_4}\big[{M}^4-2
{M}^2
\big({M^{\prime 2}}+{q^2}\big)+\big({M^{\prime 2}}-{q^2}\big)^2
\big]\big\},
\end{eqnarray}
\begin{eqnarray}
{\mathtt{f}_{3}^{\prime}}(q^2)&=&\frac{{M}{M^{\prime}}}{\left[({M}-{M^{\prime}})^2-{q^2}\right]^3
\left[({M}+{M^{\prime}})^2-{q^2}\right]^2}\times\nonumber\\
&&\big\{{M^{\prime}}\left[{H_1}
\left(({M}-{M^{\prime}})^2-{q^2}\right)\left({M}^2-4
{M}{M^{\prime}}+{M^{\prime 2}}-{q^2}\right)
-4{H_3}{M^{\prime}}
\left(2{M}^2+{M}{M^{\prime}}+2{M^{\prime 2}}-2
{q^2}\right)\right]\nonumber\\
&&+2{H_2}\big[{M}^4-2{M}^3
{M^{\prime}}+2{M}^2\left(6{M^{\prime 2}}-{q^2}\right)+2{M}
{M^{\prime}}
\left({q^2}-{M^{\prime 2}}\right)+\left({M^{\prime 2}}-{q^2}\right)^2
\big]\nonumber\\
&&-{H_4}\left[({M}-{M^{\prime}})^2-{q^2}\right]
\left({M}^2-{M}{M^{\prime}}+{M^{\prime 2}}-{q^2}\right)
\left[({M}+{M^{\prime}})^2-{q^2}\right]\big\},
\\
{\mathtt{f}_{4}^{\prime}}(q^2)&=&\frac{1}{2\left[({M}-{M^{\prime}})^2-{q^2}\right]^2
\left[({M}+{M^{\prime}})^2-{q^2}\right]}
\big\{{M^{\prime}}\left[{H_1}
\left({q^2}-({M}-{M^{\prime}})^2\right)+2{H_3}
{M^{\prime}}\right]\nonumber\\
&&-2{H_2}\left({M}^2-{M}
{M^{\prime}}+{M^{\prime 2}}-{q^2}\right)
+{H_4}\big[{M}^4-2
{M}^2
\left({M^{\prime 2}}+{q^2}\right)+\left({M^{\prime 2}}-{q^2}\right)^2
\big]\big\},
\end{eqnarray}
where $H_i$ is defined as follows,
\begin{eqnarray}
	H_{i} & = & \int\{d^{3}p_{2}\}\frac{\phi^{\prime}(x^{\prime},k_{\perp}^{\prime})\phi(x,k_{\perp})}{2\sqrt{p_{1}^{+}p_{1}^{\prime+}(p_{1}\cdot\bar{P}+m_{1}M_{0})(p_{1}^{\prime}\cdot\bar{P}^{\prime}+m_{1}^{\prime}M_{0}^{\prime})}}\nonumber \\
	&  & \quad\times\sum_{S_{z}^{\prime}\lambda_{2}}{\rm Tr}\Big\{ u_{\beta}(\bar{P}^{\prime},S_{z}^{\prime})\bar{u}_{\alpha}(\bar{P}^{\prime},S_{z}^{\prime})\bar{\Gamma}^{\prime\alpha}_{A}(\slashed p_{1}^{\prime}+m_{1}^{\prime})\gamma_{\mu}(\slashed p_{1}+m_{1})\Gamma_{A}(\bar{\slashed P}+M_{0})(\bar{\Gamma}_{5}^{\mu\beta})_{i}\Big\}.\label{eq:Fi_L}
\end{eqnarray}
With the same method, one can obtain the form
factors $\mathtt{g}_{1,2,3,4}^{\prime}$. $\mathtt{g}_{1}^{\prime}$ and $\mathtt{g}_{3}^{\prime}$ are similar to $\mathtt{f}_{1}^{\prime}$ and $\mathtt{f}_{3}^{\prime}$ respectively except for $M^{\prime}\to -M^{\prime}$ and $H_{i}\to K_{i}$, and $\mathtt{g}_{2}^{\prime}$ and $\mathtt{g}_{4}^{\prime}$ are similar to $\mathtt{f}_{2}^{\prime}$ and $\mathtt{f}_{4}^{\prime}$ respectively except for $M^{\prime}\to -M^{\prime}$ and $H_{i}\to -K_{i}$. For example, we have
\begin{eqnarray}
{\mathtt{g}_{1}^{\prime}}(q^2)&=&-\frac{{M}{M^{\prime}}}{2\left[{M}^4-2{M}^2
\left({M^{\prime 2}}+{q^2}\right)+\left({M^{\prime 2}}-{q^2}\right)^2
\right]^2}
\Big\{4{M^{\prime}}\left[{K_1}
\left(({M}+{M^{\prime}})^2-{q^2}\right)-{K_3}
{M^{\prime}}\right]\nonumber\\
&&-2{K_2}\left({M}^2+4{M}
{M^{\prime}}+{M^{\prime 2}}-{q^2}\right)
+{K_4}\left[{M}^4-2
{M}^2
\left({M^{\prime 2}}+{q^2}\right)+\left({M^{\prime 2}}-{q^2}\right)^2
\right]\Big\},
\end{eqnarray}
where $K_i$ is defined by the following Eq.~(\ref{eq:Gi_1}),
\begin{eqnarray}
K_{i} & = & \int\{d^{3}p_{2}\}\frac{\phi^{\prime}(x^{\prime},k_{\perp}^{\prime})\phi(x,k_{\perp})}{2\sqrt{p_{1}^{+}p_{1}^{\prime+}(p_{1}\cdot\bar{P}+m_{1}M_{0})(p_{1}^{\prime}\cdot\bar{P}^{\prime}+m_{1}^{\prime}M_{0}^{\prime})}}\nonumber \\
&  & \times\sum_{S_{z}^{\prime}\lambda_{2}}{\rm Tr}\Big\{ u_{\beta}(\bar{P}^{\prime},S_{z}^{\prime})\bar{u}_{\alpha}(\bar{P}^{\prime},S_{z}^{\prime})\bar{\Gamma}^{\prime\alpha}_{A}(\slashed p_{1}^{\prime}+m_{1}^{\prime})\gamma_{\mu}\gamma_{5}(\slashed p_{1}+m_{1})\Gamma_{A}(\bar{\slashed P}+M_{0})(\bar{\Gamma}^{\mu\beta})_{i}\Big\},\label{eq:Gi_1}
\end{eqnarray}
with $(\bar{\Gamma}^{\mu\beta})_{i}=\{\gamma^{\mu}\bar{P}^{\beta},\bar{P}^{\prime\mu}\bar{P}^{\beta},\bar{P}^{\mu}\bar{P}^{\beta},g^{\mu\beta}\}$.
Note that $\mathtt{f}_{1,2,3,4}^{\prime T}$ and $\mathtt{g}_{1,2,3,4}^{\prime T}$ should not be independent. Multiplying the Eq.~(\ref{eq:matrix_element_32t}) by $q^{\mu}$ leads to
\begin{eqnarray}
\bar{u}_{\alpha}(P^{\prime},S_{z}^{\prime})\Big[(-M-M^{\prime})P^{\alpha}\frac{\mathtt{f}_{1}^{\prime T}(q^{2})}{M}+\frac{\mathtt{f}_{2}^{\prime T}(q^{2})}{M^{2}}P^{\alpha}P\cdot q +\frac{\mathtt{f}_{3}^{\prime T}(q^{2})}{MM^{\prime}}P^{\alpha}P^{\prime}\cdot q+\mathtt{f}_{4}^{\prime T}(q^{2})q^{\alpha}\Big]\gamma_{5}u(P,S_{z})& = &0 ,\nonumber \\
 \bar{u}_{\alpha}(P^{\prime},S_{z}^{\prime})\Big[(M-M^{\prime})P^{\alpha}\frac{\mathtt{g}_{1}^{\prime T}(q^{2})}{M}+\frac{\mathtt{g}_{2}^{\prime T}(q^{2})}{M^{2}}P^{\alpha}P\cdot q +\frac{\mathtt{g}_{3}^{\prime T}(q^{2})}{MM^{\prime}}P^{\alpha}P^{\prime}\cdot q+\mathtt{g}_{4}^{\prime T}(q^{2})q^{\alpha}\Big]u(P,S_{z})&  =&0,\label{eq:matrix_element_32t1}
\end{eqnarray}
and the following two relations can be arrived,
\begin{equation}
\frac{P^{\alpha}}{M}\mathtt{f}_{1}^{\prime T}(q^2)=\frac{1}{(M+M^{\prime})}\Big[\frac{\mathtt{f}_{2}^{\prime T}(q^{2})}{M^{2}}P^{\alpha}P\cdot q +\frac{\mathtt{f}_{3}^{\prime T}(q^{2})}{MM^{\prime}}P^{\alpha}P^{\prime}\cdot q+\mathtt{f}_{4}^{\prime T}(q^{2})q^{\alpha}\Big],\label{eq:f1pT}
\end{equation}
\begin{equation}
\frac{P^{\alpha}}{M}\mathtt{g}_{1}^{\prime T}(q^2)=-\frac{1}{(M-M^{\prime})}\Big[\frac{\mathtt{g}_{2}^{\prime T}(q^{2})}{M^{2}}P^{\alpha}P\cdot q +\frac{\mathtt{f}_{3}^{\prime T}(q^{2})}{MM^{\prime}}P^{\alpha}P^{\prime}\cdot q+\mathtt{g}_{4}^{\prime T}(q^{2})q^{\alpha}\Big].\label{eq:g1pT}
\end{equation}
With the above relations, $\langle{\cal B}^{\prime}(P^{\prime},S_{z}^{\prime})|\bar{q}_{1}i\sigma_{\mu\nu}\frac{q^{\nu}}{M}(1+\gamma_{5})Q_{1}|{\cal B}(P,S_{z})\rangle$
can be parameterized with the form factors $\mathtt{f}_{2,3,4}^{\prime T}$ and $\mathtt{g}_{2,3,4}^{\prime T}$.
These form factors $\mathtt{f}_{2,3,4}^{\prime T}$ and $\mathtt{g}_{2,3,4}^{\prime T}$ can be extracted in the same way as we have conducted on the form factors $\mathtt{f}_{1,2,3,4}^{\prime}$ and $\mathtt{g}_{1,2,3,4}^{\prime}$~\cite{Ke:2017eqo}.
Multiply $\bar{u}(P,S_{z})(\bar{\Gamma}_{5}^{\mu\beta})_{i}u_{\beta}(P^{\prime},S_{z}^{\prime})$
and $\bar{u}(P,S_{z})(\Gamma_{5}^{\mu\beta})_{i}u_{\beta}(P^{\prime},S_{z}^{\prime})$
on the ``$\langle{\cal B}^{\prime}_{f}(P^{\prime},S^{\prime}=\frac{1}{2},S_{z}^{\prime})
|\bar{q}_{1}i\sigma_{\mu\nu}\frac{q^{\nu}}{M}Q_{1}|{\cal B}_{i}(P,S=\frac{1}{2},S_{z})\rangle$"
part of Eq.~(\ref{eq:matrix_element_onehalf_tensor}) and Eq.~(\ref{eq:matrix_element_32t}), respectively.
At the same time, the approximation $P^{(\prime)}\to\bar{P}^{(\prime)}$ need to be taken for the integral.
After summing the polarizations of the initial and final baryon states up, we can get the three equations as follows,
\begin{eqnarray}
&&\int\{d^{3}p_{2}\}\frac{\phi^{\prime}(x^{\prime},k_{\perp}^{\prime})\phi(x,k_{\perp})}{2\sqrt{p_{1}^{+}p_{1}^{\prime+}(p_{1}\cdot\bar{P}+m_{1}M_{0})(p_{1}^{\prime}\cdot\bar{P}^{\prime}+m_{1}^{\prime}M_{0}^{\prime})}}\nonumber \\
&&\qquad\times\sum_{S_{z}^{\prime}\lambda_{2}}{\rm Tr}\Big\{ u_{\beta}(\bar{P}^{\prime},S_{z}^{\prime})\bar{u}_{\alpha}(\bar{P}^{\prime},S_{z}^{\prime})\bar{\Gamma}^{\prime\alpha}_{A}(\slashed p_{1}^{\prime}+m_{1}^{\prime})i\sigma_{\mu\nu}\frac{q^{\nu}}{M}(\slashed p_{1}+m_{1})\Gamma_{A}(\bar{\slashed P}+M_{0})(\bar{\Gamma}_{5}^{\mu\beta})_{i}\Big\}\nonumber\\
&&= {\rm Tr}\Big\{ u_{\beta}(P^{\prime},S_{z}^{\prime})\bar{u}_{\alpha}(P^{\prime},S_{z}^{\prime})
\Big[\frac{\gamma^{\mu}}{M+M^{\prime}}\Big(\frac{\mathtt{f}_{2,A}^{\prime T}(q^{2})}{M^{2}}P^{\alpha}P\cdot q
+\frac{\mathtt{f}_{3,A}^{\prime T}(q^{2})}{MM^{\prime}}P^{\alpha}P^{\prime}\cdot q
+\mathtt{f}_{4,A}^{\prime T}(q^{2})q^{\alpha}\Big)\nonumber\\
&&\qquad\qquad+\frac{\mathtt{f}_{2,A}^{\prime T}(q^{2})}{M^{2}}P^{\alpha}P^{\mu}
+\frac{\mathtt{f}_{3,A}^{\prime T}(q^{2})}{MM^{\prime}}P^{\alpha}P^{\prime\mu}
+\mathtt{f}_{4,A}^{\prime T}(q^{2})g^{\alpha\mu}\Big]
\gamma_{5} (\slashed P+M)(\Gamma_{5}^{\mu\beta})_{i}\Big\}
,\label{eq:FiT_solving}
\end{eqnarray}
with $(\bar{\Gamma}_{5}^{\mu\beta})_{i}=\{\gamma^{\mu}\bar{P}^{\beta},\bar{P}^{\prime\mu}\bar{P}^{\beta},g^{\mu\beta}\}\gamma_{5}$ and $(\Gamma_{5}^{\mu\beta})_{i}=\{\gamma^{\mu}P^{\beta},P^{\prime\mu}P^{\beta},g^{\mu\beta}\}\gamma_{5}$.

Then the expressions of form factors $\mathtt{f}_{2,3,4}^{\prime T}$
could be got by solving the above three equations and are shown with the following formulas:
\begin{eqnarray}
{\mathtt{f}_{2}^{\prime T}}(q^2)&=&\frac{{M}^2{M^{\prime 2}}
\left\{\left[({M}-{M^{\prime}})^2-{q^2}\right]\left[2{H^T_1}
{M^{\prime}}+{H^T_3}
\left(({M}+{M^{\prime}})^2-{q^2}\right)\right]-4{H^T_2}
\left(2{M}^2+{M}{M^{\prime}}-3{M^{\prime 2}}-2
{q^2}\right)\right\}}{\left[({M}-{M^{\prime}})^2-{q^2}\right]^3\left[({M}+{M^{\prime}})^2-{q^2}\right]^2},
\\
{\mathtt{f}_{3}^{\prime T}}(q^2)&=&\frac{{M}{M^{\prime}}}{\left[({M}-
{M^{\prime}})^2-{q^2}\right]^3
\left[({M}+{M^{\prime}})^2-{q^2}\right]^2}\times\\
&&\Big\{\left[({M}-{M^{\prime}})^2-{q^2}\right]\left[{H^T_1}{M^{\prime}}
\left({M}^2-4{M}
{M^{\prime}}+{M^{\prime 2}}-{q^2}\right)-{H^T_3}\left({M}^2-{M}
{M^{\prime}}+{M^{\prime 2}}-{q^2}\right)
\left(({M}+{M^{\prime}})^2-{q^2}\right)\right]\nonumber\\
&&2{H^T_2}
\left[{M}^4-2{M}^3{M^{\prime}}+{M}^2\left(8
{M^{\prime 2}}-2{q^2}\right)+2{M}{M^{\prime}}\left({q^2}-2
{M^{\prime 2}}\right)-3{M^{\prime}}^4+2{M^{\prime 2}}
{q^2}+{q^4}\right]\Big\},\nonumber
\\
{\mathtt{f}_{4}^{\prime T}}(q^2)&=&\frac{\left[({M}-{M^{\prime}})^2-{q^2}\right]
\left[{H^T_3}
\left(({M}+{M^{\prime}})^2-{q^2}\right)-{H^T_1}
{M^{\prime}}\right]+2{H^T_2}[{M}
({M^{\prime}}-{M})+{q^2}]}{2
\left[({M}-{M^{\prime}})^2-{q^2}\right]^2
\left[({M}+{M^{\prime}})^2-{q^2}\right]},
\end{eqnarray}
where $H_i^{T}$ is defined in Eq.~(\ref{eq:FiT_L}),
\begin{eqnarray}
H_{i}^{T} & = & \int\{d^{3}p_{2}\}\frac{\phi^{\prime}(x^{\prime},k_{\perp}^{\prime})\phi(x,k_{\perp})}{2\sqrt{p_{1}^{+}p_{1}^{\prime+}(p_{1}\cdot\bar{P}+m_{1}M_{0})(p_{1}^{\prime}\cdot\bar{P}^{\prime}+m_{1}^{\prime}M_{0}^{\prime})}}\nonumber \\
&  & \times\sum_{S_{z}^{\prime}\lambda_{2}}{\rm Tr}\Big\{ u_{\beta}(\bar{P}^{\prime},S_{z}^{\prime})\bar{u}_{\alpha}(\bar{P}^{\prime},S_{z}^{\prime})\bar{\Gamma}^{\prime\alpha}_{A}(\slashed p_{1}^{\prime}+m_{1}^{\prime})i\sigma_{\mu\nu}\frac{q^{\nu}}{M}(\slashed p_{1}+m_{1})\Gamma_{A}(\bar{\slashed P}+M_{0})(\bar{\Gamma}_{5}^{\mu\beta})_{i}\Big\}.\label{eq:FiT_L}
\end{eqnarray}
With the same method, one can obtain the form
factors $g_{2,3,4}^{\prime T}$. $\mathtt{g}_{2}^{\prime T}$ and $\mathtt{g}_{4}^{\prime T}$ are similar to $\mathtt{f}_{2}^{\prime T}$ and $\mathtt{f}_{4}^{\prime T}$ respectively except for $M^{\prime}\to -M^{\prime}$ and $H_{i}^{T}\to -K_{i}^{T}$. $\mathtt{g}_{3}^{\prime T}$ is similar to $\mathtt{f}_{3}^{\prime T}$ except for $M^{\prime}\to -M^{\prime}$ and $H_{i}^{T}\to K_{i}^{T}$. For example, we have
\begin{eqnarray}
{\mathtt{g}_{2}^{\prime T}}(q^2)&=&-\frac{{M}^2{M^{\prime 2}}
}{\left[({M}-{M^{\prime}})^2-{q^2}\right]^2\left[({M}+{M^{\prime}})^2-{q^2}\right]^3}\times
\nonumber\\
&&\left\{\left[({M}+{M^{\prime}})^2-{q^2}\right]\left[{K^T_3}
\left(({M}-{M^{\prime}})^2-{q^2}\right)-2{K^T_1}
{M^{\prime}}\right]+4{K^T_2}\left(-2{M}^2+{M}{M^{\prime}}+3
{M^{\prime 2}}+2{q^2}\right)\right\},
\end{eqnarray}
where $K_i^{T}$ is defined by the following Eq.~(\ref{eq:GiT_L}),
\begin{eqnarray}
K_{i}^{T} & = & \int\{d^{3}p_{2}\}\frac{\phi^{\prime}(x^{\prime},k_{\perp}^{\prime})\phi(x,k_{\perp})}{2\sqrt{p_{1}^{+}p_{1}^{\prime+}(p_{1}\cdot\bar{P}+m_{1}M_{0})(p_{1}^{\prime}\cdot\bar{P}^{\prime}+m_{1}^{\prime}M_{0}^{\prime})}}\nonumber \\
&  & \times\sum_{S_{z}^{\prime}\lambda_{2}}{\rm Tr}\Big\{ u_{\beta}(\bar{P}^{\prime},S_{z}^{\prime})\bar{u}_{\alpha}(\bar{P}^{\prime},S_{z}^{\prime})\bar{\Gamma}^{\prime\alpha}_{A}(\slashed p_{1}^{\prime}+m_{1}^{\prime})i\sigma_{\mu\nu}\frac{q^{\nu}}{M}\gamma_5(\slashed p_{1}+m_{1})\Gamma_{A}(\bar{\slashed P}+M_{0})(\bar{\Gamma}^{\mu\beta})_{i}\Big\},\label{eq:GiT_L}
\end{eqnarray}
with $(\bar{\Gamma}^{\mu\beta})_{i}=\{\gamma^{\mu}\bar{P}^{\beta},\bar{P}^{\prime\mu}\bar{P}^{\beta},g^{\mu\beta}\}$.

\subsection{Flavor-spin wave functions}
\label{sec:flavorspin}
In the above subsection, we have presented the explicit expressions of form factors. However, a physical form factor should be a linear combination of the transition form factors with a scalar and an axial-vector diquark spectator.
\begin{equation}
{\rm [form~factor]}^{\rm physical}(q^2)=
 c_{S}\times {\rm [form~factor]}_{S}^{\rm in~Subsec.A}+c_{A}\times {\rm [form~factor]}_{A}^{\rm in~Subsec.A},\label{eq:physical_ff}
\end{equation}
and here $c_{S}$ and $c_{A}$ are the overlapping factors which are derived from the flavor spin wave functions of the initial and final baryon states with $S$ and $A$ corresponding to  the scalar and the axial vector diquark spectator of these doubly heavy baryons decays. The  hadronic  matrix elements can be written as
\begin{equation}
\langle B^{\prime}|\Gamma_{\mu}|B\rangle=c_{S}\langle q_{1}[Q_{2}q]_{S}|\Gamma_{\mu}|Q_{1}[Q_{2}q]_{S}\rangle+c_{A}\langle q_{1}\{Q_{2}q\}_{A}|\Gamma_{\mu}|Q_{1}\{Q_{2}q\}_{A}\rangle,\label{eq:csca}
\end{equation}
and the form factors $f_{i,S}$  and $f_{i,A}$ extracted from  Eq.~(\ref{eq:solve_fi}) are involved with the transition matrix elements $\langle q_{1}[Q_{2}q]_{S}|\Gamma_{\mu}|Q_{1}[Q_{2}q]_{S}\rangle$
and $\langle q_{1}\{Q_{2}q\}_{A}|\Gamma_{\mu}|Q_{1}\{Q_{2}q\}_{A}\rangle$, respectively.
Here the current $\Gamma_{\mu}$ is  $\Gamma_{\mu}=\bar q_1\gamma_{\mu}(1-\gamma_5)Q_1$ or $\bar q_1\sigma_{\mu\nu}\frac{q^{\nu}}{M}(1+\gamma_5)Q_1$.
Eqs.~(\ref{eq:physical_ff}) and (\ref{eq:csca}) are for the  $1/2\to1/2$ transitions. For the transition $1/2\to 3/2$, the diquark in the final state baryons can not be a scalar state, so the transition matrix element $\langle q_{1}[Q_{2}q]_{S}|\Gamma_{\mu}|Q_{1}[Q_{2}q]_{S}\rangle$ is zero and the physical form factor is given as
\begin{equation}
{\rm [form~factor]}^{\rm physical}(q^2)=c_{A}\times {\rm [form~factor]}_{A}^{\rm in~Subsec.A}.\label{eq:physical_ff3}
\end{equation}
In this subsection, via performing the inner
product of the flavor-spin wave functions of the initial and final states, the overlapping factors $c_{S}$ and $c_{A}$
in Eqs.~(\ref{eq:physical_ff}) and (\ref{eq:physical_ff3}) can be calculated easily.
For shortage of the paper, the detail calculation of the wave functions for
the initial and final baryons is arranged in the Appendix~\ref{app:wave_functions}.
The flavor spin wave functions for the doubly charmed SU(3) triplets $\Xi_{cc}^{++}$, $\Xi_{cc}^{+}$ and $\Omega_{cc}^{+}$ are\begin{align}
\mathcal{B}_{cc}=\frac{1}{\sqrt{2}}\Big[\Big(-\frac{\sqrt{3}}{2}c^{1}(c^{2}q)_{S}+\frac{1}{2}c^{1}(c^{2}q)_{A}\Big)+(c^{1}\leftrightarrow c^{2})\Big],
\end{align}
here the two charm quarks noted by $c^{1}$ and $c^{2}$ are symmetric.
The flavor spin wave functions of the doubly bottomed SU(3) triplets $\Xi_{bb}^{0}$, $\Xi_{bb}^{-}$ and $\Omega_{bb}^{-}$
can be obtained through the replacement $c\to b$. While the bottom-charm baryons could form two sets of SU(3) triplets, ($\Xi_{bc},\Omega_{bc}$) and ($\Xi_{bc}^{\prime},\Omega_{bc}^{\prime}$).
The flavor spin wave functions of bottom-charm baryons ($\Xi_{bc},\Omega_{bc}$) can be given as
\begin{align}
\mathcal{B}_{bc} & =  -\frac{\sqrt{3}}{2}b(cq)_{S}+\frac{1}{2}b(cq)_{A}  =  -\frac{\sqrt{3}}{2}c(bq)_{S}+\frac{1}{2}c(bq)_{A},
,\quad q=u,~d,~s,
\end{align}
while the flavor spin wave functions of bottom-charm baryons ($\Xi_{bc}^{\prime},\Omega_{bc}^{\prime}$) are given as
\begin{align}
\mathcal{B}_{bc}^{\prime} & =  -\frac{1}{2}b(cq)_{S}-\frac{\sqrt{3}}{2}b(cq)_{A}  = \frac{1}{2}c(bq)_{S}+\frac{\sqrt{3}}{2}c(bq)_{A},\quad q=u,~d,~s.
\end{align}
The flavor-spin wave functions of the anti-triplet singly charmed baryons can be shown as follows,
\begin{align}
\Lambda_{c}^{+} & = -\frac{1}{2}d(cu)_{S}+\frac{\sqrt{3}}{2}d(cu)_{A}  = \frac{1}{2}u(cd)_{S}-\frac{\sqrt{3}}{2}u(cd)_{A},\nonumber \\
\Xi_{c}^{+} & =  -\frac{1}{2}s(cu)_{S}+\frac{\sqrt{3}}{2}s(cu)_{A}   = \frac{1}{2}u(cs)_{S}-\frac{\sqrt{3}}{2}u(cs)_{A},\nonumber\\
\Xi_{c}^{0} & =  -\frac{1}{2}s(cd)_{S}+\frac{\sqrt{3}}{2}s(cd)_{A}  =  \frac{1}{2}d(cs)_{S}-\frac{\sqrt{3}}{2}d(cs)_{A}.\label{eq:flavor_spin_anti-triplet}
\end{align}
while the flavor spin wave functions of the sextet of singly charmed baryons are demonstrated as
\begin{align}
\Sigma_{c}^{++} & =\frac{1}{\sqrt{2}}\Big[\frac{\sqrt{3}}{2}u^{1}(cu^{2})_{S}+\frac{1}{2}u^{1}(cu^{2})_{A}+(u^{1}\leftrightarrow u^{2})\Big],\nonumber \\
\Sigma_{c}^{+} & =  \frac{\sqrt{3}}{2}d(cu)_{S}+\frac{1}{2}d(cu)_{A}  =   \frac{\sqrt{3}}{2}u(cd)_{S}+\frac{1}{2}u(cd)_{A},\nonumber\\
\Sigma_{c}^{0} & =\frac{1}{\sqrt{2}}\Big[\frac{\sqrt{3}}{2}d^{1}(cd^{2})_{S}+\frac{1}{2}d^{1}(cd^{2})_{A}+(d^{1}\leftrightarrow d^{2})\Big],\nonumber \\
\Xi_{c}^{\prime+} & = \frac{\sqrt{3}}{2}s(cu)_{S}+\frac{1}{2}s(cu)_{A}  = \frac{\sqrt{3}}{2}u(cs)_{S}+\frac{1}{2}u(cs)_{A},\nonumber \\
\Xi_{c}^{\prime0} & = \frac{\sqrt{3}}{2}s(cd)_{S}+\frac{1}{2}s(cd)_{A}   = \frac{\sqrt{3}}{2}d(cs)_{S}+\frac{1}{2}d(cs)_{A},\nonumber \\
\Omega_{c}^{0} & =\frac{1}{\sqrt{2}}\Big[\frac{\sqrt{3}}{2}s^{2}(cs^{1})_{S}+\frac{1}{2}s^{2}(cs^{1})_{A}+(s^{1}\leftrightarrow s^{2})\Big].\label{eq:flavor_spin_sextet}
\end{align}
Then we can get the wave functions of the singly bottom baryons by changing $c$ in Eqs.~(\ref{eq:flavor_spin_anti-triplet})-(\ref{eq:flavor_spin_sextet}) to $b$.
While for the baryons ${\cal B}^{*}$ with spin-$3/2$ in the final states,
their flavor spin wave function are given as follows,
\begin{eqnarray}
&&{\cal B}_{Qqq^{\prime}}^{*}=q(Qq^{\prime})_{A}=q^{\prime}(Qq)_{A},~{\cal B}_{Qqq}^{*}=\sqrt{2}q(Qq)_{A},\\
&&{\cal B}_{QQ^{\prime}q}^{*}=Q(Q^{\prime}q)_{A}=Q^{\prime}(Qq)_{A},~{\cal B}_{QQq}^{*}=\sqrt{2}Q(Qq)_{A},
\end{eqnarray}
with $q^{(\prime)}=u,d,s$, and $Q^{(\prime)}=c,b$.

With the above wave functions of doubly heavy baryons and singly heavy baryons,
the overlapping factors $c_{S,A}$ for each transition can be got.
The corresponding results of the overlapping factors $c_{S,A}$ for
the $1/2\to1/2$ transitions induced by the charged current
and the FCNC in Eq.~(\ref{eq:csca})
are collected in Tab. \ref{Tab:overlapping_factors_22}.
For the $1/2\to3/2$ transitions induced by the charged current and the FCNC,
the numerical results of the overlapping factors $c_{A}$ are listed in Tab.~\ref{Tab:overlapping_factors_23}.
Under SU(3) symmetry the doubly heavy baryons can be formed into triplets and the singly heavy baryons can  be formed into an anti-triplet and a sextet. The overlapping factors $c_{S,A}$ can be calculated with SU(3) approach, and the detail calculation can be found in the Appendix~\ref{su3approach}. Using the SU(3) approach, one gets the same numerical results of $c_{S(A)}$ as those listed in Tabs. \ref{Tab:overlapping_factors_22} and \ref{Tab:overlapping_factors_23}.
Then for a spin $1/2$ finial state with a scalar and an axial-vector diquark, the physical form factors are then obtained by
\begin{equation}
	f^{\frac{1}{2}\to\frac{1}{2}}_{i}=c_{S}f_{i,S}+c_{A}f_{i,A},\quad
	g^{\frac{1}{2}\to\frac{1}{2}}_{i}=c_{S}g_{i,S}+c_{A}g_{i,A},\quad f^{\frac{1}{2}\to\frac{1}{2},T}_{i}=c_{S}f_{i,S}^{T}+c_{A}f_{i,A}^{T},\quad g^{\frac{1}{2}\to\frac{1}{2},T}_{i}=c_{S}g_{i,S}^{T}+c_{A}g_{i,A}^{T},\label{eq:physical_ff22}
\end{equation}
where these form factors $f_{i,S(A)}$, $g_{i,S(A)}$, $f_{i,S(A)}^{T}$
or $g_{i,S(A)}^{T}$ are defined by Eqs.~(\ref{eq:matrix_element_2}) and (\ref{eq:matrix_element_2p}).
However, for a spin $3/2$ finial state with only an axial-vector diquark, the physical form factors are then obtained by
\begin{equation}
	f^{\frac{1}{2}\to\frac{3}{2}}_{i}=c_{A}\mathtt{f}_{i},\quad
	g^{\frac{1}{2}\to\frac{3}{2}}_{i}=c_{A}\mathtt{g}_{i},\quad f^{\frac{1}{2}\to\frac{3}{2},T}_{i}=c_{A}\mathtt{f}_{i}^{T},\quad g^{\frac{1}{2}\to\frac{3}{2},T}_{i}=c_{A}\mathtt{g}_{i}^{T},\label{eq:physical_ff23}
\end{equation}
where these form factors $\mathtt{f}_{i}$, $\mathtt{g}_{i}$, ${\mathtt{f}}_{i}^{T}$
or $\mathtt{g}_{i}^{T}$ are defined in Eqs.~(\ref{eq:matrix_element_32nVA})-(\ref{eq:matrix_element_32nT}).


\begin{table}
\caption{Numerical results of the overlapping factors for the $1/2\to1/2$ transitions induced by $c\to d,s$, $b\to u,c$ and $c\to u$, $b\to d,s$. For example, the physical form factor of transition $\Xi_{cc}^{++}\to \Lambda_{c}^{+}$, $f^{\frac{1}{2}\to\frac{1}{2}}_{1}=c_{S}f_{1,S}+c_{A}f_{1,A}$ can be calculated with $c_{S}=\sqrt{6}/4$ and $c_{A}=\sqrt{6}/4$.}
\label{Tab:overlapping_factors_22}
\begin{tabular}{c|c|c|c|c|c|c|c|c}
\hline \hline
transitions& $c_{S}$ & $c_{A}$	&transitions& $c_{S}$ & $c_{A}$& transitions&$c_{S}$ & $c_{A}$\tabularnewline\hline
$\Xi_{cc}^{++}(ccu)\to\Lambda_{c}^{+}(dcu)$ & $\frac{\sqrt{6}}{4}$ & $\frac{\sqrt{6}}{4}$&$\Xi_{bc}^{+}(cbu)\to\Lambda_{b}^{0}(dbu)$ & $\frac{\sqrt{3}}{4}$ & $\frac{\sqrt{3}}{4}$&$\Xi_{bc}^{\prime+}(cbu)\to\Lambda_{b}^{0}(dbu)$ & $-\frac{1}{4}$ & $\frac{3}{4}$\tabularnewline\hline
$\Xi_{cc}^{++}(ccu)\to\Sigma_{c}^{+}(dcu)$ & $-\frac{3\sqrt{2}}{4}$ & $\frac{\sqrt{2}}{4}$&$\Xi_{bc}^{+}(cbu)\to\Sigma_{b}^{0}(dbu)$ & $-\frac{3}{4}$ & $\frac{1}{4}$&	$\Xi_{bc}^{\prime+}(cbu)\to\Sigma_{b}^{0}(dbu)$ & $\frac{\sqrt{3}}{4}$ & $\frac{\sqrt{3}}{4}$\tabularnewline\hline
$\Xi_{cc}^{++}(ccu)\to\Xi_{c}^{+}(scu)$ & $\frac{\sqrt{6}}{4}$ & $\frac{\sqrt{6}}{4}$&$\Xi_{bc}^{+}(cbu)\to\Xi_{b}^{0}(sbu)$ & $\frac{\sqrt{3}}{4}$ & $\frac{\sqrt{3}}{4}$&$\Xi_{bc}^{\prime+}(cbu)\to\Xi_{b}^{0}(sbu)$ & $-\frac{1}{4}$ & $\frac{3}{4}$\tabularnewline\hline
$\Xi_{cc}^{++}(ccu)\to\Xi_{c}^{\prime+}(scu)$ & $-\frac{3\sqrt{2}}{4}$ & $\frac{\sqrt{2}}{4}$&	$\Xi_{bc}^{+}(cbu)\to\Xi_{b}^{\prime0}(sbu)$ & $-\frac{3}{4}$ & $\frac{1}{4}$&$\Xi_{bc}^{\prime+}(cbu)\to\Xi_{b}^{\prime0}(sbu)$ & $\frac{\sqrt{3}}{4}$ & $\frac{\sqrt{3}}{4}$\tabularnewline\hline
$\Xi_{cc}^{+}(ccd)\to\Sigma_{c}^{0}(dcd)$ & $-\frac{3}{2}$ & $\frac{1}{2}$&	$\Xi_{bc}^{0}(cbd)\to\Sigma_{b}^{-}(dbd)$ & $-\frac{3\sqrt{2}}{4}$ & $\frac{\sqrt{2}}{4}$&$\Xi_{bc}^{\prime0}(cbd)\to\Sigma_{b}^{-}(dbd)$ & $\frac{\sqrt{6}}{4}$ & $\frac{\sqrt{6}}{4}$\tabularnewline\hline
$\Xi_{cc}^{+}(ccd)\to\Xi_{c}^{0}(scd)$ & $\frac{\sqrt{6}}{4}$ & $\frac{\sqrt{6}}{4}$&$\Xi_{bc}^{0}(cbd)\to\Xi_{b}^{-}(sbd)$ & $\frac{\sqrt{3}}{4}$ & $\frac{\sqrt{3}}{4}$&$\Xi_{bc}^{\prime0}(cbd)\to\Xi_{b}^{-}(sbd)$ & $-\frac{1}{4}$ & $\frac{3}{4}$\tabularnewline\hline
$\Xi_{cc}^{+}(ccd)\to\Xi_{c}^{\prime0}(scd)$ & $-\frac{3\sqrt{2}}{4}$ & $\frac{\sqrt{2}}{4}$&$\Xi_{bc}^{0}(cbd)\to\Xi_{b}^{\prime-}(sbd)$ & $-\frac{3}{4}$ & $\frac{1}{4}$&$\Xi_{bc}^{\prime0}(cbd)\to\Xi_{b}^{\prime-}(sbd)$ & $\frac{\sqrt{3}}{4}$ & $\frac{\sqrt{3}}{4}$\tabularnewline\hline
$\Omega_{cc}^{+}(ccs)\to\Xi_{c}^{0}(dcs)$ & $-\frac{\sqrt{6}}{4}$ & $-\frac{\sqrt{6}}{4}$&$\Omega_{bc}^{0}(cbs)\to\Xi_{b}^{-}(dbs)$ & $-\frac{\sqrt{3}}{4}$ & $-\frac{\sqrt{3}}{4}$&	$\Omega_{bc}^{\prime0}(cbs)\to\Xi_{b}^{-}(dbs)$ & $\frac{1}{4}$ & $-\frac{3}{4}$\tabularnewline\hline
$\Omega_{cc}^{+}(ccs)\to\Xi_{c}^{\prime0}(dcs)$ & $-\frac{3\sqrt{2}}{4}$ & $\frac{\sqrt{2}}{4}$&$\Omega_{bc}^{0}(cbs)\to\Xi_{b}^{\prime-}(dbs)$ & $-\frac{3}{4}$ & $\frac{1}{4}$&$\Omega_{bc}^{\prime0}(cbs)\to\Xi_{b}^{\prime-}(dbs)$ & $\frac{\sqrt{3}}{4}$ & $\frac{\sqrt{3}}{4}$\tabularnewline\hline
$\Omega_{cc}^{+}(ccs)\to\Omega_{c}^{0}(scs)$ & $-\frac{3}{2}$ & $\frac{1}{2}$&$\Omega_{bc}^{0}(cbs)\to\Omega_{b}^{-}(sbs)$ & $-\frac{3\sqrt{2}}{4}$ & $\frac{\sqrt{2}}{4}$&	$\Omega_{bc}^{\prime0}(cbs)\to\Omega_{b}^{-}(sbs)$ & $\frac{\sqrt{6}}{4}$ & $\frac{\sqrt{6}}{4}$\tabularnewline
\hline \hline
$\Xi_{bb}^{0}(bbu)\to\Sigma_{b}^{+}(ubu)$ & $-\frac{3}{2}$ & $\frac{1}{2}$&$\Xi_{bc}^{+}(bcu)\to\Sigma_{c}^{++}(ucu)$ & $-\frac{3\sqrt{2}}{4}$ & $\frac{\sqrt{2}}{4}$&$\Xi_{bc}^{\prime+}(bcu)\to\Sigma_{c}^{++}(ucu)$ & $-\frac{\sqrt{6}}{4}$ & $-\frac{\sqrt{6}}{4}$\tabularnewline\hline
$\Xi_{bb}^{0}(bbu)\to\Xi_{bc}^{+}(cbu)$ & $\frac{3\sqrt{2}}{4}$ & $\frac{\sqrt{2}}{4}$&$\Xi_{bc}^{+}(bcu)\to\Xi_{cc}^{++}(ccu)$ & $\frac{3\sqrt{2}}{4}$ & $\frac{\sqrt{2}}{4}$&$\Xi_{bc}^{\prime+}(bcu)\to\Xi_{cc}^{++}(ccu)$ & $\frac{\sqrt{6}}{4}$ & $-\frac{\sqrt{6}}{4}$\tabularnewline\hline
$\Xi_{bb}^{0}(bbu)\to\Xi_{bc}^{\prime+}(cbu)$ & $-\frac{\sqrt{6}}{4}$ & $\frac{\sqrt{6}}{4}$&$\Xi_{bc}^{0}(bcd)\to\Lambda_{c}^{+}(ucd)$ & $-\frac{\sqrt{3}}{4}$ & $-\frac{\sqrt{3}}{4}$&$\Xi_{bc}^{\prime0}(bcd)\to\Lambda_{c}^{+}(ucd)$ & $-\frac{1}{4}$ & $\frac{3}{4}$\tabularnewline\hline
$\Xi_{bb}^{-}(bbd)\to\Lambda_{b}^{0}(ubd)$ & $-\frac{\sqrt{6}}{4}$ & $-\frac{\sqrt{6}}{4}$&	$\Xi_{bc}^{0}(bcd)\to\Sigma_{c}^{+}(ucd)$ & $-\frac{3}{4}$ & $\frac{1}{4}$&$\Xi_{bc}^{\prime0}(bcd)\to\Sigma_{c}^{+}(ucd)$ & $-\frac{\sqrt{3}}{4}$ & $-\frac{\sqrt{3}}{4}$\tabularnewline\hline
$\Xi_{bb}^{-}(bbd)\to\Sigma_{b}^{0}(ubd)$ & $-\frac{3\sqrt{2}}{4}$ & $\frac{\sqrt{2}}{4}$&	$\Xi_{bc}^{0}(bcd)\to\Xi_{cc}^{+}(ccd)$ & $\frac{3\sqrt{2}}{4}$ & $\frac{\sqrt{2}}{4}$&$\Xi_{bc}^{\prime0}(bcd)\to\Xi_{cc}^{+}(ccd)$ & $\frac{\sqrt{6}}{4}$ & $-\frac{\sqrt{6}}{4}$\tabularnewline\hline
$\Xi_{bb}^{-}(bbd)\to\Xi_{bc}^{0}(cbd)$ & $\frac{3\sqrt{2}}{4}$ & $\frac{\sqrt{2}}{4}$&$\Omega_{bc}^{0}(bcs)\to\Xi_{c}^{+}(ucs)$ & $-\frac{\sqrt{3}}{4}$ & $-\frac{\sqrt{3}}{4}$&$\Omega_{bc}^{\prime0}(bcs)\to\Xi_{c}^{+}(ucs)$ & $-\frac{1}{4}$ & $\frac{3}{4}$\tabularnewline\hline
$\Xi_{bb}^{-}(bbd)\to\Xi_{bc}^{\prime0}(cbd)$ & $-\frac{\sqrt{6}}{4}$ & $\frac{\sqrt{6}}{4}$&$\Omega_{bc}^{0}(bcs)\to\Xi_{c}^{\prime+}(ucs)$ & $-\frac{3}{4}$ & $\frac{1}{4}$&$\Omega_{bc}^{\prime0}(bcs)\to\Xi_{c}^{\prime+}(ucs)$ & $-\frac{\sqrt{3}}{4}$ & $-\frac{\sqrt{3}}{4}$\tabularnewline\hline
$\Omega_{bb}^{-}(bbs)\to\Xi_{b}^{0}(ubs)$ & $-\frac{\sqrt{6}}{4}$ & $-\frac{\sqrt{6}}{4}$&	$\Omega_{bc}^{0}(bcs)\to\Omega_{cc}^{+}(ccs)$ & $\frac{3\sqrt{2}}{4}$ & $\frac{\sqrt{2}}{4}$&$\Omega_{bc}^{\prime0}(bcs)\to\Omega_{cc}^{+}(ccs)$ & $\frac{\sqrt{6}}{4}$ & $-\frac{\sqrt{6}}{4}$\tabularnewline\hline
$\Omega_{bb}^{-}(bbs)\to\Xi_{b}^{\prime0}(ubs)$ & $-\frac{3\sqrt{2}}{4}$ & $\frac{\sqrt{2}}{4}$&$\Omega_{bb}^{-}(bbs)\to\Omega_{bc}^{0}(cbs)$ & $\frac{3\sqrt{2}}{4}$ & $\frac{\sqrt{2}}{4}$&$\Omega_{bb}^{-}(bbs)\to\Omega_{bc}^{\prime0}(cbs)$ & $-\frac{\sqrt{6}}{4}$ & $\frac{\sqrt{6}}{4}$\tabularnewline
\hline	\hline
$\Xi_{cc}^{++}(ccu)\to\Sigma_{c}^{++}(ucu)$  & $-\frac{3}{2}$  & $\frac{1}{2}$  &$\Xi_{bc}^{+}(cbu)\to\Sigma_{b}^{+}(ubu)$  & $-\frac{3\sqrt{2}}{4}$  & $\frac{\sqrt{2}}{4}$
&$\Xi_{bc}^{\prime+}(cbu)\to\Sigma_{b}^{+}(ubu)$  & $\frac{\sqrt{6}}{4}$  &$\frac{\sqrt{6}}{4} $\tabularnewline
\hline
$\Xi_{cc}^{+}(ccd)\to\Lambda_{c}^{+}(ucd)$  & $-\frac{\sqrt{6}}{4}$  & $-\frac{\sqrt{6}}{4}$  &$\Xi_{bc}^{0}(cbd)\to\Lambda_{b}^{+}(ubd)$  & $-\frac{\sqrt{3}}{4}$  & $-\frac{\sqrt{3}}{4}$
&$\Xi_{bc}^{\prime0}(cbd)\to\Lambda_{b}^{0}(ubd)$  & $\frac{1}{4}$  &$-\frac{3}{4} $\tabularnewline
\hline
$\Xi_{cc}^{+}(ccd)\to\Sigma_{c}^{+}(ucd)$  & $-\frac{3\sqrt{2}}{4}$  & $\frac{\sqrt{2}}{4}$  &$\Xi_{bc}^{0}(cbd)\to\Sigma_{b}^{0}(ubd)$  & $-\frac{3}{4}$  & $\frac{1}{4}$
&$\Xi_{bc}^{\prime0}(cbd)\to\Sigma_{b}^{0}(ubd)$  & $\frac{\sqrt{3}}{4}$  &$\frac{\sqrt{3}}{4} $\tabularnewline
\hline
$\Omega_{cc}^{+}(ccs)\to\Xi_{c}^{+}(ucs)$  & $-\frac{\sqrt{6}}{4}$  & $\frac{\sqrt{6}}{4}$  &$\Omega_{bc}^{0}(cbs)\to\Xi_{b}^{0}(ubs)$  & $-\frac{\sqrt{3}}{4}$  & $-\frac{\sqrt{3}}{4}$
&$\Omega_{bc}^{\prime0}(cbs)\to\Xi_{b}^{0}(ubs)$  & $\frac{1}{4}$  &$-\frac{3}{4} $\tabularnewline
\hline
$\Omega_{cc}^{+}(ccs)\to\Xi_{c}^{\prime+}(ucs)$  & $-\frac{3\sqrt{2}}{4}$  & $\frac{\sqrt{2}}{4}$  &$\Omega_{bc}^{0}(cbs)\to\Xi_{b}^{\prime0}(ubs)$  & $-\frac{3}{4}$  & $\frac{1}{4}$
&$\Omega_{bc}^{\prime0}(cbs)\to\Xi_{b}^{\prime0}(ubs)$  & $\frac{\sqrt{3}}{4}$  &$\frac{\sqrt{3}}{4} $\tabularnewline
\hline\hline
$\Xi_{bb}^{0}(bbu)\to\Xi_{b}^{0}(sbu)$  & $\frac{\sqrt{6}}{4}$  & $\frac{\sqrt{6}}{4}$  &$\Xi_{bc}^{+}(bcu)\to\Xi_{c}^{+}(scu)$  & $\frac{\sqrt{3}}{4}$  & $\frac{\sqrt{3}}{4}$
&$\Xi_{bc}^{\prime+}(bcu)\to\Xi_{c}^{+}(scu)$  & $\frac{1}{4}$  &$-\frac{3}{4} $\tabularnewline\hline
$\Xi_{bb}^{-}(bbd)\to\Xi_{b}^{-}(sbd)$ & $\frac{\sqrt{6}}{4}$  & $\frac{\sqrt{6}}{4}$& $\Xi_{bc}^{0}(bcd)\to\Xi_{c}^{0}(scd)$  & $\frac{\sqrt{3}}{4}$  & $\frac{\sqrt{3}}{4}$
&$\Xi_{bc}^{\prime0}(bcd)\to\Xi_{c}^{0}(scd)$  &  $\frac{1}{4}$  & $-\frac{3}{4} $\tabularnewline\hline
$\Xi_{bb}^{0}(bbu)\to\Xi_{b}^{\prime0}(sbu)$  &  $-\frac{3\sqrt{2}}{4}$  & $\frac{\sqrt{2}}{4}$
& $\Xi_{bc}^{+}(bcu)\to\Xi_{c}^{\prime+}(scu)$  & $-\frac{3}{4}$& $\frac{1}{4}$
&$\Xi_{bc}^{\prime+}(bcu)\to\Xi_{c}^{\prime+}(scu)$  & $-\frac{\sqrt{3}}{4}$ & $-\frac{\sqrt{3}}{4}$\tabularnewline\hline
$\Xi_{bb}^{-}(bbd)\to\Xi_{b}^{\prime-}(sbd)$ & $-\frac{3\sqrt{2}}{4}$ & $\frac{\sqrt{2}}{4}$
&$\Xi_{bc}^{0}(bcd)\to\Xi_{c}^{\prime0}(scd)$  &  $-\frac{3}{4}$  & $\frac{1}{4}$
&$\Xi_{bc}^{\prime0}(bcd)\to\Xi_{c}^{\prime0}(scd)$  &  $-\frac{\sqrt{3}}{4}$  & $-\frac{\sqrt{3}}{4}$\tabularnewline\hline
$\Omega_{bb}^{-}(bbs)\to\Omega_{b}^{-}(sbs)$  & $-\frac{3}{2}$  & $\frac{1}{2}$ & $\Omega_{bc}^{0}(bcs)\to\Omega_{c}^{0}(scs)$  & $-\frac{3\sqrt{2}}{4}$  & $\frac{\sqrt{2}}{4}$ &$\Omega_{bc}^{\prime0}(bcs)\to\Omega_{c}^{0}(scs)$  & $-\frac{\sqrt{6}}{4}$& $-\frac{\sqrt{6}}{4}$ \tabularnewline\hline
$\Xi_{bb}^{0}(bbu)\to\Lambda_{b}^{0}(dbu)$  & $\frac{\sqrt{6}}{4}$  & $\frac{\sqrt{6}}{4}$
 & $\Xi_{bc}^{+}(bcu)\to\Lambda_{c}^{+}(dcu)$  & $\frac{\sqrt{3}}{4}$  & $\frac{\sqrt{3}}{4}$& $\Xi_{bc}^{\prime+}(bcu)\to\Lambda_{c}^{+}(dcu)$  & $\frac{1}{4}$  & $-\frac{3}{4}$\tabularnewline\hline
$\Omega_{bb}^{-}(bbs)\to\Xi_{b}^{-}(dbs)$  & $-\frac{\sqrt{6}}{4}$  & $-\frac{\sqrt{6}}{4}$
&$\Omega_{bc}^{0}(bcs)\to\Xi_{c}^{0}(dcs)$  & $-\frac{\sqrt{3}}{4}$  & $-\frac{\sqrt{3}}{4}$ & $\Omega_{bc}^{\prime0}(bcs)\to\Xi_{c}^{0}(dcs)$  & $-\frac{1}{4}$  & $\frac{3}{4}$\tabularnewline\hline
$\Xi_{bb}^{0}(bbu)\to\Sigma_{b}^{0}(dbu)$  & $-\frac{3\sqrt{2}}{4}$  & $\frac{\sqrt{2}}{4}$&$\Xi_{bc}^{+}(bcu)\to\Sigma_{c}^{+}(dcu)$  & $-\frac{3}{4}$  & $\frac{1}{4}$& $\Xi_{bc}^{\prime+}(bcu)\to\Sigma_{c}^{+}(dcu)$  & $-\frac{\sqrt{3}}{4}$  & $-\frac{\sqrt{3}}{4}$\tabularnewline\hline
 $\Xi_{bb}^{-}(bbd)\to\Sigma_{b}^{-}(dbd)$  & $-\frac{3}{2}$  & $\frac{1}{2}$&$\Xi_{bc}^{0}(bcd)\to\Sigma_{c}^{0}(dcd)$  & $-\frac{3\sqrt{2}}{4}$  & $\frac{\sqrt{2}}{4}$& $\Xi_{bc}^{\prime0}(bcd)\to\Sigma_{c}^{0}(dcd)$  & $-\frac{\sqrt{6}}{4}$  & $-\frac{\sqrt{6}}{4}$\tabularnewline\hline
$\Omega_{bb}^{-}(bbs)\to\Xi_{b}^{\prime-}(dbs)$  & $-\frac{3\sqrt{2}}{4}$  & $\frac{\sqrt{2}}{4}$&$\Omega_{bc}^{0}(bcs)\to\Xi_{c}^{\prime0}(dcs)$  & $-\frac{3}{4}$  & $\frac{1}{4}$& $\Omega_{bc}^{\prime0}(bcs)\to\Xi_{c}^{\prime0}(dcs)$  & $-\frac{\sqrt{3}}{4}$  & $-\frac{\sqrt{3}}{4}$\tabularnewline\hline	\hline
\end{tabular}
\end{table}	

\begin{table}
\caption{Numerical results of the overlapping factors for the $1/2\to3/2$ transitions induced by $c\to d,s$, $b\to u,c$ and $c\to u$, $b\to d,s$. For example, the physical form factor of transition $\Xi_{cc}^{++}\to \Sigma_{c}^{*+}$, $f^{\frac{1}{2}\to\frac{3}{2}}_{1}=c_{A}\mathtt{f}_{1}$ can be calculated with $c_{A}=1/\sqrt{2}$.}
\label{Tab:overlapping_factors_23}
\begin{tabular}{c|c|c|c|c|c}
\hline\hline
transitions  & $c_{A}$  & transitions  & $c_{A}$  & transitions  & $c_{A}$ \tabularnewline
\hline
$\Xi_{cc}^{++}(ccu)\to\Sigma_{c}^{*+}(dcu)$  & $\frac{1}{\sqrt{2}}$  & $\Xi_{bc}^{+}(cbu)\to\Sigma_{b}^{*0}(dbu)$  & $\frac{1}{2}$  & $\Xi_{bc}^{\prime+}(cbu)\to\Sigma_{b}^{*0}(dbu)$  & $\frac{\sqrt{3}}{2}$ \tabularnewline
\hline
$\Xi_{cc}^{+}(ccd)\to\Sigma_{c}^{*0}(dcd)$  & $1$  & $\Xi_{bc}^{0}(cbd)\to\Sigma_{b}^{*-}(dbd)$  & $\frac{\sqrt{2}}{2}$  & $\Xi_{bc}^{\prime0}(cbd)\to\Sigma_{b}^{*-}(dbd)$  & $\frac{\sqrt{6}}{2}$ \tabularnewline
\hline
$\Omega_{cc}^{+}(ccs)\to\Xi_{c}^{\prime*0}(dcs)$  & $\frac{1}{\sqrt{2}}$  & $\Omega_{bc}^{0}(cbs)\to\Xi_{b}^{\prime*-}(dbs)$  & $\frac{1}{2}$  & $\Omega_{bc}^{\prime0}(cbs)\to\Xi_{b}^{\prime*-}(dbs)$  & $\frac{\sqrt{3}}{2}$ \tabularnewline
\hline
$\Xi_{cc}^{++}(ccu)\to\Xi_{c}^{\prime*+}(scu)$  & $\frac{1}{\sqrt{2}}$  & $\Xi_{bc}^{+}(cbu)\to\Xi_{b}^{\prime*0}(sbu)$  & $\frac{1}{2}$  & $\Xi_{bc}^{\prime+}(cbu)\to\Xi_{b}^{\prime*0}(sbu)$  & $\frac{\sqrt{3}}{2}$ \tabularnewline
\hline
$\Xi_{cc}^{+}(ccd)\to\Xi_{c}^{\prime*0}(scd)$  & $\quad\frac{1}{\sqrt{2}}$  & $\Xi_{bc}^{0}(cbd)\to\Xi_{b}^{\prime*-}(sbd)$  & $\frac{1}{2}$  & $\Xi_{bc}^{\prime0}(cbd)\to\Xi_{b}^{\prime*-}(sbd)$  & $\frac{\sqrt{3}}{2}$ \tabularnewline
\hline
$\Omega_{cc}^{+}(ccs)\to\Omega_{c}^{*0}(scs)$  & $1$  & $\Omega_{bc}^{0}(cbs)\to\Omega_{b}^{*-}(sbs)$  & $\frac{\sqrt{2}}{2}$  & $\Omega_{bc}^{\prime0}(cbs)\to\Omega_{b}^{*-}(sbs)$  & $\frac{\sqrt{6}}{2}$ \tabularnewline
\hline \hline
$\Xi_{bb}^{0}(bbu)\to\Sigma_{b}^{*+}(ubu)$  & $1$ & $\Xi_{bc}^{+}(bcu)\to\Sigma_{c}^{*++}(ucu)$  & $\frac{\sqrt{2}}{2}$ & $\Xi_{bc}^{\prime+}(bcu)\to\Sigma_{c}^{*++}(ucu)$  & $-\frac{\sqrt{6}}{2}$\tabularnewline
\hline
$\Xi_{bb}^{-}(bbd)\to\Sigma_{b}^{*0}(ubd)$  & $\frac{1}{\sqrt{2}}$ & $\Xi_{bc}^{0}(bcd)\to\Sigma_{c}^{*+}(ucd)$  & $\frac{1}{2}$ & $\Xi_{bc}^{\prime0}(bcd)\to\Sigma_{c}^{*+}(ucd)$  & $-\frac{\sqrt{3}}{2}$\tabularnewline
\hline
$\Omega_{bb}^{-}(bbs)\to\Xi_{b}^{\prime*0}(ubs)$  & $\frac{1}{\sqrt{2}}$ & $\Omega_{bc}^{0}(bcs)\to\Xi_{c}^{\prime*+}(ucs)$  & $\frac{1}{2}$ & $\Omega_{bc}^{\prime0}(bcs)\to\Xi_{c}^{\prime*+}(ucs)$  & $-\frac{\sqrt{3}}{2}$\tabularnewline
\hline
$\Xi_{bb}^{0}(bbu)\to\Xi_{bc}^{*+}(cbu)$  & $\quad\frac{1}{\sqrt{2}}$ & $\Xi_{bc}^{+}(bcu)\to\Xi_{cc}^{*++}(ccu)$  & $\frac{\sqrt{2}}{2}$ & $\Xi_{bc}^{\prime+}(bcu)\to\Xi_{cc}^{*++}(ccu)$  & $-\frac{\sqrt{6}}{2}$\tabularnewline
\hline
$\Xi_{bb}^{-}(bbd)\to\Xi_{bc}^{*0}(cbd)$  & $\frac{1}{\sqrt{2}}$ & $\Xi_{bc}^{0}(bcd)\to\Xi_{cc}^{*+}(ccd)$  & $\frac{\sqrt{2}}{2}$ & $\Xi_{bc}^{\prime0}(bcd)\to\Xi_{cc}^{*+}(ccd)$  & $-\frac{\sqrt{6}}{2}$\tabularnewline
\hline
$\Omega_{bb}^{-}(bbs)\to\Omega_{bc}^{*0}(cbs)$  & $\frac{1}{\sqrt{2}}$ & $\Omega_{bc}^{0}(bcs)\to\Omega_{cc}^{*+}(ccs)$  & $\frac{\sqrt{2}}{2}$ & $\Omega_{bc}^{\prime0}(bcs)\to\Omega_{cc}^{*+}(ccs)$  & $-\frac{\sqrt{6}}{2}$\tabularnewline
		\hline	\hline
		$\Xi_{cc}^{++}(ccu)\to\Sigma_{c}^{*++}(ucu)$ & $1$ & $\Xi_{bc}^{+}(cbu)\to\Sigma_{b}^{*+}(ubu)$ & $\frac{1}{\sqrt{2}}$ &$\Xi_{bc}^{\prime+}(cbu)\to\Sigma_{b}^{*+}(ubu)$ & $\frac{\sqrt{6}}{2}$ \tabularnewline
		\hline
		$\Xi_{cc}^{+}(ccd)\to\Sigma_{c}^{*+}(ucd)$ & $\frac{1}{\sqrt{2}}$ & $\Xi_{bc}^{0}(cbd)\to\Sigma_{b}^{*0}(ubd)$ & $\frac{1}{2}$ &$\Xi_{bc}^{\prime0}(cbd)\to\Sigma_{b}^{*0}(ubd)$ & $\frac{\sqrt{3}}{2}$ \tabularnewline
		\hline
		$\Omega_{cc}^{+}(ccs)\to\Xi_{c}^{\prime*+}(ucs)$ & $\frac{1}{\sqrt{2}}$ & $\Omega_{bc}^{0}(cbs)\to\Xi_{b}^{\prime*0}(ubs)$ & $\frac{1}{2}$ &$\Omega_{bc}^{\prime0}(cbs)\to\Xi_{b}^{\prime*0}(ubs)$ & $\frac{\sqrt{3}}{2}$ \tabularnewline
		\hline\hline
		$\Xi_{bb}^{0}(bbu)\to\Xi_{b}^{\prime*0}(sbu)$ & $\frac{1}{\sqrt{2}}$ & $\Xi_{bc}^{+}(bcu)\to\Xi_{c}^{\prime*+}(scu)$ & $\frac{1}{2}$ &$\Xi_{bc}^{\prime+}(bcu)\to\Xi_{c}^{\prime*+}(scu)$ & $-\frac{\sqrt{3}}{2}$ \tabularnewline
		\hline
		$\Xi_{bb}^{-}(bbd)\to\Xi_{b}^{\prime*-}(sbd)$ & $\frac{1}{\sqrt{2}}$ &$\Xi_{bc}^{0}(bcd)\to\Xi_{c}^{\prime*0}(scd)$ & $\frac{1}{2}$ & $\Xi_{bc}^{\prime0}(bcd)\to\Xi_{c}^{\prime*0}(scd)$ & $-\frac{\sqrt{3}}{2}$ \tabularnewline
		\hline
		$\Omega_{bb}^{-}(bbs)\to\Omega_{b}^{*-}(sbs)$ & $1$ &$\Omega_{bc}^{0}(bcs)\to\Omega_{c}^{*0}(scs)$ & $\frac{\sqrt{2}}{2}$ &$\Omega_{bc}^{\prime0}(bcs)\to\Omega_{c}^{*0}(scs)$ & $-\frac{\sqrt{6}}{2}$  \tabularnewline
		\hline
         $\Xi_{bb}^{0}(bbu)\to\Sigma_{b}^{*0}(dbu)$ & $\frac{1}{\sqrt{2}}$&
		 $\Xi_{bc}^{+}(bcu)\to\Sigma_{c}^{*+}(dcu)$ & $\frac{1}{2}$& $\Xi_{bc}^{\prime+}(bcu)\to\Sigma_{c}^{*+}(dcu)$ & $-\frac{\sqrt{3}}{2}$\tabularnewline
		\hline
		 $\Xi_{bb}^{-}(bbd)\to\Sigma_{b}^{*-}(dbd)$ & $1$&$\Xi_{bc}^{0}(bcd)\to\Sigma_{c}^{*0}(dcd)$ & $\frac{\sqrt{2}}{2}$& $\Xi_{bc}^{\prime0}(bcd)\to\Sigma_{c}^{*0}(dcd)$ & $-\frac{\sqrt{6}}{2}$\tabularnewline
		\hline$\Omega_{bb}^{-}(bbs)\to\Xi_{b}^{\prime*-}(dbs)$ & $\frac{1}{\sqrt{2}}$& $\Omega_{bc}^{0}(bcs)\to\Xi_{c}^{\prime*+}(dcs)$ & $\frac{1}{2}$
		& $\Omega_{bc}^{\prime0}(bcs)\to\Xi_{c}^{\prime*+}(dcs)$ & $-\frac{\sqrt{3}}{2}$\tabularnewline
		\hline	\hline
	\end{tabular}
\end{table}

\section{Numerical results of form factors}
The masses of quarks are taken from Refs.~\cite{Lu:2007sg,Wang:2007sxa,Wang:2008xt,
Wang:2008ci,Wang:2009mi,Chen:2009qk,Li:2010bb,
Verma:2011yw,Shi:2016gqt},
\begin{eqnarray}
 m_u=m_d= 0.25~{\rm GeV}, \;\; m_s=0.37~{\rm GeV}, \;\; m_c=1.4~{\rm GeV}, \;\; m_b=4.8~{\rm GeV}.\label{eq:mass_quark}
\end{eqnarray}
The masses of diquark are approximatively taken as,
\begin{eqnarray}
m_{[cq]}=m_{c}+m_{q}~\text{and}~m_{[bq]}=m_{b}+m_{q}~\text{with}~q=u,d,s.\label{eq:mass_diquark}
\end{eqnarray}
The masses of all baryons, lifetime of parent baryons and shape parameters $\beta$ in Eq.~(\ref{eq:Gauss}) are collected in Tab.~\ref{Tab:para_doubly_heavy}
~\cite{1707.01621,Brown:2014ena,Aaij:2018wzf,Cheng:2018mwu,Karliner:2014gca,Kiselev:2001fw,Olive:2016xmw}.
\begin{table}[!htb]
\caption{Masses of all baryons (in unit of GeV), lifetimes (in unit of fs) of parent baryons and the shape parameters $\beta$'s in the Gaussian-type wave functions Eq.~(\ref{eq:Gauss})~\cite{1707.01621,Brown:2014ena,Aaij:2018wzf,Cheng:2018mwu,Karliner:2014gca,Kiselev:2001fw,Olive:2016xmw}.}
\label{Tab:para_doubly_heavy} %
\begin{tabular}{c|c|c|c|c|c|c|c|c|c}
\hline \hline
baryons  & $\Xi_{cc}^{++}$  & $\Xi_{cc}^{+}$  & $\Omega_{cc}^{+}$  & $\Xi_{bc}^{+}$  & $\Xi_{bc}^{0}$  & $\Omega_{bc}^{0}$  & $\Xi_{bb}^{0}$  & $\Xi_{bb}^{-}$  & $\Omega_{bb}^{-}$ \tabularnewline
\hline
masses  & $3.621$  & $3.621$   & $3.738$   & $6.943$   & $6.943$   & $6.998$   & $10.143$  & $10.143$   & $10.273$\tabularnewline
\hline
lifetimes  & $256$  & $45$  & $180$   & $244$   & $93$   & $220$  & $370$   & $370$  & $800$\tabularnewline
\hline \hline
baryons&~$\Lambda_{c}^{+}$~  & ~$\Sigma_{c}^{++}$~  & ~$\Sigma_{c}^{+}$~  & ~$\Sigma_{c}^{0}$~  & ~$\Xi_{c}^{+}$~  & ~$\Xi_{c}^{\prime+}$~  & ~$\Xi_{c}^{0}$~  & ~$\Xi_{c}^{\prime0}$~  & ~$\Omega_{c}^{0}$~ \tabularnewline\hline
masses&$2.286$  & $2.454$  & $2.453$  & $2.454$  & $2.468$  & $2.576$  & $2.471$  & $2.578$  & $2.695$ \tabularnewline\hline
baryons&$\Lambda_{b}^{0}$  & $\Sigma_{b}^{+}$  & $\Sigma_{b}^{0}$  & $\Sigma_{b}^{-}$  & $\Xi_{b}^{0}$ & $\Xi_{b}^{\prime0}$ & $\Xi_{b}^{-}$  & $\Xi_{b}^{\prime-}$  & $\Omega_{b}^{-}$\tabularnewline\hline
masses&$5.620$  & $5.811$  & $5.814$  & $5.816$  & $5.793$ & $5.935$ & $5.795$  & $5.935$  & $6.046$\tabularnewline
\hline\hline
baryons&$\Sigma_{c}^{*++}$ & $\Sigma_{c}^{*+}$  & $\Sigma_{c}^{*0}$ & $\Xi_{c}^{\prime*+}$ & $\Xi_{c}^{\prime*0}$  & $\Omega_{c}^{*0}$ & $\Xi_{cc}^{*++}$ & $\Xi_{cc}^{*+}$  & $\Omega_{cc}^{*+}$\tabularnewline\hline
masses&$2.518$  & $2.518$  & $2.518$  & $2.646$  & $2.646$  & $2.766$  & $3.692$  & $3.692$  & $3.822$ \tabularnewline\hline
baryons&$\Sigma_{b}^{*+}$ & $\Sigma_{b}^{*0}$  & $\Sigma_{b}^{*-}$ & $\Xi_{b}^{\prime*0}$ & $\Xi_{b}^{\prime*-}$  & $\Omega_{b}^{*-}$ & $\Xi_{bc}^{*+}$ & $\Xi_{bc}^{*0}$  & $\Omega_{bc}^{*0}$\tabularnewline\hline
masses&$5.832$  & $5.833$  & $5.835$  & $5.949$  & $5.955$  & $6.085$  & $6.985$  & $6.985$  & $7.059$ \tabularnewline
\hline\hline
~~~$\beta_{u[cq]}$ & $\beta_{d[cq]}$ & $\beta_{s[cq]}$ & $\beta_{c[cq]}$ & $\beta_{b[cq]}$& $\beta_{u[bq]}$ & $\beta_{d[bq]}$ & $\beta_{s[bq]}$ & $\beta_{c[bq]}$ & $\beta_{b[bq]}$~~~\tabularnewline
\hline
$0.470$ & $0.470$ & $0.535$ & $0.753$ & $0.886$ &$0.562$ & $0.562$ & $0.623$ & $0.886$ & $1.472$\tabularnewline
\hline \hline
\end{tabular}
\end{table}
Using these analytical expression of form factors shown in Subsec.~\ref{subsec_lightfrontquarkmodel} and the input parameters listed in Eqs.~(\ref{eq:mass_quark})-(\ref{eq:mass_diquark}) and Tab.~\ref{Tab:para_doubly_heavy}, one can calculate the form factors with the scalar or axial vector diquarks in Eqs.~(\ref{scalar diquark}), (\ref{eq:momentum_wave_function_1/2gamma}) and (\ref{axial-vector diquarkprime}). While for each form factors are functions of $q^2$, in order to obtain the dependence of form factors on the momentum $q^2$, we take the following parametrization scheme for $b\to u,d,s,c$ processes,
\begin{align}
&F(q^{2})=\frac{F(0)}{1-\frac{q^{2}}{m_{{\rm fit}}^{2}}+\delta\left(\frac{q^{2}}{m_{{\rm fit}}^{2}}\right)^{2}}, \label{eq:main_fit_formula_bdecay}
\end{align}
here $F(0)$ is the numerical result of form factor at $q^2=0$, $m_{\rm fit}$ and $\delta$ are two parameters waiting for fitting from numerical result of form factor at different $q^2$ values.
When the fitting result of $m_{\rm fit}$ is an imaginary result using the above parametrization scheme, we need to take the modified parametrization scheme as follows,
\begin{align}
&F(q^{2})=\frac{F(0)}{1+\frac{q^{2}}{m_{{\rm fit}}^{2}}+\delta\left(\frac{q^{2}}{m_{{\rm fit}}^{2}}\right)^{2}}, \label{eq:auxiliary_fit_formula_bdecay}
\end{align}
In the tables, we mark the imaginary results with superscripts  ``$*$".  While for c quark decay process, the single pole structure is assumed
\begin{align}
&F(q^{2})=\frac{F(0)}{1-\frac{q^{2}}{m_{{\rm pole}}^{2}}},\label{eq:main_fit_formula_cdecay}
\end{align}
for $c\to u,~d,~s$ decays, $m_{{\rm pole}}$s are respectively $1.87$, $1.87$, $1.97$ GeV.
\begin{itemize}
\item
For the charged current transition ${1}/{2}\to{1}/{2}$, the results for  form factors with a scalar diquark or an axial-vector diquark spectator are shown in Tabs. \ref{Tab:ff_ccc}, \ref{Tab:ff_bbb} and \ref{Tab:ff_bcc}.
As shown in Eq.~(\ref{eq:matrix_element_2}), the numerical results of the form factors can be used to calculate the physical hadronic transition matrix elements.
We take $\Xi_{bb}^{0}\to\Sigma_{b}^{+}$ as an example to show the $q^2$-dependence of form factors in Fig.~\ref{fig:Fxibbsigmab}. There is no singular point for the form factors $f_{1,2,3}$ and $g_{1,2,3}$ in the integration interval shown in Fig.~\ref{fig:Fxibbsigmab}.
\begin{figure}
\includegraphics[width=0.8\columnwidth]{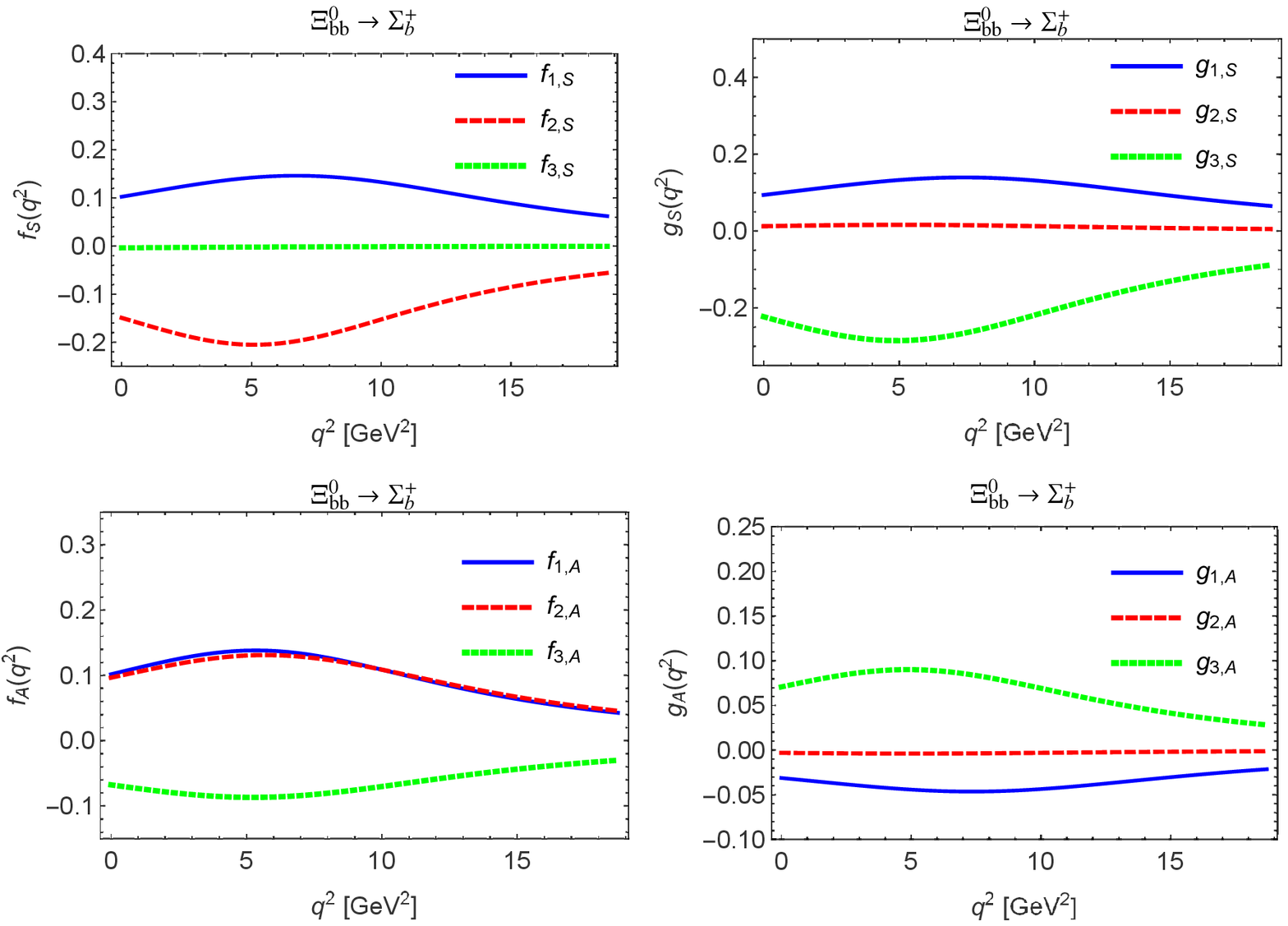}
\caption{$q^2$ dependence of the form factors for the transition $\Xi_{bb}^{0}\to\Sigma_{b}^{+}$. The two graphs in the first line correspond to form factors with scalar diquarks, the two graphs in the second correspond to form factors with axial-vector diquarks. The numerical result of $F(0)$, $\delta$ and $m_{fit}$ are shown in Tab.~\ref{Tab:ff_bbb}.}
\label{fig:Fxibbsigmab}
\end{figure}
\item  For the FCNC transition ${1}/{2}\to{1}/{2}$, the results for  form factors with a scalar diquark or an axial-vector diquark spectator are shown in Tabs. \ref{Tab:ff_cucc_axial},~\ref{Tab:ff_bsbb_axial} and \ref{Tab:ff_bdbb_axial}. With the help of the results of the form factors and Eqs.~(\ref{eq:matrix_element_2})-(\ref{eq:matrix_element_2p}), one can calculate the physical hadronic transition matrix elements.  $\Xi_{bb}^{0}\to\Lambda_{b}^{0}$ is taken as an example to show the $q^2$-dependence of these form factors which are depicted in Fig.~\ref{fig:Fxibblambdab}.
As one can see, these form factors are stable and no divergence exists in the integration interval.
\begin{figure}
\includegraphics[width=0.8\columnwidth]{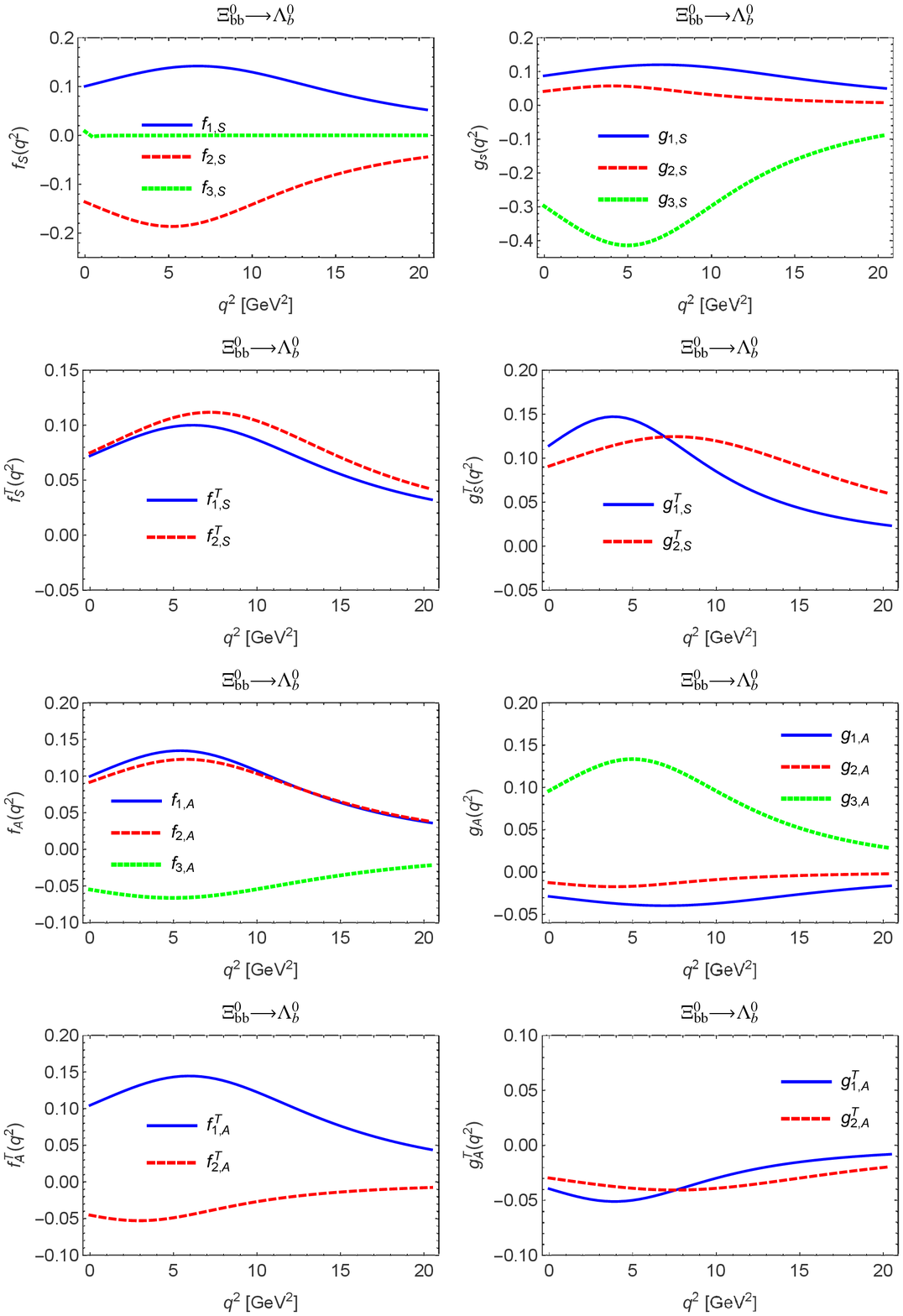}
\caption{$q^2$ dependence of the form factors for $\Xi_{bb}^{0}\to\Lambda_{b}^{0}$ . The first four graphs correspond to form factors with scalar diquark, the last four graphs correspond to form factors with axial-vector diquark. Here the numerical results of $F(0)$, $\delta$ and $m_{fit}$ are shown in Tab.~\ref{Tab:ff_bsbb_axial}.}
\label{fig:Fxibblambdab}
\end{figure}
\item  For the charged current transition ${1}/{2}\to{3}/{2}$, the results for  form factors with an axial-vector diquark spectator are shown in Tabs. \ref{Tab:ff32_cdscc} and \ref{Tab:ff32_bucbb}. As shown in Eq.~(\ref{eq:matrix_element_32nVA}), the numerical results of the form factors can be used to calculate the physical hadronic transition matrix elements. In Fig.~\ref{fig:Fxibbsigmastar} we use $\Xi_{bb}^{0}\to\Sigma_{b}^{*0}$ as an example to show the $q^2$-dependence of form factors. As shown in Fig.~\ref{fig:Fxibbsigmastar} these form factors are stable, which indicates our fitting result in the Tabs. \ref{Tab:ff32_cdscc} and \ref{Tab:ff32_bucbb} are reliable.
\begin{figure}
\includegraphics[width=0.8\columnwidth]{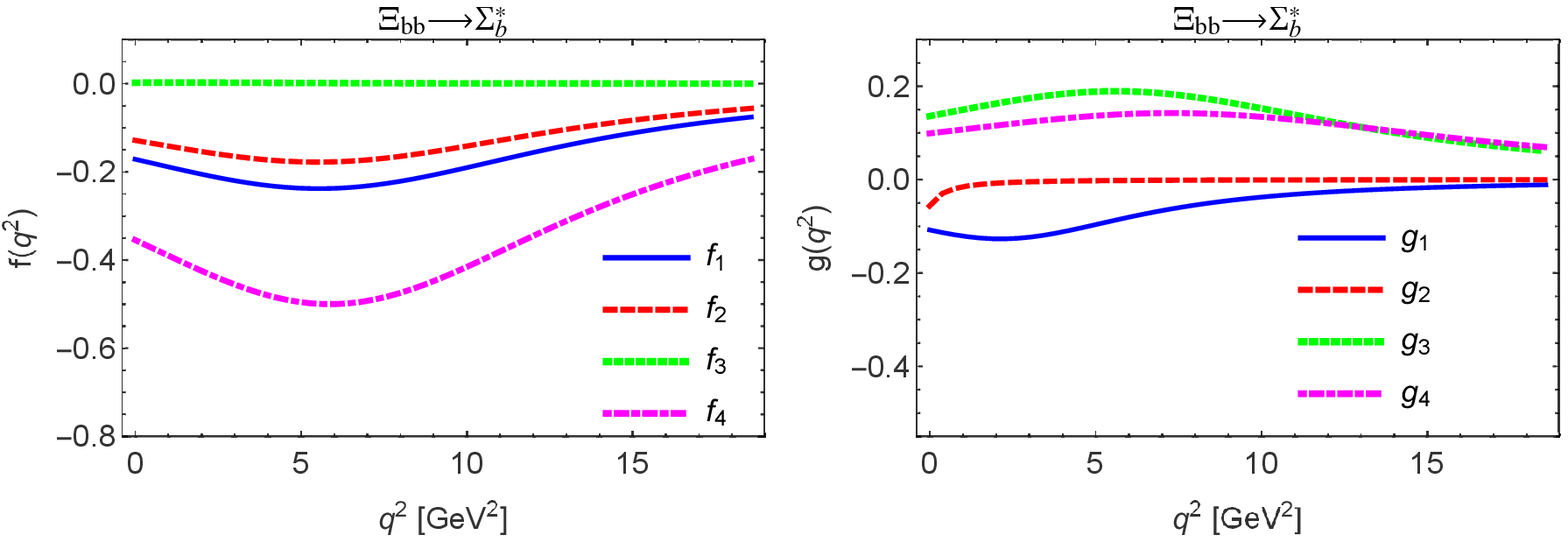}
\caption{$q^2$ dependence of the transition $\Xi_{bb}\to\Sigma_{b}^{*}$ form factors. The numerical result of the parameters $F(0)$, $\delta$ and $m_{fit}$ are shown in Tab.~\ref{Tab:ff32_bucbb}.}
\label{fig:Fxibbsigmastar}
\end{figure}
\item For the FCNC transition ${1}/{2}\to{3}/{2}$, the results for  form factors with an axial-vector diquark are shown in Tabs. \ref{Tab:fcnc32_cu}, \ref{Tab:fcnc32_bd} and~\ref{Tab:fcnc32_bs}. As shown in Eqs.~(\ref{eq:matrix_element_32nVA}) and (\ref{eq:matrix_element_32nT}), the numerical results of the form factors can be used to calculate the physical hadronic transition matrix elements. To describe the dependence of form factors on $q^2$, we take the transition $\Omega_{bb}\to\Xi_{b}^{\prime*-}$  as an example shown in Fig.~\ref{fig:FOmegabbXibS}. The curves are all approaching to zero at large $q^2$ stably.
\begin{figure}
\includegraphics[width=0.8\columnwidth]{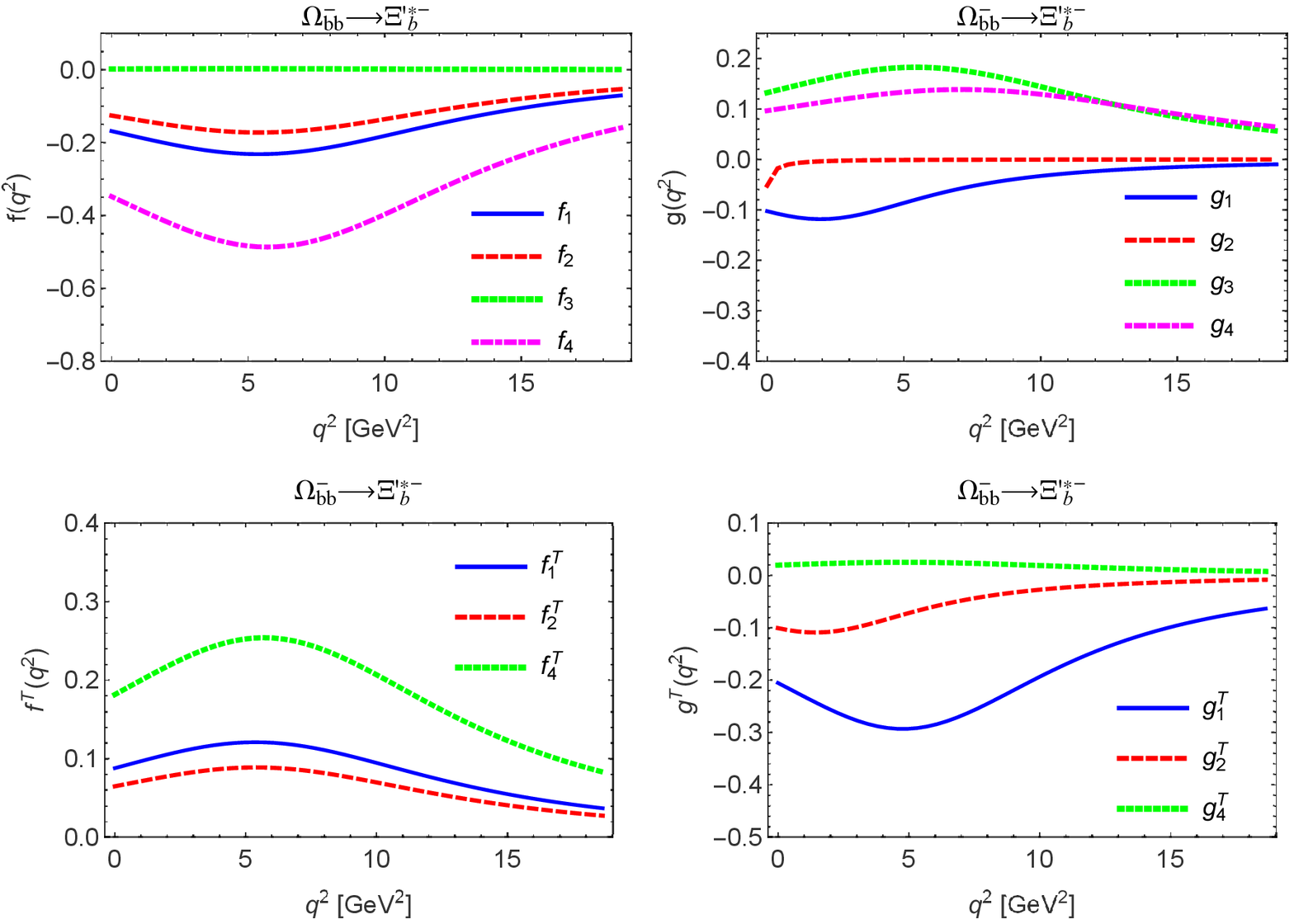}
\caption{$q^2$ dependence of the transition $\Omega_{bb}^{-}\to\Xi_{b}^{\prime*-}$ form factors. The numerical result of the parameters $F(0)$, $\delta$ and $m_{fit}$ are shown in Tab.~\ref{Tab:fcnc32_bd}.}
\label{fig:FOmegabbXibS}
\end{figure}
\end{itemize}
\begin{table}\scriptsize
 \caption{
 Numerical results for the transition $1/2\to 1/2$ form factors $f_{i,S(A)}$ and $g_{i,S(A)}$ at $q^2=0$ of $c\to d,s$ processes. The parametrization scheme in Eq.~(\ref{eq:main_fit_formula_cdecay}) is introduced for these form factors,
and the values of the singly pole $m_{\rm pole}$s are taken as $1.87, ~1.97~{\rm GeV}$ for $c\to d,s$, respectively.}\label{Tab:ff_ccc}
\begin{tabular}{c|c|c|c|c|c|c|c}
\hline\hline
$F$ &$F(0)$ &$F$ &$F(0)$ &$F$ &$F(0)$ &$F$ &$F(0)$\\ \hline
$f_{1,S}^{\Xi_{cc}^{++}\to\Lambda_{c}^{+}}$ &$0.495$ &
$g_{1,S}^{\Xi_{cc}^{++}\to\Lambda_{c}^{+}}$ &$0.332$ &
$f_{1,A}^{\Xi_{cc}^{++}\to\Lambda_{c}^{+}}$ &$0.489$ &
$g_{1,A}^{\Xi_{cc}^{++}\to\Lambda_{c}^{+}}$ &$-0.111$ \\
$f_{2,S}^{\Xi_{cc}^{++}\to\Lambda_{c}^{+}}$ &$-0.621$ &
$g_{2,S}^{\Xi_{cc}^{++}\to\Lambda_{c}^{+}}$ &$1.004$ &
$f_{2,A}^{\Xi_{cc}^{++}\to\Lambda_{c}^{+}}$ &$0.290$ &
$g_{2,A}^{\Xi_{cc}^{++}\to\Lambda_{c}^{+}}$ &$-0.325$ \\
$f_{3,S}^{\Xi_{cc}^{++}\to\Lambda_{c}^{+}}$ &$0.832$ &
$g_{3,S}^{\Xi_{cc}^{++}\to\Lambda_{c}^{+}}$ &$-2.957$ &
$f_{3,A}^{\Xi_{cc}^{++}\to\Lambda_{c}^{+}}$ &$0.648$ &
$g_{3,A}^{\Xi_{cc}^{++}\to\Lambda_{c}^{+}}$ &$0.943$ \\ \hline
$f_{1,S}^{\Xi_{cc}\to\Sigma_{c}}$ &$0.536$ &
$g_{1,S}^{\Xi_{cc}\to\Sigma_{c}}$ &$0.422$ &
$f_{1,A}^{\Xi_{cc}\to\Sigma_{c}}$ &$0.529$ &
$g_{1,A}^{\Xi_{cc}\to\Sigma_{c}}$ &$-0.141$ \\
$f_{2,S}^{\Xi_{cc}\to\Sigma_{c}}$ &$-0.732$ &
$g_{2,S}^{\Xi_{cc}\to\Sigma_{c}}$ &$0.561$ &
$f_{2,A}^{\Xi_{cc}\to\Sigma_{c}}$ &$0.427$ &
$g_{2,A}^{\Xi_{cc}\to\Sigma_{c}}$ &$-0.177$ \\
$f_{3,S}^{\Xi_{cc}\to\Sigma_{c}}$ &$0.620$ &
$g_{3,S}^{\Xi_{cc}\to\Sigma_{c}}$ &$-0.808$ &
$f_{3,A}^{\Xi_{cc}\to\Sigma_{c}}$ &$0.423$ &
$g_{3,A}^{\Xi_{cc}\to\Sigma_{c}}$ &$0.215$ \\ \hline
$f_{1,S}^{\Xi_{cc}\to\Xi_{c}}$ &$0.588$ &
$g_{1,S}^{\Xi_{cc}\to\Xi_{c}}$ &$0.424$ &
$f_{1,A}^{\Xi_{cc}\to\Xi_{c}}$ &$0.582$ &
$g_{1,A}^{\Xi_{cc}\to\Xi_{c}}$ &$-0.141$ \\
\hline
\hline
\end{tabular}
\end{table}
\begin{table}\footnotesize \label{Tab:ff_bbb}
 \caption{Numerical results for the transition $1/2\to 1/2$ form factors $f^{(T)}_{i,S(A)}$ and $g^{(T)}_{i,S(A)}$ of doubly bottom baryon ${\cal B}_{bb}$ decay with $b\to u,c$ processes. The parametrization scheme in Eq.~(\ref{eq:auxiliary_fit_formula_bdecay}) is introduced for these form factors with asterisk, and Eq.~(\ref{eq:main_fit_formula_bdecay}) for all the other ones.}
\begin{tabular}{c|c|c|c|c|c|c|c|c|c|c|c}
\hline\hline
$F$ &$F(0)$ &$m_{\rm{fit}}$ &$\delta$ &$F$ &$F(0)$ &$m_{\rm{fit}}$ &$\delta$
&$F$ &$F(0)$ &$m_{\rm{fit}}$ &$\delta$\\ \hline
$f_{1,S}^{\Xi_{bb}\to\Sigma_{b}}$ &$
0.102$ &$3.35$ &$0.84$ &
$f_{2,S}^{\Xi_{bb}\to\Sigma_{b}}$ &$
-0.149$ &$3.06$ &$0.92$ &
$f_{3,S}^{\Xi_{bb}\to\Sigma_{b}}$ &$
-0.004^{*}$ &$3.13^{*}$ &$0.99^{*}$ \\
$g_{1,S}^{\Xi_{bb}\to\Sigma_{b}}$ &$
0.094$ &$3.36$ &$0.76$ &
$g_{2,S}^{\Xi_{bb}\to\Sigma_{b}}$ &$
0.013$ &$3.36$ &$1.11$ &
$g_{3,S}^{\Xi_{bb}\to\Sigma_{b}}$ &$
-0.223$ &$3.35$ &$1.14$ \\
$f_{1,A}^{\Xi_{bb}\to\Sigma_{b}}$ &$
0.102$ &$3.18$ &$0.94$ &
$f_{2,A}^{\Xi_{bb}\to\Sigma_{b}}$ &$
0.096$ &$3.28$ &$0.95$ &
$f_{3,A}^{\Xi_{bb}\to\Sigma_{b}}$ &$
-0.068$ &$3.44$ &$1.13$ \\
$g_{1,A}^{\Xi_{bb}\to\Sigma_{b}}$ &$
-0.031$ &$3.35$ &$0.76$ &
$g_{2,A}^{\Xi_{bb}\to\Sigma_{b}}$ &$
-0.003$ &$3.46$ &$1.21$ &
$g_{3,A}^{\Xi_{bb}\to\Sigma_{b}}$ &$
0.071$ &$3.37$ &$1.16$ \\ \hline
$f_{1,S}^{\Xi_{bb}\to\Xi_{bc}^{(\prime)}}$ &$
0.471$ &$4.23$ &$0.76$ &
$f_{2,S}^{\Xi_{bb}\to\Xi_{bc}^{(\prime)}}$ &$
-0.659$ &$3.15$ &$0.46$ &
$f_{3,S}^{\Xi_{bb}\to\Xi_{bc}^{(\prime)}}$ &$
0.096$ &$3.36$ &$6.76$ \\
$g_{1,S}^{\Xi_{bb}\to\Xi_{bc}^{(\prime)}}$ &$
0.458$ &$3.54$ &$0.19$ &
$g_{2,S}^{\Xi_{bb}\to\Xi_{bc}^{(\prime)}}$ &$
-0.087^{*}$ &$0.97^{*}$ &$0.06^{*}$&
$g_{3,S}^{\Xi_{bb}\to\Xi_{bc}^{(\prime)}}$ &$
0.573$ &$1.34$ &$-0.33$ \\
$f_{1,A}^{\Xi_{bb}\to\Xi_{bc}^{(\prime)}}$ &$
0.469$ &$3.88$ &$0.74$ &
$f_{2,A}^{\Xi_{bb}\to\Xi_{bc}^{(\prime)}}$ &$
0.318$ &$5.18$ &$2.26$ &
$f_{3,A}^{\Xi_{bb}\to\Xi_{bc}^{(\prime)}}$ &$
-0.079^{*}$ &$5.06^{*}$ &$3.69^{*}$ \\
\hline
\hline

\end{tabular}
\end{table}
\begin{table}\footnotesize\label{Tab:ff_bcc}
 \caption{Same with Tab.~\ref{Tab:ff_bbb} expect for bottom charm  baryon ${\cal B}_{bc^{(\prime)}}$ decay.}
\begin{tabular}{c|c|c|c|c|c|c|c|c|c|c|c}
\hline\hline
$F$ &$F(0)$ &$m_{\rm{fit}}$ &$\delta$ &$F$ &$F(0)$ &$m_{\rm{fit}}$ &$\delta$
&$F$ &$F(0)$ &$m_{\rm{fit}}$ &$\delta$\\ \hline
$f_{1,S}^{\Xi_{bc}^{(\prime)}\to\Sigma_{c}}$ &$
0.143$ &$3.76$ &$0.66$ &
$f_{2,S}^{\Xi_{bc}^{(\prime)}\to\Sigma_{c}}$ &$
-0.067$ &$3.23$ &$0.72$ &
$f_{3,S}^{\Xi_{bc}^{(\prime)}\to\Sigma_{c}}$ &$
0.001^{*}$ &$1.12^{*}$ &$0.11^{*}$ \\
$g_{1,S}^{\Xi_{bc}^{(\prime)}\to\Sigma_{c}}$ &$
0.123$ &$4.17$ &$0.85$ &
$g_{2,S}^{\Xi_{bc}^{(\prime)}\to\Sigma_{c}}$ &$
0.046$ &$2.81$ &$0.76$ &
$g_{3,S}^{\Xi_{bc}^{(\prime)}\to\Sigma_{c}}$ &$
-0.197$ &$3.07$ &$0.68$ \\
$f_{1,A}^{\Xi_{bc}^{(\prime)}\to\Sigma_{c}}$ &$
0.138$ &$3.34$ &$0.77$ &
$f_{2,A}^{\Xi_{bc}^{(\prime)}\to\Sigma_{c}}$ &$
0.147$ &$3.42$ &$0.71$ &
$f_{3,A}^{\Xi_{bc}^{(\prime)}\to\Sigma_{c}}$ &$
-0.095$ &$3.54$ &$0.78$ \\
$g_{1,A}^{\Xi_{bc}^{(\prime)}\to\Sigma_{c}}$ &$
-0.041$ &$4.07$ &$0.81$ &
$g_{2,A}^{\Xi_{bc}^{(\prime)}\to\Sigma_{c}}$ &$
-0.011$ &$2.75$ &$0.95$ &
$g_{3,A}^{\Xi_{bc}^{(\prime)}\to\Sigma_{c}}$ &$
0.057$ &$3.04$ &$0.68$ \\ \hline
$f_{1,S}^{\Xi_{bc}^{(\prime)}\to\Xi_{cc}}$ &$
0.546$ &$5.01$ &$0.64$ &
$f_{2,S}^{\Xi_{bc}^{(\prime)}\to\Xi_{cc}}$ &$
-0.245$ &$3.63$ &$0.45$ &
$f_{3,S}^{\Xi_{bc}^{(\prime)}\to\Xi_{cc}}$ &$
0.049$ &$2.84$ &$1.49$ \\
$g_{1,S}^{\Xi_{bc}^{(\prime)}\to\Xi_{cc}}$ &$
0.512$ &$5.12$ &$0.51$ &
$g_{2,S}^{\Xi_{bc}^{(\prime)}\to\Xi_{cc}}$ &$
0.086$ &$4.19$ &$2.20$ &
$g_{3,S}^{\Xi_{bc}^{(\prime)}\to\Xi_{cc}}$ &$
-0.492$ &$6.16$ &$5.80$ \\
$f_{1,A}^{\Xi_{bc}^{(\prime)}\to\Xi_{cc}}$ &$
0.536$ &$4.17$ &$0.61$ &
$f_{2,A}^{\Xi_{bc}^{(\prime)}\to\Xi_{cc}}$ &$
0.488$ &$4.33$ &$0.59$ &
$f_{3,A}^{\Xi_{bc}^{(\prime)}\to\Xi_{cc}}$ &$
-0.167$ &$5.52$ &$1.81$ \\
\hline
\hline
\end{tabular}
\end{table}

\begin{table}\footnotesize\label{Tab:ff_cucc_axial}
\caption{Numerical results for the transition $1/2\to 1/2$ form factors $f_{i,S(A)}$ and $g_{i,S(A)}$ at $q^2=0$ of $c\to u$ processes. The parametrization scheme in Eq.~(\ref{eq:main_fit_formula_cdecay}) is introduced for these form factors,
and the value of the singly pole $m_{\rm pole}$ is taken as $1.87~{\rm GeV}$ for $c\to u$.}
\begin{tabular}{c|c|c|c|c|c|c|c}
\hline\hline
$F$ &$F(0)$ &$F$ &$F(0)$ &$F$ &$F(0)$ &$F$ &$F(0)$\\ \hline
$f_{1,S}^{\Xi_{cc}\to\Sigma_{c}}$ &$0.536$ &
$g_{1,S}^{\Xi_{cc}\to\Sigma_{c}}$ &$0.422$ &
$f_{1,A}^{\Xi_{cc}\to\Sigma_{c}}$ &$0.529$ &
$g_{1,A}^{\Xi_{cc}\to\Sigma_{c}}$ &$-0.141$ \\
$f_{2,S}^{\Xi_{cc}\to\Sigma_{c}}$ &$-0.732$ &
$g_{2,S}^{\Xi_{cc}\to\Sigma_{c}}$ &$0.558$ &
$f_{2,A}^{\Xi_{cc}\to\Sigma_{c}}$ &$0.428$ &
$g_{2,A}^{\Xi_{cc}\to\Sigma_{c}}$ &$-0.176$ \\
$f_{3,S}^{\Xi_{cc}\to\Sigma_{c}}$ &$0.619$ &
$g_{3,S}^{\Xi_{cc}\to\Sigma_{c}}$ &$-0.788$ &
$f_{3,A}^{\Xi_{cc}\to\Sigma_{c}}$ &$0.422$ &
$g_{3,A}^{\Xi_{cc}\to\Sigma_{c}}$ &$0.208$ \\
$f_{1,S}^{T\Xi_{cc}\to\Sigma_{c}}$ &$0.181$ &
$g_{1,S}^{T\Xi_{cc}\to\Sigma_{c}}$ &$1.958$ &
$f_{1,A}^{T\Xi_{cc}\to\Sigma_{c}}$ &$0.282$ &
$g_{1,A}^{T\Xi_{cc}\to\Sigma_{c}}$ &$-0.670$ \\
$f_{2,S}^{T\Xi_{cc}\to\Sigma_{c}}$ &$0.482$ &
$g_{2,S}^{T\Xi_{cc}\to\Sigma_{c}}$ &$0.415$ &
$f_{2,A}^{T\Xi_{cc}\to\Sigma_{c}}$ &$-0.150$ &
$g_{2,A}^{T\Xi_{cc}\to\Sigma_{c}}$ &$-0.135$ \\ \hline
$f_{1,S}^{\Xi_{cc}^{+}\to\Lambda_{c}^{+}}$ &$0.495$ &
$g_{1,S}^{\Xi_{cc}^{+}\to\Lambda_{c}^{+}}$ &$0.332$ &
$f_{1,A}^{\Xi_{cc}^{+}\to\Lambda_{c}^{+}}$ &$0.489$ &
$g_{1,A}^{\Xi_{cc}^{+}\to\Lambda_{c}^{+}}$ &$-0.111$ \\
$f_{2,S}^{\Xi_{cc}^{+}\to\Lambda_{c}^{+}}$ &$-0.621$ &
$g_{2,S}^{\Xi_{cc}^{+}\to\Lambda_{c}^{+}}$ &$1.004$ &
$f_{2,A}^{\Xi_{cc}^{+}\to\Lambda_{c}^{+}}$ &$0.290$ &
$g_{2,A}^{\Xi_{cc}^{+}\to\Lambda_{c}^{+}}$ &$-0.325$ \\
\hline
\hline
\end{tabular}
\end{table}

\begin{table}\footnotesize\label{Tab:ff_bdbb_axial}
\caption{Numerical results for the transition $1/2\to 1/2$ form factors $f^{(T)}_{i,S(A)}$ and $g^{(T)}_{i,S(A)}$ of $b\to d$ processes. The parametrization scheme in Eq.~(\ref{eq:auxiliary_fit_formula_bdecay}) is introduced for these form factors with asterisk, and Eq.~(\ref{eq:main_fit_formula_bdecay}) for all the other ones.}
\setlength{\tabcolsep}{0.4mm}{
\begin{tabular}{c|c|c|c|c|c|c|c|c|c|c|c|c|c|c|c}
\hline\hline
$F$ &$F(0)$ &$m_{\rm{fit}}$ &$\delta$ &$F$ &$F(0)$ &$m_{\rm{fit}}$ &$\delta$&
$F$ &$F(0)$ &$m_{\rm{fit}}$ &$\delta$&$F$ &$F(0)$ &$m_{\rm{fit}}$ &$\delta$
\\ \hline
$f_{1,S}^{\Xi_{bb}^{0}\to\Lambda_{b}^{0}}$ &$
0.100$ &$3.40$ &$0.86$ &
$g_{1,S}^{\Xi_{bb}^{0}\to\Lambda_{b}^{0}}$ &$
0.087$ &$3.57$ &$0.91$ &
$f_{1,A}^{\Xi_{bb}^{0}\to\Lambda_{b}^{0}}$ &$
0.100$ &$3.22$ &$0.96$ &
$g_{1,A}^{\Xi_{bb}^{0}\to\Lambda_{b}^{0}}$ &$
-0.029$ &$3.56$ &$0.91$ \\
$f_{2,S}^{\Xi_{bb}^{0}\to\Lambda_{b}^{0}}$ &$
-0.136$ &$3.10$ &$0.93$ &
$g_{2,S}^{\Xi_{bb}^{0}\to\Lambda_{b}^{0}}$ &$
0.041$ &$2.70$ &$0.89$ &
$f_{2,A}^{\Xi_{bb}^{0}\to\Lambda_{b}^{0}}$ &$
0.092$ &$3.36$ &$0.99$ &
$g_{2,A}^{\Xi_{bb}^{0}\to\Lambda_{b}^{0}}$ &$
-0.013$ &$2.68$ &$0.92$ \\
$f_{3,S}^{\Xi_{bb}^{0}\to\Lambda_{b}^{0}}$ &$
0.008$ &$0.32$ &$-0.01$ &
$g_{3,S}^{\Xi_{bb}^{0}\to\Lambda_{b}^{0}}$ &$
-0.298$ &$2.99$ &$0.89$ &
$f_{3,A}^{\Xi_{bb}^{0}\to\Lambda_{b}^{0}}$ &$
-0.055$ &$3.79$ &$1.45$ &
$g_{3,A}^{\Xi_{bb}^{0}\to\Lambda_{b}^{0}}$ &$
0.096$ &$2.98$ &$0.89$ \\
$f_{1,S}^{T\Xi_{bb}^{0}\to\Lambda_{b}^{0}}$ &$
0.072$ &$3.32$ &$0.89$ &
$g_{1,S}^{T\Xi_{bb}^{0}\to\Lambda_{b}^{0}}$ &$
0.114$ &$2.93$ &$1.12$ &
$f_{1,A}^{T\Xi_{bb}^{0}\to\Lambda_{b}^{0}}$ &$
0.104$ &$3.27$ &$0.90$ &
$g_{1,A}^{T\Xi_{bb}^{0}\to\Lambda_{b}^{0}}$ &$
-0.040$ &$2.94$ &$1.10$ \\
$f_{2,S}^{T\Xi_{bb}^{0}\to\Lambda_{b}^{0}}$ &$
0.075$ &$3.30$ &$0.76$ &
$g_{2,S}^{T\Xi_{bb}^{0}\to\Lambda_{b}^{0}}$ &$
0.091$ &$3.73$ &$0.93$ &
$f_{2,A}^{T\Xi_{bb}^{0}\to\Lambda_{b}^{0}}$ &$
-0.045$ &$3.20$ &$1.74$ &
$g_{2,A}^{T\Xi_{bb}^{0}\to\Lambda_{b}^{0}}$ &$
-0.030$ &$3.74$ &$0.93$ \\
\hline
$f_{1,S}^{\Xi_{bb}\to\Sigma_{b}}$ &$
0.102$ &$3.35$ &$0.83$ &
$g_{1,S}^{\Xi_{bb}\to\Sigma_{b}}$ &$
0.094$ &$3.36$ &$0.76$ &
$f_{1,A}^{\Xi_{bb}\to\Sigma_{b}}$ &$
0.102$ &$3.18$ &$0.94$ &
$g_{1,A}^{\Xi_{bb}\to\Sigma_{b}}$ &$
-0.031$ &$3.35$ &$0.76$ \\
\hline
\hline
\end{tabular}}
\end{table}
\begin{table}\footnotesize\label{Tab:ff_bsbb_axial}
\caption{Same with Tab.~\ref{Tab:ff_bdbb_axial} except for $b\to s$ process.}
\setlength{\tabcolsep}{0.4mm}{
\begin{tabular}{c|c|c|c|c|c|c|c|c|c|c|c|c|c|c|c}
\hline\hline
$F$ &$F(0)$ &$m_{\rm{fit}}$ &$\delta$ &$F$ &$F(0)$ &$m_{\rm{fit}}$ &$\delta$&
$F$ &$F(0)$ &$m_{\rm{fit}}$ &$\delta$&$F$ &$F(0)$ &$m_{\rm{fit}}$ &$\delta$
\\ \hline
$f_{1,S}^{\Xi_{bb}\to\Xi_{b}}$ &$
0.141$ &$3.56$ &$0.81$ &
$g_{1,S}^{\Xi_{bb}\to\Xi_{b}}$ &$
0.122$ &$3.73$ &$0.85$ &
$f_{1,A}^{\Xi_{bb}\to\Xi_{b}}$ &$
0.140$ &$3.35$ &$0.89$ &
$g_{1,A}^{\Xi_{bb}\to\Xi_{b}}$ &$
-0.041$ &$3.71$ &$0.84$ \\
$f_{2,S}^{\Xi_{bb}\to\Xi_{b}}$ &$
-0.189$ &$3.16$ &$0.83$ &
$g_{2,S}^{\Xi_{bb}\to\Xi_{b}}$ &$
0.056$ &$2.79$ &$0.82$ &
$f_{2,A}^{\Xi_{bb}\to\Xi_{b}}$ &$
0.123$ &$3.55$ &$0.96$ &
$g_{2,A}^{\Xi_{bb}\to\Xi_{b}}$ &$
-0.017$ &$2.77$ &$0.85$ \\
$f_{3,S}^{\Xi_{bb}\to\Xi_{b}}$ &$
0.016$ &$0.34$ &$-0.02$ &
$g_{3,S}^{\Xi_{bb}\to\Xi_{b}}$ &$
-0.406$ &$3.14$ &$0.86$ &
$f_{3,A}^{\Xi_{bb}\to\Xi_{b}}$ &$
-0.066$ &$4.27$ &$1.85$ &
$g_{3,A}^{\Xi_{bb}\to\Xi_{b}}$ &$
0.130$ &$3.13$ &$0.86$ \\
$f_{1,S}^{T\Xi_{bb}\to\Xi_{b}}$ &$
0.091$ &$3.43$ &$0.82$ &
$g_{1,S}^{T\Xi_{bb}\to\Xi_{b}}$ &$
0.156$ &$2.84$ &$0.95$ &
$f_{1,A}^{T\Xi_{bb}\to\Xi_{b}}$ &$
0.134$ &$3.38$ &$0.83$ &
$g_{1,A}^{T\Xi_{bb}\to\Xi_{b}}$ &$
-0.054$ &$2.85$ &$0.93$ \\
$f_{2,S}^{T\Xi_{bb}\to\Xi_{b}}$ &$
0.108$ &$3.42$ &$0.70$ &
$g_{2,S}^{T\Xi_{bb}\to\Xi_{b}}$ &$
0.128$ &$3.96$ &$0.96$ &
$f_{2,A}^{T\Xi_{bb}\to\Xi_{b}}$ &$
-0.061$ &$3.25$ &$1.54$ &
$g_{2,A}^{T\Xi_{bb}\to\Xi_{b}}$ &$
-0.042$ &$3.97$ &$0.97$ \\\hline
$f_{1,S}^{\Xi_{bb}\to\Xi_{b}^{\prime}}$ &$
0.143$ &$3.52$ &$0.79$ &
$g_{1,S}^{\Xi_{bb}\to\Xi_{b}^{\prime}}$ &$
0.130$ &$3.53$ &$0.70$ &
$f_{1,A}^{\Xi_{bb}\to\Xi_{b}^{\prime}}$ &$
0.142$ &$3.31$ &$0.87$ &
$g_{1,A}^{\Xi_{bb}\to\Xi_{b}^{\prime}}$ &$
-0.043$ &$3.51$ &$0.70$ \\
$f_{2,S}^{\Xi_{bb}\to\Xi_{b}^{\prime}}$ &$
-0.202$ &$3.13$ &$0.81$ &
$g_{2,S}^{\Xi_{bb}\to\Xi_{b}^{\prime}}$ &$
0.024$ &$3.45$ &$1.24$ &
$f_{2,A}^{\Xi_{bb}\to\Xi_{b}^{\prime}}$ &$
0.129$ &$3.46$ &$0.91$ &
$g_{2,A}^{\Xi_{bb}\to\Xi_{b}^{\prime}}$ &$
-0.007$ &$3.51$ &$1.35$ \\
\hline
\hline
\end{tabular}}
\end{table}
\begin{table}\footnotesize\label{Tab:ff32_cdscc}
\caption{Numerical results for the transition $1/2\to 3/2$ form factors $\mathtt{f}_{i}$ and $\mathtt{g}_{i}$ at $q^2=0$ of $c\to d,s$ processes. The parametrization scheme in Eq.~(\ref{eq:main_fit_formula_cdecay}) is introduced for these form factors,
and the value of the singly pole $m_{\rm pole}$s are taken as $1.87, ~1.97~{\rm GeV}$ for $c\to d,s$, respectively.}
\begin{tabular}{c|c|c|c|c|c|c|c}
\hline\hline
$F$ &$F(0)$&$F$ &$F(0)$ &$F$ &$F(0)$&$F$ &$F(0)$ \\ \hline
$\mathtt{f}_{1}^{\Xi_{cc}^{++}\to\Sigma_{c}^{*+}}$ &$-0.979$ &
$\mathtt{f}_{2}^{\Xi_{cc}^{++}\to\Sigma_{c}^{*+}}$ &$-0.645$ &
$\mathtt{f}_{3}^{\Xi_{cc}^{++}\to\Sigma_{c}^{*+}}$ &$0.047$ &
$\mathtt{f}_{4}^{\Xi_{cc}^{++}\to\Sigma_{c}^{*+}}$ &$-1.969$ \\
$\mathtt{g}_{1}^{\Xi_{cc}^{++}\to\Sigma_{c}^{*+}}$ &$-5.792$ &
$\mathtt{g}_{2}^{\Xi_{cc}^{++}\to\Sigma_{c}^{*+}}$ &$-3.602$ &
$\mathtt{g}_{3}^{\Xi_{cc}^{++}\to\Sigma_{c}^{*+}}$ &$0.947$ &
$\mathtt{g}_{4}^{\Xi_{cc}^{++}\to\Sigma_{c}^{*+}}$ &$0.393$ \\\hline
$\mathtt{f}_{1}^{\Xi_{cc}^{+}\to\Sigma_{c}^{*0}}$ &$-0.979$ &
$\mathtt{f}_{2}^{\Xi_{cc}^{+}\to\Sigma_{c}^{*0}}$ &$-0.645$ &
$\mathtt{f}_{3}^{\Xi_{cc}^{+}\to\Sigma_{c}^{*0}}$ &$0.047$ &
$\mathtt{f}_{4}^{\Xi_{cc}^{+}\to\Sigma_{c}^{*0}}$ &$-1.969$ \\
$\mathtt{g}_{1}^{\Xi_{cc}^{+}\to\Sigma_{c}^{*0}}$ &$-5.792$ &
$\mathtt{g}_{2}^{\Xi_{cc}^{+}\to\Sigma_{c}^{*0}}$ &$-3.602$ &
$\mathtt{g}_{3}^{\Xi_{cc}^{+}\to\Sigma_{c}^{*0}}$ &$0.947$ &
$\mathtt{g}_{4}^{\Xi_{cc}^{+}\to\Sigma_{c}^{*0}}$ &$0.393$ \\\hline
$\mathtt{f}_{1}^{\Omega_{cc}^{+}\to\Xi_{c}^{\prime*0}}$ &$-1.017$ &
$\mathtt{f}_{2}^{\Omega_{cc}^{+}\to\Xi_{c}^{\prime*0}}$ &$-0.665$ &
$\mathtt{f}_{3}^{\Omega_{cc}^{+}\to\Xi_{c}^{\prime*0}}$ &$0.046$ &
$\mathtt{f}_{4}^{\Omega_{cc}^{+}\to\Xi_{c}^{\prime*0}}$ &$-2.045$ \\
$\mathtt{g}_{1}^{\Omega_{cc}^{+}\to\Xi_{c}^{\prime*0}}$ &$-6.244$ &
$\mathtt{g}_{2}^{\Omega_{cc}^{+}\to\Xi_{c}^{\prime*0}}$ &$-3.856$ &
$\mathtt{g}_{3}^{\Omega_{cc}^{+}\to\Xi_{c}^{\prime*0}}$ &$0.980$ &
$\mathtt{g}_{4}^{\Omega_{cc}^{+}\to\Xi_{c}^{\prime*0}}$ &$0.389$ \\\hline
$\mathtt{f}_{1}^{\Xi_{cc}^{++}\to\Xi_{c}^{\prime*+}}$ &$-1.148$ &
$\mathtt{f}_{2}^{\Xi_{cc}^{++}\to\Xi_{c}^{\prime*+}}$ &$-0.714$ &
$\mathtt{f}_{3}^{\Xi_{cc}^{++}\to\Xi_{c}^{\prime*+}}$ &$0.049$ &
$\mathtt{f}_{4}^{\Xi_{cc}^{++}\to\Xi_{c}^{\prime*+}}$ &$-2.297$ \\
\hline
\hline
\end{tabular}
\end{table}
\begin{table}\footnotesize\label{Tab:ff32_bucbb}
\caption{Numerical results for the transition $1/2\to 3/2$ form factors $\mathtt{f}_{i}$ and $\mathtt{g}_{i}$ of $b\to u,c$ processes. The parametrization scheme shown with Eq.~(\ref{eq:auxiliary_fit_formula_bdecay}) is introduced for these form factors with asterisk, and Eq.~(\ref{eq:main_fit_formula_bdecay}) for all the other ones.}
\begin{tabular}{c|c|c|c|c|c|c|c|c|c|c|c|c|c|c|c}
\hline \hline
$F$  & $F(0)$  & $m_{{\rm {fit}}}$  & $\delta$  & $F$  & $F(0)$  & $m_{{\rm {fit}}}$  & $\delta$ & $F$  & $F(0)$  & $m_{{\rm {fit}}}$  & $\delta$  & $F$  & $F(0)$  & $m_{{\rm {fit}}}$  & $\delta$\tabularnewline
\hline
$\mathtt{f}_{1}^{\Xi_{bb}\to\Sigma_{b}^{*}}$  & $-0.172$  & $3.15$  & $0.89$  & $\mathtt{g}_{1}^{\Xi_{bb}\to\Sigma_{b}^{*}}$  & $-0.108$  & $2.70$  & $1.84$  & $\mathtt{f}_{1}^{\Xi_{bc}^{(\prime)}\to\Sigma_{c}^{*}}$  & $-0.108$  & $3.24$  & $0.57$  & $\mathtt{g}_{1}^{\Xi_{bc}^{(\prime)}\to\Sigma_{c}^{*}}$  & $0.021$  & $3.86$  & $1.02$ \tabularnewline
$\mathtt{f}_{2}^{\Xi_{bb}\to\Sigma_{b}^{*}}$  & $-0.129$  & $3.17$  & $0.90$  & $\mathtt{g}_{2}^{\Xi_{bb}\to\Sigma_{b}^{*}}$  & $-0.056^{*}$  & $0.51^{*}$  & $0.06^{*}$  & $\mathtt{f}_{2}^{\Xi_{bc}^{(\prime)}\to\Sigma_{c}^{*}}$  & $-0.145$  & $3.26$  & $0.57$  & $\mathtt{g}_{2}^{\Xi_{bc}^{(\prime)}\to\Sigma_{c}^{*}}$  & $0.120$  & $3.45$  & $0.60$ \tabularnewline
$\mathtt{f}_{3}^{\Xi_{bb}\to\Sigma_{b}^{*}}$  & $0.002$  & $2.95$  & $0.91$  & $\mathtt{g}_{3}^{\Xi_{bb}\to\Sigma_{b}^{*}}$  & $0.136$  & $3.15$  & $0.89$  & $\mathtt{f}_{3}^{\Xi_{bc}^{(\prime)}\to\Sigma_{c}^{*}}$  & $0.003$  & $3.07$  & $0.55$  & $\mathtt{g}_{3}^{\Xi_{bc}^{(\prime)}\to\Sigma_{c}^{*}}$  & $0.096$  & $3.23$  & $0.57$ \tabularnewline
$\mathtt{f}_{4}^{\Xi_{bb}\to\Sigma_{b}^{*}}$  & $-0.355$  & $3.18$  & $0.86$  & $\mathtt{g}_{4}^{\Xi_{bb}\to\Sigma_{b}^{*}}$  & $0.099$  & $3.46$  & $0.82$  & $\mathtt{f}_{4}^{\Xi_{bc}^{(\prime)}\to\Sigma_{c}^{*}}$  & $-0.270$  & $3.45$  & $0.53$  & $\mathtt{g}_{4}^{\Xi_{bc}^{(\prime)}\to\Sigma_{c}^{*}}$  & $0.174$  & $3.76$  & $0.54$ \tabularnewline
\hline\hline
\end{tabular}
\end{table}

\begin{table}\footnotesize\label{Tab:fcnc32_cu}

\caption{
Numerical results for the transition $1/2\to 3/2$ form factors $\mathtt{f}_{i}$ and $\mathtt{g}_{i}$ at $q^2=0$ of $c\to u$ processes. The parametrization scheme in Eq.~(\ref{eq:main_fit_formula_cdecay}) is introduced for these form factors,
and the value of the singly pole $m_{\rm pole}$ is $1.87~{\rm GeV}$.}
\begin{tabular}{c|c|c|c|c|c|c|c}
\hline\hline
$F$  & $F(0)$  & $F$  & $F(0)$  & $F$  & $F(0)$  & $F$  & $F(0)$ \tabularnewline
\hline
$\mathtt{f}_{1}^{\Xi_{cc}\to\Sigma_{c}^{*}}$  & $-0.979$  & $\mathtt{g}_{1}^{\Xi_{cc}\to\Sigma_{c}^{*}}$  & $-5.792$  & $\mathtt{f}_{1}^{\Xi_{bc}^{(\prime)}\to\Sigma_{b}^{*}}$  & $-0.733$  & $\mathtt{g}_{1}^{\Xi_{bc}^{(\prime)}\to\Sigma_{b}^{*}}$  & $-2.193$ \tabularnewline
$\mathtt{f}_{2}^{\Xi_{cc}\to\Sigma_{c}^{*}}$  & $-0.645$  & $\mathtt{g}_{2}^{\Xi_{cc}\to\Sigma_{c}^{*}}$  & $-3.602$  & $\mathtt{f}_{2}^{\Xi_{bc}^{(\prime)}\to\Sigma_{b}^{*}}$  & $-1.380$  & $\mathtt{g}_{2}^{\Xi_{bc}^{(\prime)}\to\Sigma_{b}^{*}}$  & $-1.195$ \tabularnewline
$\mathtt{f}_{3}^{\Xi_{cc}\to\Sigma_{c}^{*}}$  & $0.047$  & $\mathtt{g}_{3}^{\Xi_{cc}\to\Sigma_{c}^{*}}$  & $0.947$  & $\mathtt{f}_{3}^{\Xi_{bc}^{(\prime)}\to\Sigma_{b}^{*}}$  & $0.095$  & $\mathtt{g}_{3}^{\Xi_{bc}^{(\prime)}\to\Sigma_{b}^{*}}$  & $0.537$ \tabularnewline
$\mathtt{f}_{4}^{\Xi_{cc}\to\Sigma_{c}^{*}}$  & $-1.969$  & $\mathtt{g}_{4}^{\Xi_{cc}\to\Sigma_{c}^{*}}$  & $0.393$  & $\mathtt{f}_{4}^{\Xi_{bc}^{(\prime)}\to\Sigma_{b}^{*}}$  & $-1.726$  & $\mathtt{g}_{4}^{\Xi_{bc}^{(\prime)}\to\Sigma_{b}^{*}}$  & $0.157$ \tabularnewline
$\mathtt{f}_{1}^{T\Xi_{cc}\to\Sigma_{c}^{*}}$  & $0.534$  & $\mathtt{g}_{1}^{T\Xi_{cc}\to\Sigma_{c}^{*}}$  & $-2.469$  & $\mathtt{f}_{1}^{T\Xi_{bc}^{(\prime)}\to\Sigma_{b}^{*}}$  & $0.221$  & $\mathtt{g}_{1}^{T\Xi_{bc}^{(\prime)}\to\Sigma_{b}^{*}}$  & $12.050$ \tabularnewline
$\mathtt{f}_{2}^{T\Xi_{cc}\to\Sigma_{c}^{*}}$  & $0.330$  & $\mathtt{g}_{2}^{T\Xi_{cc}\to\Sigma_{c}^{*}}$  & $-1.394$  & $\mathtt{f}_{2}^{T\Xi_{bc}^{(\prime)}\to\Sigma_{b}^{*}}$  & $0.379$  & $\mathtt{g}_{2}^{T\Xi_{bc}^{(\prime)}\to\Sigma_{b}^{*}}$  & $6.959$ \tabularnewline
\hline \hline
\end{tabular}
\end{table}

\begin{table}\footnotesize
\caption{Numerical results for the transition $1/2\to 3/2$ form factors $\mathtt{f}_{i}$ and $\mathtt{g}_{i}$ of $b\to d$ processes. The parametrization scheme shown with Eq.~(\ref{eq:auxiliary_fit_formula_bdecay}) is introduced for these form factors with asterisk, and Eq.~(\ref{eq:main_fit_formula_bdecay}) for all the other ones.}\label{Tab:fcnc32_bd}
\begin{tabular}{c|c|c|c|c|c|c|c|c|c|c|c|c|c|c|c}
\hline\hline
$F$  & $F(0)$  & $m_{{\rm {fit}}}$  & $\delta$  & $F$  & $F(0)$  & $m_{{\rm {fit}}}$  & $\delta$ & $F$  & $F(0)$  & $m_{{\rm {fit}}}$  & $\delta$  & $F$  & $F(0)$  & $m_{{\rm {fit}}}$  & $\delta$\tabularnewline
\hline
$\mathtt{f}_{1}^{\Xi_{bb}\to\Sigma_{b}^{*}}$  & $-0.172$  & $3.15$  & $0.89$  & $\mathtt{g}_{1}^{\Xi_{bb}\to\Sigma_{b}^{*}}$  & $-0.108$  & $2.70$  & $1.84$  & $\mathtt{f}_{1}^{\Xi_{bc}^{(\prime)}\to\Sigma_{c}^{*}}$  & $-0.108$  & $3.23$  & $0.54$  & $\mathtt{g}_{1}^{\Xi_{bc}^{(\prime)}\to\Sigma_{c}^{*}}$  & $0.021$  & $3.86$  & $1.01$\tabularnewline
$\mathtt{f}_{2}^{\Xi_{bb}\to\Sigma_{b}^{*}}$  & $-0.129$  & $3.17$  & $0.90$  & $\mathtt{g}_{2}^{\Xi_{bb}\to\Sigma_{b}^{*}}$  & $-0.057^{*}$  & $0.54^{*}$  & $0.06^{*}$  & $\mathtt{f}_{2}^{\Xi_{bc}^{(\prime)}\to\Sigma_{c}^{*}}$  & $-0.145$  & $3.25$  & $0.54$  & $\mathtt{g}_{2}^{\Xi_{bc}^{(\prime)}\to\Sigma_{c}^{*}}$  & $0.120$  & $3.44$  & $0.58$ \tabularnewline
$\mathtt{f}_{3}^{\Xi_{bb}\to\Sigma_{b}^{*}}$  & $0.002$  & $2.95$  & $0.91$  & $\mathtt{g}_{3}^{\Xi_{bb}\to\Sigma_{b}^{*}}$  & $0.136$  & $3.15$  & $0.89$  & $\mathtt{f}_{3}^{\Xi_{bc}^{(\prime)}\to\Sigma_{c}^{*}}$  & $0.003$  & $3.06$  & $0.51$  & $\mathtt{g}_{3}^{\Xi_{bc}^{(\prime)}\to\Sigma_{c}^{*}}$  & $0.096$  & $3.21$  & $0.54$ \tabularnewline
$\mathtt{f}_{4}^{\Xi_{bb}\to\Sigma_{b}^{*}}$  & $-0.355$  & $3.18$  & $0.86$  & $\mathtt{g}_{4}^{\Xi_{bb}\to\Sigma_{b}^{*}}$  & $0.099$  & $3.46$  & $0.82$  & $\mathtt{f}_{4}^{\Xi_{bc}^{(\prime)}\to\Sigma_{c}^{*}}$  & $-0.270$  & $3.44$  & $0.51$  & $\mathtt{g}_{4}^{\Xi_{bc}^{(\prime)}\to\Sigma_{c}^{*}}$  & $0.174$  & $3.75$  & $0.52$ \tabularnewline
\hline \hline
\end{tabular}
\end{table}

\begin{table}\footnotesize
\caption{Same with Tab.~\ref{Tab:fcnc32_bd} except for $b\to s$ process.}\label{Tab:fcnc32_bs}
\begin{tabular}{c|c|c|c|c|c|c|c|c|c|c|c|c|c|c|c}
\hline \hline
$F$  & $F(0)$  & $m_{{\rm {fit}}}$  & $\delta$  & $F$  & $F(0)$  & $m_{{\rm {fit}}}$  & $\delta$ & $F$  & $F(0)$  & $m_{{\rm {fit}}}$  & $\delta$  & $F$  & $F(0)$  & $m_{{\rm {fit}}}$  & $\delta$\tabularnewline
\hline
$\mathtt{f}_{1}^{\Xi_{bb}\to\Xi_{b}^{\prime*}}$  & $-0.232$  & $3.16$  & $0.62$  & $\mathtt{g}_{1}^{\Xi_{bb}\to\Xi_{b}^{\prime*}}$  & $-0.142$  & $2.30$  & $0.87$  & $\mathtt{f}_{1}^{\Xi_{bc}^{(\prime)}\to\Xi_{c}^{\prime*}}$  & $-0.153$  & $3.42$  & $0.54$  & $\mathtt{g}_{1}^{\Xi_{bc}^{(\prime)}\to\Xi_{c}^{\prime*}}$  & $0.040$  & $4.00$  & $1.07$ \tabularnewline
$\mathtt{f}_{2}^{\Xi_{bb}\to\Xi_{b}^{\prime*}}$  & $-0.172$  & $3.20$  & $0.64$  & $\mathtt{g}_{2}^{\Xi_{bb}\to\Xi_{b}^{\prime*}}$  & $-0.065^{*}$  & $0.81^{*}$  & $0.28^{*}$  & $\mathtt{f}_{2}^{\Xi_{bc}^{(\prime)}\to\Xi_{c}^{\prime*}}$  & $-0.196$  & $3.45$  & $0.55$  & $\mathtt{g}_{2}^{\Xi_{bc}^{(\prime)}\to\Xi_{c}^{\prime*}}$  & $0.182$  & $3.63$  & $0.61$ \tabularnewline
$\mathtt{f}_{3}^{\Xi_{bb}\to\Xi_{b}^{\prime*}}$  & $0.003$  & $2.91$  & $0.57$  & $\mathtt{g}_{3}^{\Xi_{bb}\to\Xi_{b}^{\prime*}}$  & $0.191$  & $3.17$  & $0.62$  & $\mathtt{f}_{3}^{\Xi_{bc}^{(\prime)}\to\Xi_{c}^{\prime*}}$  & $0.005$  & $3.25$  & $0.48$  & $\mathtt{g}_{3}^{\Xi_{bc}^{(\prime)}\to\Xi_{c}^{\prime*}}$  & $0.141$  & $3.40$  & $0.54$ \tabularnewline
$\mathtt{f}_{4}^{\Xi_{bb}\to\Xi_{b}^{\prime*}}$  & $-0.479$  & $3.21$  & $0.61$  & $\mathtt{g}_{4}^{\Xi_{bb}\to\Xi_{b}^{\prime*}}$  & $0.142$  & $3.59$  & $0.66$  & $\mathtt{f}_{4}^{\Xi_{bc}^{(\prime)}\to\Xi_{c}^{\prime*}}$  & $-0.375$  & $3.66$  & $0.51$  & $\mathtt{g}_{4}^{\Xi_{bc}^{(\prime)}\to\Xi_{c}^{\prime*}}$  & $0.251$  & $4.05$  & $0.54$ \tabularnewline
\hline \hline
\end{tabular}
\end{table}

\section{Semi-leptonic weak decays}
For the charged current process $c\to d,s~l^{+}\nu_{l}$, the effective Hamiltonian is
\begin{equation}
{\cal H}_{{\rm eff}}(c\to d,s~l^{+}\nu_{l})=
\frac{G_{F}}{\sqrt{2}}\Big(V_{cd}^{*}[\bar{d}\gamma_{\mu}(1-\gamma_5)c][\bar{\nu}_{l}\gamma_{\mu}(1-\gamma_5)l]+V_{cs}^{*}[\bar{s}\gamma_{\mu}(1-\gamma_5)c][\bar{\nu}_{l}\gamma_{\mu}(1-\gamma_5)l]\Big)\label{eq:efhvc},
\end{equation}
and for $b\to u,c~l^{-}\bar{\nu}_{l}$, the effective Hamiltonian has been given as
\begin{equation}
{\cal H}_{{\rm eff}}(b\to u,c~l^{-}\bar{\nu}_{l})=
\frac{G_{F}}{\sqrt{2}}\Big(V_{ub}[\bar{u}\gamma_{\mu}(1-\gamma_5)b][\bar{l}\gamma_{\mu}(1-\gamma_5)\nu_{l}]+V_{cb}[\bar{c}\gamma_{\mu}(1-\gamma_5)b][\bar{l}\gamma_{\mu}(1-\gamma_5)\nu_{l}]\Big)\label{eq:efhvb}.
	\end{equation}
While for the FCNC process $b\to sl^{+}l^{-}$, the effective Hamiltonian can be given as
\begin{equation}
{\cal H}_{{\rm eff}}(b\to sl^{+}l^{-})=-\frac{G_{F}}{\sqrt{2}}V_{tb}V_{ts}^{*}\sum_{i=1}^{10}C_{i}(\mu)O_{i}(\mu)\label{eq:efhfcnc}.
\end{equation}
In Eqs.~(\ref{eq:efhvc}), (\ref{eq:efhvb}) and (\ref{eq:efhfcnc}), the Fermi constant $G_F$ and the CKM matrix elements are taken from Ref.~\cite{Tanabashi:2018oca}:
\begin{eqnarray}
&&G_F=1.166\times 10^{-5}{\rm GeV}^{-2},\quad|V_{cd}|=0.218,\quad|V_{cs}|=0.997,\quad|V_{cb}|=0.0422,\nonumber\\
&&~|V_{ub}|=0.00394,\quad|V_{ts}|=0.0394,\quad|V_{td}|=0.0081,\quad|V_{tb}|=1.019.
\end{eqnarray}
The reader interested in the explicit forms of operators $O_{i}$ in Eq.~(\ref{eq:efhfcnc}) can consult Ref.~\cite{Buchalla:1995vs}. Wilson coefficients $C_{i}$ for each operators $O_{i}$ are calculated in the leading logarithmic approximation, with $m_{W}=80.4$~GeV and $\mu=m_{b,{\rm pole}}$~\cite{Buchalla:1995vs} and can be given as follows,
\begin{eqnarray}
&&C_{1}=1.107,\quad C_{2}=-0.248,\quad C_{3}=-0.011,\quad C_{4}=-0.026, \nonumber\\
&&C_{5}=-0.007,\quad C_{6}=-0.031,\quad C_{7}^{{\rm eff}}=-0.313,\quad C_{9}=4.344,\quad C_{10}=-4.669,
\end{eqnarray}
For the FCNC process ${\cal B}_{b}\to{\cal B}_{s}^{\prime}l^{+}l^{-}$, the amplitude can be obtained in following form,
\begin{eqnarray}
{\cal M}({\cal B}\to{\cal B}^{\prime}l^{+}l^{-}) & = & -\frac{G_{F}}{\sqrt{2}}V_{tb}V_{ts}^{*}\frac{\alpha_{{\rm em}}}{2\pi}\Big\{\Big(C_{9}^{{\rm eff}}(q^{2})\langle{\cal B}^{\prime}|\bar{s}\gamma_{\mu}(1-\gamma_{5})b|{\cal B}\rangle-2m_{b}C_{7}^{{\rm eff}}\langle{\cal B}^{\prime}|\bar{s}i\sigma_{\mu\nu}\frac{q^{\nu}}{q^{2}}(1+\gamma_{5})b|{\cal B}\rangle\Big)\bar{l}\gamma^{\mu}l\nonumber \\
&  & \qquad\qquad\qquad\quad+\,C_{10}\langle{\cal B}^{\prime}|\bar{s}\gamma_{\mu}(1-\gamma_{5})b|{\cal B}\rangle\bar{l}\gamma^{\mu}\gamma_{5}l\Big\}.\label{eq:the_amplitude}
\end{eqnarray}
In Refs.~\cite{Li:2009rc,Lu:2011jm,Giri:2005mt}, the signs before $C_{7}^{{\rm eff}}$ are different. In this paper, we take the same sign with the ones in Refs.~\cite{Li:2009rc,Lu:2011jm}, which is different from the one in Ref.~\cite{Giri:2005mt}. In the above Eq.~(\ref{eq:the_amplitude}), $C_{7}^{{\rm eff}}$ and $C_{9}^{{\rm eff}}$ are obtained as ~\cite{Buras:1994dj}
\begin{eqnarray}
C_{7}^{{\rm eff}} & = & C_{7}-C_{5}/3-C_{6},\nonumber \\
C_{9}^{{\rm eff}}(q^{2}) & = & C_{9}(\mu)+h(\hat{m}_{c},\hat{s})C_{0}-\frac{1}{2}h(1,\hat{s})(4C_{3}+4C_{4}+3C_{5}+C_{6})\nonumber \\
&  & -\frac{1}{2}h(0,\hat{s})(C_{3}+3C_{4})+\frac{2}{9}(3C_{3}+C_{4}+3C_{5}+C_{6}),
\end{eqnarray}
where $\hat{s}=q^{2}/m_{b}^{2}$, $C_{0}=C_{1}+3C_{2}+3C_{3}+C_{4}+3C_{5}+C_{6}$,
and $\hat{m}_{c}=m_{c}/m_{b}$.
The auxiliary functions $h$ have been given as
\begin{eqnarray}
h(z,\hat{s}) & = & -\frac{8}{9}\ln\frac{m_{b}}{\mu}-\frac{8}{9}\ln z+\frac{8}{27}+\frac{4}{9}x-\frac{2}{9}(2+x)|1-x|^{1/2}\times\begin{cases}
\left(\ln\left|\frac{\sqrt{1-x}+1}{\sqrt{1-x}-1}\right|-i\pi\right), & x\equiv\frac{4z^{2}}{\hat{s}}<1\\
2\arctan\frac{1}{\sqrt{x-1}}, & x\equiv\frac{4z^{2}}{\hat{s}}>1
\end{cases},\nonumber \\
h(0,\hat{s}) & = & -\frac{8}{9}\ln\frac{m_{b}}{\mu}-\frac{4}{9}\ln\hat{s}+\frac{8}{27}+\frac{4}{9}i\pi.
\end{eqnarray}

While for the FCNC process $b\to d l^+l^-$, the corresponding effective Hamiltonian and amplitude can be got by taking a  replacement $s\to d$ similarly.



\subsection{Decay widths}
\subsubsection{the charged current transition}

\begin{figure}
\includegraphics[width=0.45\columnwidth]{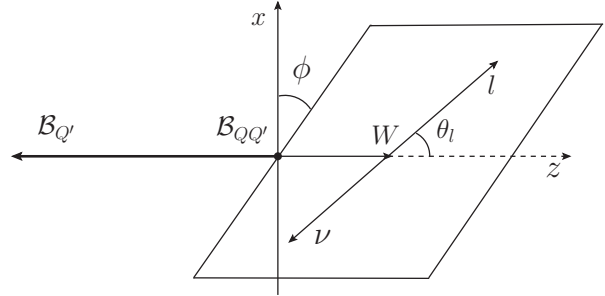}
\caption{ Kinematics for the charged current induced decay mode. }
\label{fig:dynamic}
\end{figure}

For the charged current induced transition, the kinematics are shown in Fig.~\ref{fig:dynamic}, and the helicity amplitudes are defined by
\begin{equation}
HV_{\lambda^{\prime},\lambda_{W}}^{\lambda}\equiv\langle{\cal B}_{f}^{(*)}(\lambda^{\prime})|\bar{q}\gamma^{\mu}Q|{\cal B}_{i}(\lambda)\rangle\epsilon_{W\mu}^{*}(\lambda_{W})\quad
{\rm and}\quad HA_{\lambda^{\prime},\lambda_{W}}^{\lambda}\equiv\langle{\cal B}_{f}^{(*)}(\lambda^{\prime})|\bar{q}\gamma^{\mu}\gamma_{5}Q|{\cal B}_{i}(\lambda)\rangle\epsilon_{W\mu}^{*}(\lambda_{W}),
\end{equation}
here $\epsilon_{\mu}$ and $q_{\mu}$ are the polarization vector and four-momentum of the virtual propagator W, and $\lambda_{W}$ means the polarization of the virtual propagator W. $\lambda$ and $\lambda^{\prime}$ are the helicities of the baryon in the initial and final baryon states, respectively. The detail derivation process of helicity amplitudes can be found in Appendix~\ref{appendix:helicity}.
These helicity amplitudes are related to the form factors by the following
expressions.
\begin{itemize}
\item The transition $B_{i}(\lambda)\to B_{f}(\lambda^{\prime})$ matrix elements are parameterized as shown in Eq.~\eqref{eq:matrix_element_2},
and the helicity amplitudes of $1/2\to1/2$ charged current transition can be expressed with following equations,
\begin{align}
HV_{\frac{1}{2},0}^{-\frac{1}{2}} & =-i\frac{\sqrt{Q_{-}}}{\sqrt{q^{2}}}\left((M+M^{\prime})f_{1}^{\frac{1}{2}\to\frac{1}{2}}-\frac{q^{2}}{M}f_{2}^{\frac{1}{2}\to\frac{1}{2}}\right),\quad
HV_{\frac{1}{2},1}^{\frac{1}{2}}  =i\sqrt{2Q_{-}}\left(-f_{1}^{\frac{1}{2}\to\frac{1}{2}}+\frac{M+M^{\prime}}{M}f_{2}^{\frac{1}{2}\to\frac{1}{2}}\right),\nonumber\\
HA_{\frac{1}{2},0}^{-\frac{1}{2}} & =-i\frac{\sqrt{Q_{+}}}{\sqrt{q^{2}}}\left((M-M^{\prime})g_{1}^{\frac{1}{2}\to\frac{1}{2}}+\frac{q^{2}}{M}g_{2}^{\frac{1}{2}\to\frac{1}{2}}\right),\quad
HA_{\frac{1}{2},1}^{\frac{1}{2}}  =i\sqrt{2Q_{+}}\left(-g_{1}^{\frac{1}{2}\to\frac{1}{2}}-\frac{M-M^{\prime}}{M}g_{2}^{\frac{1}{2}\to\frac{1}{2}}\right),\label{eq:helicty22v}
\end{align}
with $Q_{\pm}=2(P\cdot P^{\prime}\pm MM^{\prime})=(M\pm M^{\prime})^{2}-q^{2}$. Here $f_i^{\frac{1}{2}\to\frac{1}{2}}$ and $g_{i}^{\frac{1}{2}\to\frac{1}{2}}$ are the physical form factors which are defined by Eq.~(\ref{eq:physical_ff22}). $M$ and $M^{\prime}$ are the  masses for the initial and final baryon. While the negative helicity amplitudes have the following relations with the corresponding positive ones,
\begin{equation}
HV_{-\lambda^{\prime},-\lambda_{W}}^{-\lambda}=HV_{\lambda^{\prime},\lambda_{W}}^{\lambda}\quad\text{and}\quad HA_{-\lambda^{\prime},-\lambda_{W}}^{-\lambda}=-HA_{\lambda^{\prime},\lambda_{W}}^{\lambda}.
\end{equation}
Then the total helicity amplitudes for (V-A) current can be shown as follows,
\begin{equation}
H_{\lambda^{\prime},\lambda_{W}}^{\lambda}=HV_{\lambda^{\prime},\lambda_{W}}^{\lambda}-HA_{\lambda^{\prime},\lambda_{W}}^{\lambda}.
\end{equation}
The polarized differential decay widths can be given as
\begin{align}
\frac{d\Gamma_{L}}{dq^{2}} & =\frac{G_{F}^{2}|V_{CKM}|^{2}}{(2\pi)^{3}}\frac{q^{2}|\vec{P}^{\prime}|}{24M^{2}}
(|H_{\frac{1}{2},0}^{-\frac{1}{2}}|^{2}+|H_{-\frac{1}{2},0}^{\frac{1}{2}}|^{2}),\label{eq:longi-1}\\
\frac{d\Gamma_{T}}{dq^{2}} & =\frac{G_{F}^{2}|V_{CKM}|^{2}}{(2\pi)^{3}}\frac{q^{2}|\vec{P}^{\prime}|}{24M^{2}}
(|H_{\frac{1}{2},1}^{\frac{1}{2}}|^{2}+|H_{-\frac{1}{2},-1}^{-\frac{1}{2}}|^{2}),\label{eq:trans-1}
\end{align}
with $|\vec{P}^{\prime}|=\sqrt{Q_{+}Q_{-}}/2M$.
\item The $B_{i}(\lambda)\to B_{f}^{*}(\lambda^{\prime})$ transition matrix element are parameterized with Eq.~\eqref{eq:matrix_element_32nVA}, and the helicity amplitudes of the transitions ${1}/{2}\to{3}/{2}$ induced by charged current can be expressed with following equations,
\begin{eqnarray}
HV_{3/2,1}^{-1/2} & = & - i\sqrt{Q_{-}}f_{4}^{\frac{1}{2}\to\frac{3}{2}},\quad
HV_{1/2,1}^{1/2}  =  i\sqrt{\frac{Q_{-}}{3}}\left[f_{4}^{\frac{1}{2}\to\frac{3}{2}}-\frac{Q_{+}}{MM^{\prime}}f_{1}^{\frac{1}{2}\to\frac{3}{2}}\right],\\
HV_{1/2,0}^{-1/2} & = &
 i\sqrt{\frac{2}{3}}\frac{\sqrt{Q_{-}}}{\sqrt{q^{2}}}
\Big[\frac{M^2-M^{\prime2}-q^2}{2M^{\prime}}f_{4}^{\frac{1}{2}\to\frac{3}{2}}-
\frac{M- M^{\prime}}{2MM^{\prime}}Q_{+}f_{1}^{\frac{1}{2}\to\frac{3}{2}}
-\frac{Q_{+}Q_{-}}{2M^{2}M^{\prime}}f_{2}^{\frac{1}{2}\to\frac{3}{2}}\Big],\label{eq:helicty23v}\\
HA_{3/2,1}^{-1/2} & = &  i\sqrt{Q_{+}}f_{4}^{\frac{1}{2}\to\frac{3}{2}},\quad
HA_{1/2,1}^{1/2}  =  i\sqrt{\frac{Q_{+}}{3}}\left[g_{4}^{\frac{1}{2}\to\frac{3}{2}}-\frac{Q_{-}}{MM^{\prime}}g_{1}^{\frac{1}{2}\to\frac{3}{2}}\right],\\
HA_{1/2,0}^{-1/2} & = &
- i\sqrt{\frac{2}{3}}\frac{\sqrt{Q_{+}}}{\sqrt{q^{2}}}
\Big[\frac{M^2-M^{\prime2}-q^2}{2M^{\prime}}g_{4}^{\frac{1}{2}\to\frac{3}{2}}+
\frac{M+ M^{\prime}}{2MM^{\prime}}Q_{-}g_{1}^{\frac{1}{2}\to\frac{3}{2}}
-\frac{Q_{+}Q_{-}}{2M^{2}M^{\prime}}g_{2}^{\frac{1}{2}\to\frac{3}{2}}\Big],\label{eq:helicty23a}
\end{eqnarray}
here $f_i^{\frac{1}{2}\to\frac{3}{2}}$ and $g_{i}^{\frac{1}{2}\to\frac{3}{2}}$ are physics form factors introduced with Eq.~(\ref{eq:physical_ff23}).
$M$ and $M^{\prime}$ are the masses of the initial and final baryon states, respectively. While the negative helicity amplitudes have the following relations with the corresponding positive ones,
\begin{equation}
HV_{-\lambda^{\prime},-\lambda_{W}}^{-\lambda}=- HV_{\lambda^{\prime},\lambda_{W}}^{\lambda}\quad \text{and} \quad HA_{-\lambda^{\prime},-\lambda_{W}}^{-\lambda}= HA_{\lambda^{\prime},\lambda_{W}}^{\lambda}.
\end{equation}
Then we can get the total helicity amplitudes,
\begin{equation}
H_{\lambda^{\prime},\lambda_{W}}^{\lambda}=
HV_{\lambda^{\prime},\lambda_{W}}^{\lambda}-HA_{\lambda^{\prime},\lambda_{W}}^{\lambda}.
\end{equation}
The polarized differential decay widths can be given as
\begin{eqnarray}
\frac{d\Gamma_{L}}{dq^2} & = & \frac{G_{F}^{2}}{(2\pi)^{3}}|V_{{\rm CKM}}|^{2}\frac{q^{2}|\vec{P}^{\prime}|}{24M^2}[|H_{1/2,0}^{-1/2}|^{2}+|H_{-1/2,0}^{1/2}|^{2}],\label{eq:longi-2}\\
\frac{d\Gamma_{T}}{dq^2} & = & \frac{G_{F}^{2}}{(2\pi)^{3}}|V_{{\rm CKM}}|^{2}\frac{q^{2}|\vec{P}^{\prime}|}{24M^2}[|H_{1/2,1}^{1/2}|^{2}
+|H_{-1/2,-1}^{-1/2}|^{2}+|H_{3/2,1}^{-1/2}|^{2}+|H_{-3/2,-1}^{1/2}|^{2}].\label{eq:trans-2}
\end{eqnarray}
\end{itemize}
\subsubsection{the FCNC transition}
For the FCNC induced transition, we adopt the helicity amplitudes as follows,
\begin{eqnarray}
H_{\lambda^{\prime},\lambda_{V}}^{{\cal V}_{l},\lambda} & \equiv & \Big(C_{9}^{{\rm eff}}(q^{2})\langle{\cal B^{(*)}}(\lambda^{\prime})|\bar{s}\gamma^{\mu}(1-\gamma_{5})b|{\cal B}(\lambda)\rangle-C_{7}^{{\rm eff}}2m_{b}\langle{\cal B}^{(*)}(\lambda^{\prime})|\bar{s}i\sigma^{\mu\nu}\frac{q_{\nu}}{q^{2}}(1+\gamma_{5})b|{\cal B}(\lambda)\rangle\Big)\epsilon_{\mu}^{*}(\lambda_{V}),\nonumber \\
H_{\lambda^{\prime},t}^{{\cal V}_{l},\lambda} & \equiv & \Big(C_{9}^{{\rm eff}}(q^{2})\langle{\cal B}^{(*)}(\lambda^{\prime})|\bar{s}\gamma^{\mu}(1-\gamma_{5})b|{\cal B}(\lambda)\rangle\Big)\frac{q_{\mu}}{\sqrt{q^{2}}},
\label{eq:HV2}
\end{eqnarray}
and
\begin{eqnarray}
H_{\lambda^{\prime},\lambda_{V}}^{{\cal A}_{l},\lambda} & \equiv & \Big(C_{10}\langle{\cal B^{(*)}}(\lambda^{\prime})|\bar{s}\gamma^{\mu}(1-\gamma_{5})b|{\cal B}(\lambda)\rangle\Big)\epsilon_{\mu}^{*}(\lambda_{V}),\nonumber \\
H_{\lambda^{\prime},t}^{{\cal A}_{l},\lambda} & \equiv & \Big(C_{10}\langle{\cal B^{(*)}}(\lambda^{\prime})|\bar{s}\gamma^{\mu}(1-\gamma_{5})b|{\cal B}(\lambda)\rangle\Big)\frac{q_{\mu}}{\sqrt{q^{2}}},
\end{eqnarray}
here $\epsilon_{\mu}$ and $q_{\mu}$ are the polarization vector and four-momentum of the virtual vector propagator V, and $\lambda_{V}$ means the polarization of the virtual vector propagator V. $\lambda$ and $\lambda^{\prime}$ are the helicities of the baryon in the initial and final baryon states, respectively.
In the following, the superscripts ``${\cal V}_{l}$" and ``${\cal A}_{l}$" denote the
corresponding leptonic counterparts $\bar{l}\gamma^{\mu}l$ and $\bar{l}\gamma^{\mu}\gamma_{5}l$, respectively.
\begin{itemize}
\item  The transition $B_{i}(\lambda)\to B_{f}(\lambda^{\prime})$ matrix elements are parameterized with   Eqs.~\eqref{eq:matrix_element_2}-\eqref{eq:matrix_element_2p},
    and the helicity amplitudes of $1/2\to1/2$ induced by FCNC transition can be obtained with following expressions,
\begin{align}
	HV_{\frac{1}{2},0}^{{\cal V}_{l},-\frac{1}{2}} & =-i\frac{\sqrt{Q_{-}}}{\sqrt{q^{2}}}\left((M+M^{\prime})F_{1}^{{\cal V}_{l}}-\frac{q^{2}}{M}F_{2}^{{\cal V}_{l}}\right),\quad
	HV_{\frac{1}{2},1}^{{\cal V}_{l},\frac{1}{2}}  =i\sqrt{2Q_{-}}\left(-F_{1}^{{\cal V}_{l}}+\frac{M+M^{\prime}}{M}F_{2}^{{\cal V}_{l}}\right),\nonumber \\
	HA_{\frac{1}{2},0}^{{\cal V}_{l},-\frac{1}{2}} & =-i\frac{\sqrt{Q_{+}}}{\sqrt{q^{2}}}\left((M-M^{\prime})G_{1}^{{\cal V}_{l}}+\frac{q^{2}}{M}G_{2}^{{\cal V}_{l}}\right),\quad
	HA_{\frac{1}{2},1}^{{\cal V}_{l},\frac{1}{2}}  =i\sqrt{2Q_{+}}\left(-G_{1}^{{\cal V}_{l}}-\frac{M-M^{\prime}}{M}G_{2}^{{\cal V}_{l}}\right),\label{eq:hv22}
	\end{align}
	and
	\begin{eqnarray}
	HV_{-\lambda^{\prime},-\lambda_{V}}^{{\cal V}_{l},-\lambda} & = & HV_{\lambda^{\prime},\lambda_{V}}^{{\cal V}_{l},\lambda},
	\quad
	HA_{-\lambda^{\prime},-\lambda_{V}}^{{\cal V}_{l},-\lambda}  = -HA_{\lambda^{\prime},\lambda_{V}}^{{\cal V}_{l},\lambda}.
	\end{eqnarray}
where the ``HV" and ``HA" are corresponding to the $\Gamma^{\mu}$ and $\Gamma^{\mu}\gamma_{5}$ parts in Eq.~(\ref{eq:HV2}), respectively.
	The total helicity amplitude can be given as
	\begin{equation}
	H_{\lambda^{\prime},\lambda_{V}}^{{\cal V}_{l},\lambda}=HV_{\lambda^{\prime},\lambda_{V}}^{{\cal V}_{l},\lambda}-HA_{\lambda^{\prime},\lambda_{V}}^{{\cal V}_{l},\lambda}.
	\end{equation}
The specific expressions of $H_{\lambda^{\prime},\lambda_{V}}^{{\cal A}_{l},\lambda}$ are similar with the ones of $H_{\lambda^{\prime},\lambda_{V}}^{V,\lambda}$, except
	\begin{eqnarray}
	F_{i}^{{\cal V}_{l}}  \to  F_{i}^{{\cal A}_{l}} \qquad \rm{and} \qquad
	G_{i}^{{\cal V}_{l}}  \to  G_{i}^{{\cal A}_{l}}.
	\end{eqnarray}
	Furthermore, the timelike polarizations of the virtual vector propagator V
for the helicity amplitudes, $H^{{\cal A}_{l}}_{t}$ are necessary for FCNC induced transitions,
	\begin{align}
	HV_{-\frac{1}{2},t}^{{\cal A}_{l},\frac{1}{2}}=HV_{\frac{1}{2},t}^{{\cal A}_{l},-\frac{1}{2}}=-i\frac{\sqrt{Q_{+}}}{\sqrt{q^{2}}}\left((M-M^{\prime})F_{1}^{{\cal A}_{l}}+\frac{q^{2}}{M}F_{3}^{{\cal A}_{l}}\right),\nonumber \\
	-HA_{-\frac{1}{2},t}^{{\cal A}_{l},\frac{1}{2}}=HA_{\frac{1}{2},t}^{{\cal A}_{l},-\frac{1}{2}}=-i\frac{\sqrt{Q_{-}}}{\sqrt{q^{2}}}\left((M+M^{\prime})G_{1}^{{\cal A}_{l}}-\frac{q^{2}}{M}G_{3}^{{\cal A}_{l}}\right)\label{eq:helicty22fcnc}
   \end{align}
	and
	\begin{equation}
	H_{\lambda^{\prime},t}^{{\cal A}_{l},\lambda}=HV_{\lambda^{\prime},t}^{{\cal A}_{l},\lambda}-HA_{\lambda^{\prime},t}^{{\cal A}_{l},\lambda}.	\label{eq:HV22A}
	\end{equation}
In the above Eq.~(\ref{eq:hv22}-\ref{eq:HV22A}), the following notations have been introduced:
	\begin{eqnarray}
	F_{1}^{{\cal V}_{l}}(q^{2}) & \equiv & C_{9}^{{\rm eff}}(q^{2})f_{1}^{\frac{1}{2}\to\frac{1}{2}}(q^{2})-C_{7}^{{\rm eff}}\frac{2m_{b}}{M^{\prime}-M}f_{1}^{\frac{1}{2}\to\frac{1}{2},T}(q^{2}),\nonumber \\
    \quad F_{2}^{{\cal V}_{l}}(q^{2})  &\equiv&  C_{9}^{{\rm eff}}(q^{2})f_{2}^{\frac{1}{2}\to\frac{1}{2}}(q^{2})-C_{7}^{{\rm eff}}\frac{2m_{b}M}{q^{2}}f_{2}^{\frac{1}{2}\to\frac{1}{2},T}(q^{2}),\nonumber \\
	G_{1}^{{\cal V}_{l}}(q^{2}) & \equiv & C_{9}^{{\rm eff}}(q^{2})g_{1}^{\frac{1}{2}\to\frac{1}{2}}(q^{2})+C_{7}^{{\rm eff}}\frac{2m_{b}}{M^{\prime}+M}g_{1}^{\frac{1}{2}\to\frac{1}{2},T}(q^{2}),\nonumber \\
	 G_{2}^{{\cal V}_{l}}(q^{2})  &\equiv & C_{9}^{{\rm eff}}(q^{2})g_{2}^{\frac{1}{2}\to\frac{1}{2}}(q^{2})+C_{7}^{{\rm eff}}\frac{2m_{b}M}{q^{2}}g_{2}^{\frac{1}{2}\to\frac{1}{2},T}(q^{2}),
	\end{eqnarray}
and
\begin{eqnarray}
	F_{i}^{{\cal A}_{l}}(q^{2}) & \equiv & C_{10}f_{i}^{\frac{1}{2}\to\frac{1}{2}}(q^{2}),\quad
	 G_{i}^{{\cal A}_{l}}(q^{2})  \equiv  C_{10}g_{i}^{\frac{1}{2}\to\frac{1}{2}}(q^{2})\quad(i=1,2,3).
\end{eqnarray}
Here $f_{i}^{\frac{1}{2}\to\frac{1}{2},(T)}(q^2)$ and $g_{i}^{\frac{1}{2}\to\frac{1}{2},(T)}(q^2)$ are the physical form factors illuminated by Eq.~(\ref{eq:physical_ff22}).	
The longitudinally and transversely polarized differential decay widths read,
\begin{eqnarray}
\frac{d\Gamma_{L}}{dq^{2}} & = &\frac{G_{F}^{2}|V_{\rm CKM}|^2\alpha_{em}^{2}|\vec{P}^{\prime}||\vec{p}_{1}|}{24(2\pi)^{5}M^{2}\sqrt{q^{2}}}\Big\{(q^{2}+2m_{l}^{2})(|H_{-\frac{1}{2},0}^{{\cal V}_{l},\frac{1}{2}}|^{2}+|H_{\frac{1}{2},0}^{{\cal V}_{l},-\frac{1}{2}}|^{2})\nonumber \\
&  & +(q^{2}-4m_{l}^{2})(|H_{-\frac{1}{2},0}^{{\cal A}_{l},\frac{1}{2}}|^{2}+|H_{\frac{1}{2},0}^{{\cal A}_{l},-\frac{1}{2}}|^{2})+6m_{l}^{2}(|H_{-\frac{1}{2},t}^{{\cal A}_{l},\frac{1}{2}}|^{2}+|H_{\frac{1}{2},t}^{{\cal A}_{l},-\frac{1}{2}}|^{2})\Big\},\label{eq:longfcnc-1}\\
\frac{d\Gamma_{T}}{dq^{2}} & = & \frac{G_{F}^{2}|V_{\rm CKM}|^2\alpha_{em}^{2}|\vec{P}^{\prime}||\vec{p}_{1}|}{24(2\pi)^{5}M^{2}\sqrt{q^{2}}}\Big\{(q^{2}+2m_{l}^{2})(|H_{\frac{1}{2},1}^{{\cal V}_{l},\frac{1}{2}}|^{2}+|H_{-\frac{1}{2},-1}^{{\cal V}_{l},-\frac{1}{2}}|^{2})\nonumber \\
&  &+(q^{2}-4m_{l}^{2})(|H_{\frac{1}{2},1}^{{\cal A}_{l},\frac{1}{2}}|^{2}+|H_{-\frac{1}{2},-1}^{{\cal A}_{l},-\frac{1}{2}}|^{2})\Big\}.\label{eq:tranfcnc-1}
\end{eqnarray}
with $V_{\rm CKM}=V_{tb}V_{ts}^{*}$ for $b\to s$ processes, $V_{\rm CKM}=V_{tb}V_{td}^{*}$ for $b\to d$ processes and
$|\vec{p}_{1}|=\frac{1}{2}\sqrt{q^2-4m_{l}^2}$.
\item The transition ${1}/{2}\to{3}/{2}$ matrix elements are parameterized with Eqs.~\eqref{eq:matrix_element_32nVA}-\eqref{eq:matrix_element_32nT},
and the helicity amplitudes induced by FCNC can be given by the following expressions,
\begin{eqnarray}
HV_{3/2,1}^{{\cal V}_{l},-1/2} & = & - i\sqrt{Q_{-}}{\cal F}_{4}^{{\cal V}_{l}},\quad HV_{1/2,1}^{{\cal V}_{l},1/2}  =
i\sqrt{\frac{Q_{-}}{3}}\left[{\cal F}_{4}^{{\cal V}_{l}}-\frac{Q_{+}}{MM^{\prime}}{\cal F}_{1}^{{\cal V}_{l}}\right],\label{eq:hv23}\\
HV_{1/2,0}^{{\cal V}_{l},-1/2} & = &  i\sqrt{\frac{2}{3}}\frac{\sqrt{Q_{-}}}{\sqrt{q^{2}}}
\Big[\frac{M^2-M^{\prime2}-q^2}{2M^{\prime}}{\cal F}_{4}^{{\cal V}_{l}}-\frac{M- M^{\prime}}{2MM^{\prime}}Q_{+}{\cal F}_{1}^{{\cal V}_{l}}-\frac{Q_{+}Q_{-}}{2M^{2}M^{\prime}}{\cal F}_{2}^{{\cal V}_{l}}\Big]\label{eq:hv232},
\\
HA_{3/2,1}^{{\cal V}_{l},-1/2} & = &  i\sqrt{Q_{+}}{\cal G}_{4}^{{\cal V}_{l}},\quad HA_{1/2,1}^{{\cal V}_{l},1/2}  =  i\sqrt{\frac{Q_{+}}{3}}\left[{\cal G}_{4}^{{\cal V}_{l}}-\frac{Q_{-}}{MM^{\prime}}{\cal G}_{1}^{{\cal V}_{l}}\right],
\\
HA_{1/2,0}^{{\cal V}_{l},-1/2} & = & - i\sqrt{\frac{2}{3}}\frac{\sqrt{Q_{+}}}{\sqrt{q^{2}}}
\Big[\frac{M^2-M^{\prime2}-q^2}{2M^{\prime}}{\cal G}_{4}^{{\cal V}_{l}}
+\frac{M+ M^{\prime}}{2MM^{\prime}}Q_{-}{\cal G}_{1}^{{\cal V}_{l}}
-\frac{Q_{+}Q_{-}}{2M^{2}M^{\prime}}{\cal G}_{2}^{{\cal V}_{l}}\Big].\label{eq:helicty23fcnc}
\end{eqnarray}
and
\begin{eqnarray}
HV_{-\lambda^{\prime},-\lambda_{W}}^{{\cal V}_{l},-\lambda}&=&- HV_{\lambda^{\prime},\lambda_{W}}^{{\cal V}_{l},\lambda},\quad
HA_{-\lambda^{\prime},-\lambda_{W}}^{{\cal V}_{l},-\lambda}=HA_{\lambda^{\prime},\lambda_{W}}^{{\cal V}_{l},\lambda}.
\end{eqnarray}
where the ``HV" and ``HA" are corresponding to the $\Gamma^{\mu}$ and $\Gamma^{\mu}\gamma_{5}$ parts in Eq.~(\ref{eq:HV2}), respectively.
Then we can get the total helicity amplitudes,
\begin{equation} H_{\lambda^{\prime},\lambda_{V}}^{{\cal V}_{l},\lambda}=HV_{\lambda^{\prime},\lambda_{V}}^{{\cal V}_{l},\lambda}-HA_{\lambda^{\prime},\lambda_{V}}^{{\cal V}_{l},\lambda}.
\end{equation}
The specific expressions of $H_{\lambda^{\prime},\lambda_{V}}^{{\cal A}_{l},\lambda}$ are similar with the ones of $H_{\lambda^{\prime},\lambda_{V}}^{V,\lambda}$, except
\begin{eqnarray}
{\cal F}_{i}^{{\cal V}_{l}}  \to  {\cal F}_{i}^{{\cal A}_{l}}\qquad \rm{and}\qquad
{\cal G}_{i}^{{\cal V}_{l}} \to  {\cal G}_{i}^{{\cal A}_{l}}.
\end{eqnarray}
Furthermore, the timelike polarizations of the virtual vector propagator V
for the helicity amplitudes, $H^{{\cal A}_{l}}_{t}$ are necessary for FCNC induced transitions,
\begin{align} &-HV_{-\frac{1}{2},t}^{{\cal A}_{l},\frac{1}{2}}=HV_{\frac{1}{2},t}^{{\cal A}_{l},-\frac{1}{2}}
=i\sqrt{\frac{2}{3}}\sqrt{Q_{+}}\frac{Q_{-}}{2MM^{\prime}}\frac{M^2-M^{\prime 2}}{M}{\cal F}_3^{{\cal A}_{l}},\nonumber \\ &HA_{-\frac{1}{2},t}^{{\cal A}_{l},\frac{1}{2}}=HA_{\frac{1}{2},t}^{{\cal A}_{l},-\frac{1}{2}}
=-i\sqrt{\frac{2}{3}}\sqrt{Q_{-}}\frac{Q_{+}}{2MM^{\prime}}\frac{M^2-M^{\prime 2}}{M}{\cal G}_3^{{\cal A}_{l}}
\end{align}
and
\begin{equation}
H_{\lambda^{\prime},t}^{{\cal A}_{l},\lambda}=HV_{\lambda^{\prime},t}^{{\cal A}_{l},\lambda}-HA_{\lambda^{\prime},t}^{{\cal A}_{l},\lambda}.\label{eq:HV23A}
\end{equation}
In Eqs.~(\ref{eq:hv23}-\ref{eq:HV23A}), the following notations are introduced.
\begin{eqnarray}
{\cal F}_{i}^{{\cal V}_{l}}(q^{2}) & \equiv & C_{9}^{{\rm eff}}(q^{2})f_{i}^{\frac{1}{2}\to\frac{3}{2}}(q^{2})-C_{7}^{{\rm eff}}\frac{2m_{b}M}{q^{2}}f_{i}^{\frac{1}{2}\to\frac{3}{2},T}(q^{2}),\nonumber\\
 {\cal G}_{i}^{{\cal V}_{l}}(q^{2}) & \equiv & C_{9}^{{\rm eff}}(q^{2})g_{i}^{\frac{1}{2}\to\frac{3}{2}}(q^{2})+C_{7}^{{\rm eff}}\frac{2m_{b}M}{q^{2}}g_{i}^{\frac{1}{2}\to\frac{3}{2},T}(q^{2}),\nonumber\\
{\cal F}_{i}^{{\cal A}_{l}}(q^{2}) & \equiv & C_{10}f_{i}^{\frac{1}{2}\to\frac{3}{2}}(q^{2}),\quad {\cal G}_{i}^{{\cal A}_{l}}(q^{2})  \equiv  C_{10}g_{i}^{\frac{1}{2}\to\frac{3}{2}}(q^{2}),\quad(i=1,2,3,4).
\end{eqnarray}
Here $f_{i}^{\frac{1}{2}\to\frac{3}{2},(T)}(q^2)$ and $g_{i}^{\frac{1}{2}\to\frac{3}{2},(T)}(q^2)$ are physics form factors illuminated by Eq.~(\ref{eq:physical_ff23}).
The longitudinally and transversely polarized differential decay widths read
\begin{eqnarray}
\frac{d\Gamma_{L}}{dq^{2}} & = &\frac{G_{F}^{2}|V_{\rm CKM}|^2\alpha_{em}^{2}|\vec{P}^{\prime}||\vec{p}_{1}|}{24(2\pi)^{5}M^{2}\sqrt{q^{2}}}
\Big\{(q^{2}+2m_{l}^{2})(|H_{-\frac{1}{2},0}^{{\cal V}_{l},\frac{1}{2}}|^{2}+|H_{\frac{1}{2},0}^{{\cal V}_{l},-\frac{1}{2}}|^{2})
\nonumber \\
&  & +(q^{2}-4m_{l}^{2})(|H_{-\frac{1}{2},0}^{{\cal A}_{l},\frac{1}{2}}|^{2}+|H_{\frac{1}{2},0}^{{\cal A}_{l},-\frac{1}{2}}|^{2})+6m_{l}^{2}(|H_{-\frac{1}{2},t}^{{\cal A}_{l},\frac{1}{2}}|^{2}+|H_{\frac{1}{2},t}^{{\cal A}_{l},-\frac{1}{2}}|^{2})\Big\},\label{eq:longfcnc-2}\\
\frac{d\Gamma_{T}}{dq^{2}} & = & \frac{G_{F}^{2}|V_{\rm CKM}|^2\alpha_{em}^{2}|\vec{P}^{\prime}||\vec{p}_{1}|}{24(2\pi)^{5}M^{2}\sqrt{q^{2}}}
\Big\{(q^{2}+2m_{l}^{2})(|H_{\frac{1}{2},1}^{{\cal V}_{l},\frac{1}{2}}|^{2}+|H_{-\frac{1}{2},-1}^{{\cal V}_{l},-\frac{1}{2}}|^{2}
+|H_{\frac{3}{2},1}^{{\cal V}_{l},-\frac{1}{2}}|^{2}+|H_{-\frac{3}{2},-1}^{{\cal V}_{l},\frac{1}{2}}|^{2})\nonumber \\
&  &+(q^{2}-4m_{l}^{2})(|H_{\frac{1}{2},1}^{{\cal A}_{l},\frac{1}{2}}|^{2}+|H_{-\frac{1}{2},-1}^{{\cal A}_{l},-\frac{1}{2}}|^{2}
+|H_{\frac{3}{2},1}^{{\cal A}_{l},-\frac{1}{2}}|^{2}+|H_{-\frac{3}{2},-1}^{{\cal A}_{l},\frac{1}{2}}|^{2})\Big\}.\label{eq:tranfcnc-2}
\end{eqnarray}
with $V_{\rm CKM}=V_{tb}V_{ts}^{*}$ for $b\to s$ processes, $V_{\rm CKM}=V_{tb}V_{td}^{*}$ for $b\to d$ processes and
$|\vec{p}_{1}|=\frac{1}{2}\sqrt{q^2-4m_{l}^2}$.
\end{itemize}
In the end, the total differential decay width can be written as
\begin{eqnarray}
\frac{d\Gamma}{dq^{2}} & = & \frac{d\Gamma_{L}}{dq^{2}}+\frac{d\Gamma_{T}}{dq^{2}},
\end{eqnarray}
then we can calculate total width using the following integral,
\begin{equation}
\Gamma=\int_{q^2_{\rm min}}^{(M-M^{\prime})^{2}}dq^{2}\frac{d\Gamma}{dq^{2}},
\end{equation}
where $q^2_{\rm min}= 0$ for these decays with charged current, while $q^2_{\rm min}=4m_{l}^2$ for other decays with FCNC.
At the same time, the ratio of the longitudinal to transverse decay rates $\Gamma_{L}/\Gamma_{T}$ can be calculated.
\subsection{Results for semi-leptonic decays}
\begin{itemize}
\item
For the transition ${1}/{2}\to{1}/{2}$ with $\rm{V-A}$ current, the integrated partial decay widths, the relevant branching ratios and $\Gamma_{L}/\Gamma_{T}$s are shown in Tab.~\ref{Tab:branchingv22}. The dependence of $q^2$ of the differential decay widths can be shown in Fig.~\ref{fig:decaywidthbbtob22}.
\begin{figure}
\includegraphics[width=0.8\columnwidth]{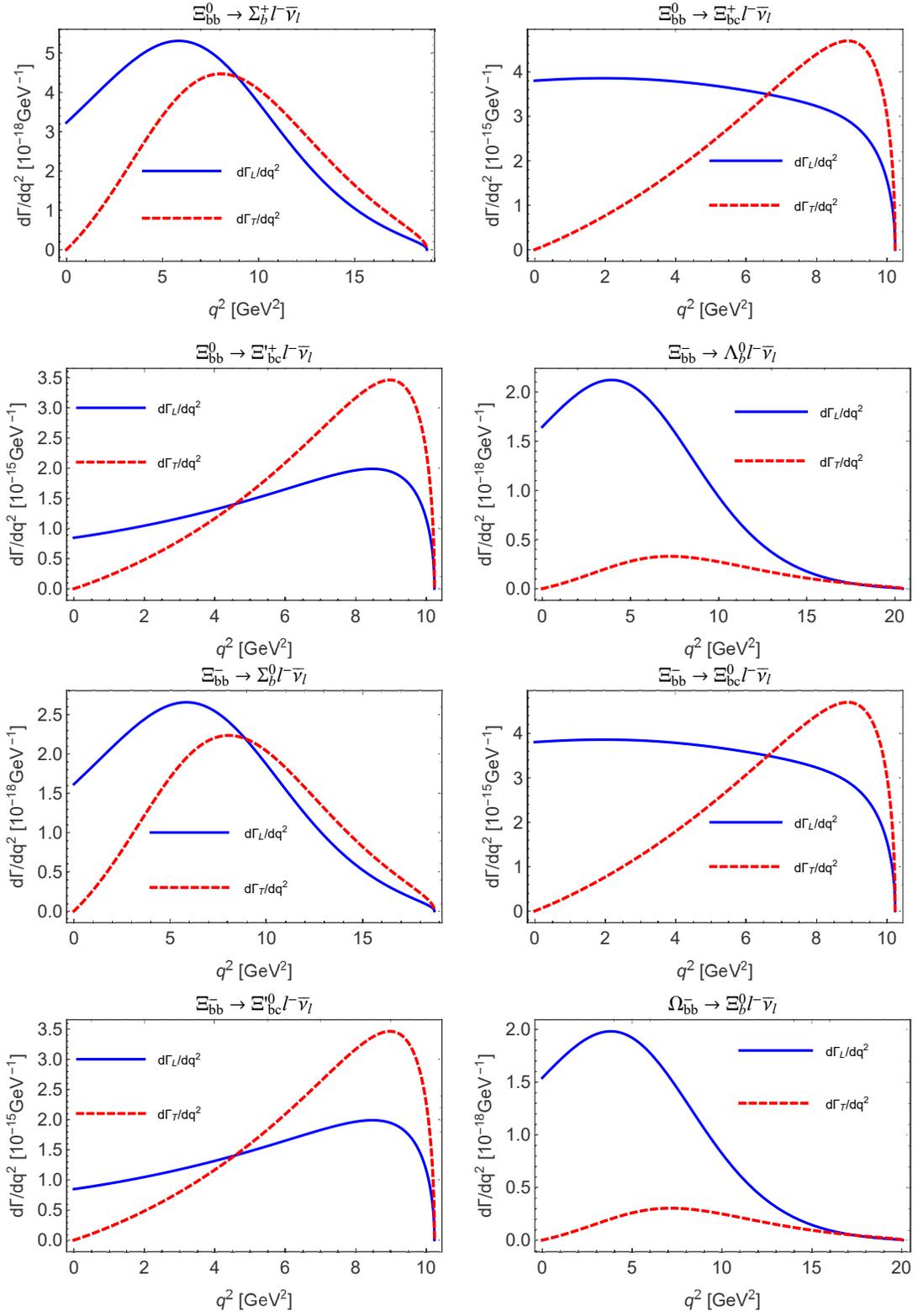}
\caption{The differential decay widths $d\Gamma_{L}/dq^2$ and $d\Gamma_{T}/dq^2$ for the processes ${\cal B}_{bb}\to{\cal B}_{b}({\cal B}_{bc})l^-\bar{\nu}_{l}$ dependence on $q^2$. Blue solid line: $d\Gamma_{L}/dq^2$ defined with Eq.~(\ref{eq:longi-1}), red dashes line: $d\Gamma_{T}/dq^2$ defined with Eq.~(\ref{eq:trans-1}).}
\label{fig:decaywidthbbtob22}
\end{figure}
\item  For the transition ${1}/{2}\to{1}/{2}$ induced by FCNC, the integrated partial decay widths, the relevant branching ratios and $\Gamma_{L}/\Gamma_{T}$s are shown in Tab.~\ref{Tab:branchingratio22}. The dependence of $q^2$ of the differential decay widths can be shown in Fig.~\ref{fig:decaywidthbbtobFCNC22}.

\begin{figure}
\includegraphics[width=0.9\columnwidth]{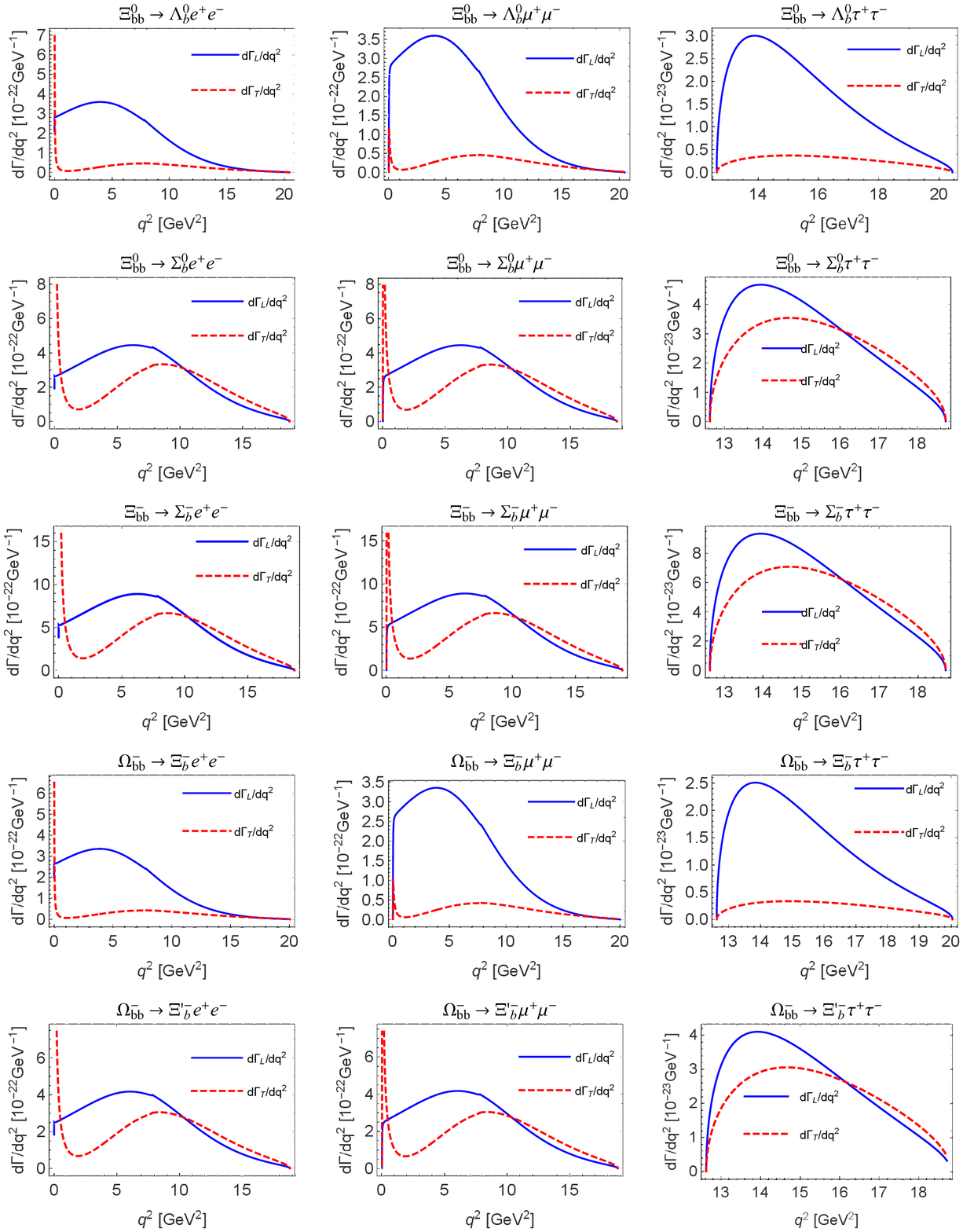}
\caption{The differential decay widths $d\Gamma_{L}/dq^2$ and $d\Gamma_{T}/dq^2$ for the processes ${\cal B}_{bb}\to{\cal B}_{b}l^+l^{-}$ dependence on $q^2$. Blue solid line: $d\Gamma_{L}/dq^2$ defined with Eq.~(\ref{eq:longfcnc-1}), red dashes line: $d\Gamma_{T}/dq^2$ defined with Eq.~(\ref{eq:tranfcnc-1}).}
\label{fig:decaywidthbbtobFCNC22}
\end{figure}

\item  For the transition ${1}/{2}\to{3}/{2}$ with $\rm{V-A}$ current , the integrated partial decay widths, the relevant branching ratios and $\Gamma_{L}/\Gamma_{T}$s are shown in Tab.~\ref{Tab:branchingv23}.
The dependence of $q^2$ of the differential decay widths can be shown in Fig.~\ref{fig:decaywidthbbtob23}.

\begin{figure}
\includegraphics[width=0.9\columnwidth]{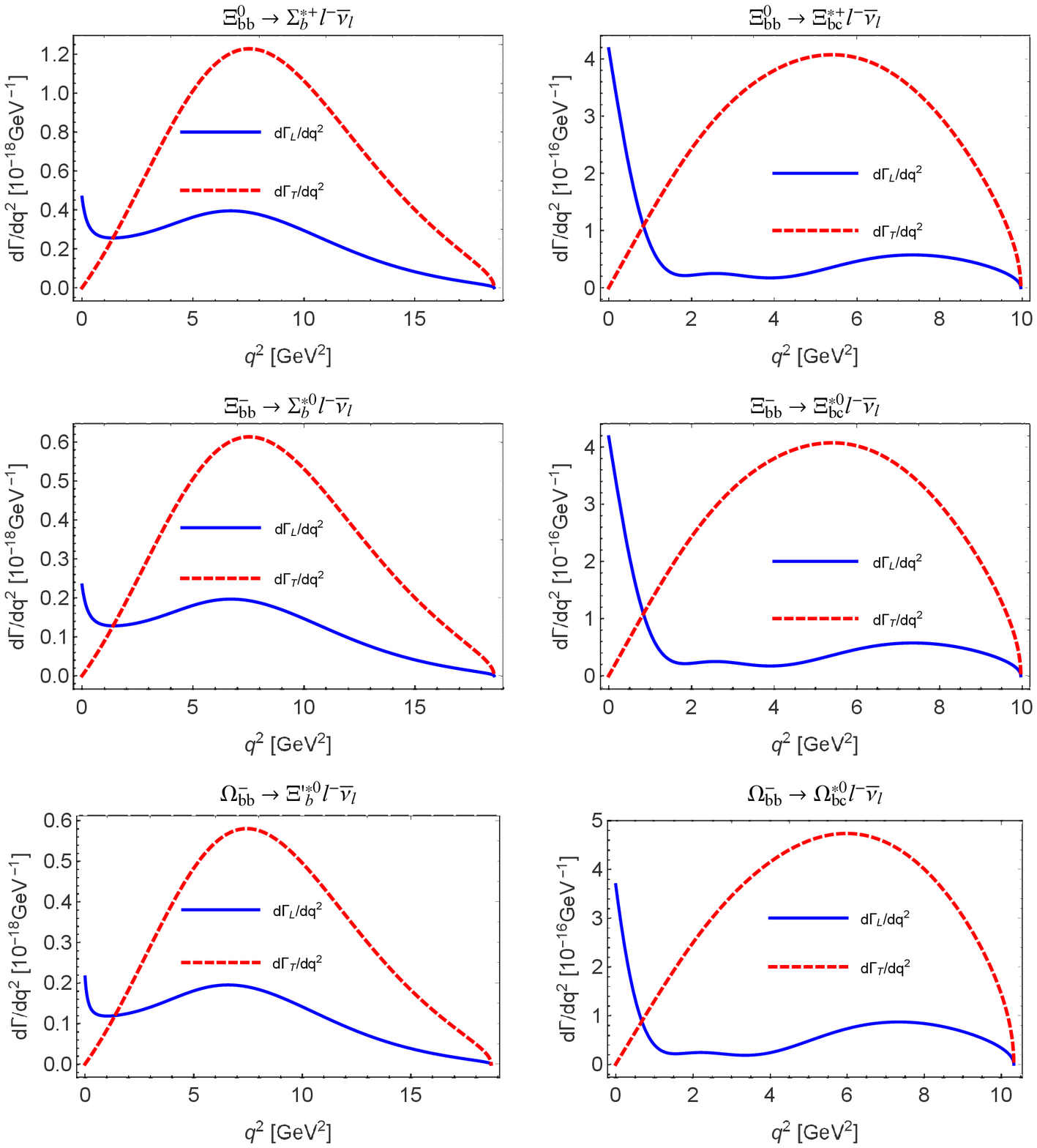}
\caption{The differential decay widths $d\Gamma_{L}/dq^2$ and $d\Gamma_{T}/dq^2$ for the processes ${\cal B}_{bb}\to{\cal B}_{b}^{*}({\cal B}_{bc}^{*})l^-\bar{\nu}_{l}$ dependence on $q^2$. Blue solid line: $d\Gamma_{L}/dq^2$ defined with Eq.~(\ref{eq:longi-2}), red dashes line: $d\Gamma_{T}/dq^2$ defined with Eq.~(\ref{eq:trans-2}).}
\label{fig:decaywidthbbtob23}
\end{figure}

\item For the transition ${1}/{2}\to{3}/{2}$ with FCNC, the integrated partial decay widths, the relevant branching ratios and $\Gamma_{L}/\Gamma_{T}$s are shown in Tab.~\ref{Tab:branching23fcnc}. The dependence of $q^2$ of the differential decay widths can be shown with Fig.~\ref{fig:decaywidthbbtobfcnc23}.
\begin{figure}
\includegraphics[width=0.9\columnwidth]{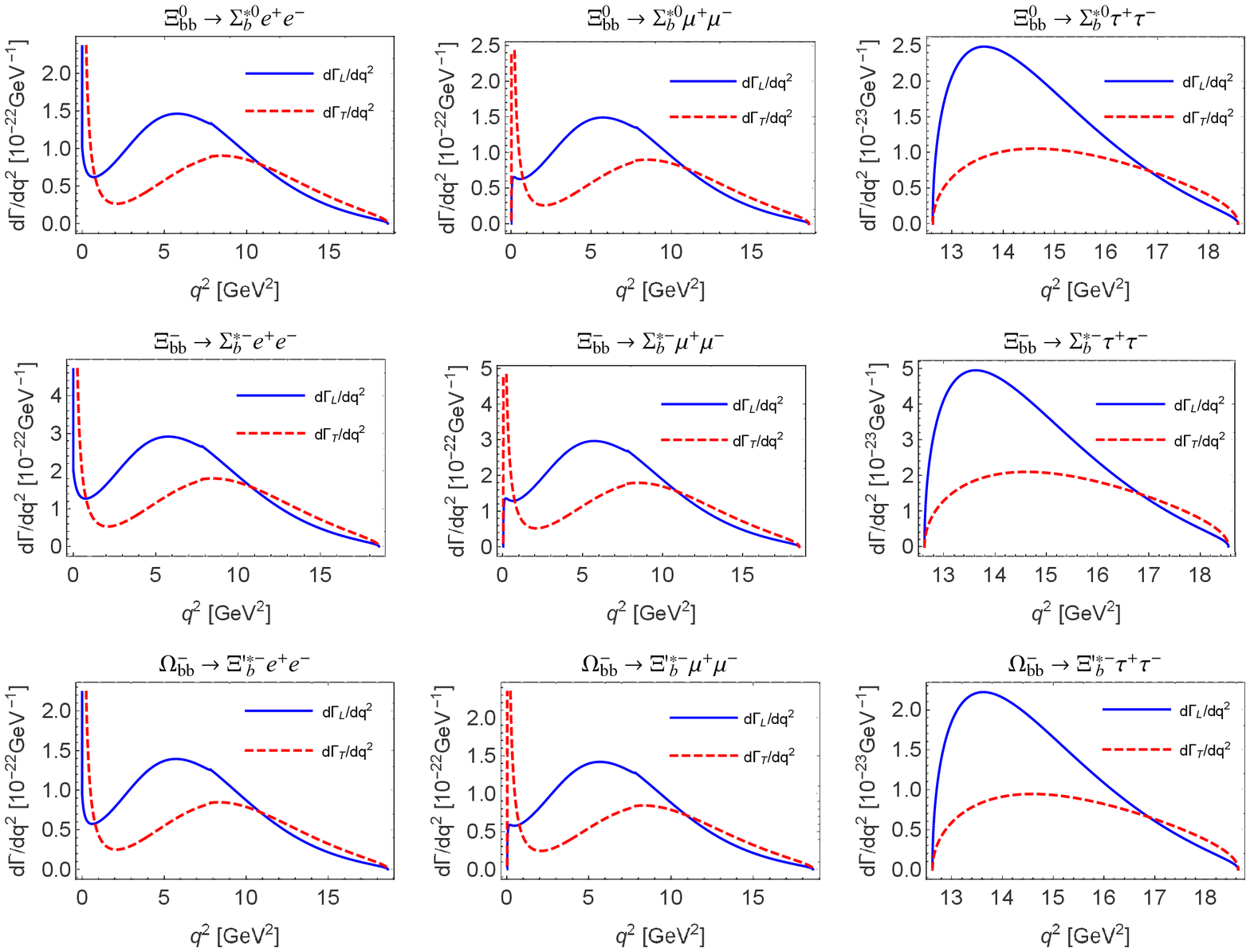}
\caption{The differential decay widths $d\Gamma_{L}/dq^2$ and $d\Gamma_{T}/dq^2$ for the processes ${\cal B}_{bb}\to{\cal B}_{b}^{*}l^+l^{-}$ dependence on $q^2$. Blue solid line: $d\Gamma_{L}/dq^2$ defined with Eq.~(\ref{eq:longfcnc-2}), red dashes line: $d\Gamma_{T}/dq^2$ defined with Eq.~(\ref{eq:tranfcnc-2}).}
\label{fig:decaywidthbbtobfcnc23}
\end{figure}
\end{itemize}
\begin{table}
\caption{The decay widths, branching ratios and $\Gamma_{L}/\Gamma_{T}$s for
the transition ${1}/{2}\to{1}/{2}$ with the charge current.}
\label{Tab:branchingv22}
\begin{tabular}{l|c|c|c|l|c|c|c}
\hline\hline
channels  & $\Gamma/\text{~GeV}$  & ${\cal B}$  & $\Gamma_{L}/\Gamma_{T}$  & channels  & $\Gamma/\text{~GeV}$  & ${\cal B}$  & $\Gamma_{L}/\Gamma_{T}$\tabularnewline
\hline
$\Xi_{cc}^{++}\to\Lambda_{c}^{+}l^{+}\nu_{l}$  & $7.97\times10^{-15}$  & $3.10\times10^{-3}$  & $2.42$ & $\Xi_{bb}^{0}\to\Sigma_{b}^{+}l^{-}\bar{\nu}_{l}$  & $1.06\times10^{-16}$  & $5.96\times10^{-5}$  & $1.27$\tabularnewline
$\Xi_{cc}^{++}\to\Sigma_{c}^{+}l^{+}\nu_{l}$  & $1.09\times10^{-14}$  & $4.25\times10^{-3}$  & $0.86$ & $\Xi_{bb}^{0}\to\Xi_{bc}^{+}l^{-}\bar{\nu}_{l}$  & $6.02\times10^{-14}$  & $3.38\times10^{-2}$  & $1.42$\tabularnewline
$\Xi_{cc}^{++}\to\Xi_{c}^{+}l^{+}\nu_{l}$  & $8.74\times10^{-14}$  & $3.40\times10^{-2}$  & $3.07$ & $\Xi_{bb}^{0}\to\Xi_{bc}^{\prime+}l^{-}\bar{\nu}_{l}$  & $3.21\times10^{-14}$  & $1.81\times10^{-2}$  & $0.84$\tabularnewline
$\Xi_{cc}^{++}\to\Xi_{c}^{\prime+}l^{+}\nu_{l}$  & $1.43\times10^{-13}$  & $5.57\times10^{-2}$  & $0.94$ & $\Xi_{bb}^{-}\to\Lambda_{b}^{0}l^{-}\bar{\nu}_{l}$  & $2.39\times10^{-17}$  & $1.35\times10^{-5}$  & $5.93$\tabularnewline
$\Xi_{cc}^{+}\to\Sigma_{c}^{0}l^{+}\nu_{l}$  & $2.17\times10^{-14}$  & $1.48\times10^{-3}$  & $0.86$ & $\Xi_{bb}^{-}\to\Sigma_{b}^{0}l^{-}\bar{\nu}_{l}$  & $5.29\times10^{-17}$  & $2.98\times10^{-5}$  & $1.27$\tabularnewline
$\Xi_{cc}^{+}\to\Xi_{c}^{0}l^{+}\nu_{l}$  & $8.63\times10^{-14}$  & $5.90\times10^{-3}$  & $3.10$ & $\Xi_{bb}^{-}\to\Xi_{bc}^{0}l^{-}\bar{\nu}_{l}$  & $6.02\times10^{-14}$  & $3.38\times10^{-2}$  & $1.42$\tabularnewline
$\Xi_{cc}^{+}\to\Xi_{c}^{\prime0}l^{+}\nu_{l}$  & $1.41\times10^{-13}$  & $9.67\times10^{-3}$  & $0.95$ & $\Xi_{bb}^{-}\to\Xi_{bc}^{\prime0}l^{-}\bar{\nu}_{l}$  & $3.21\times10^{-14}$  & $1.81\times10^{-2}$  & $0.84$\tabularnewline
$\Omega_{cc}^{+}\to\Xi_{c}^{0}l^{+}\nu_{l}$  & $5.87\times10^{-15}$  & $1.60\times10^{-3}$  & $2.94$ & $\Omega_{bb}^{-}\to\Xi_{b}^{0}l^{-}\bar{\nu}_{l}$  & $2.18\times10^{-17}$  & $2.65\times10^{-5}$  & $5.98$\tabularnewline
$\Omega_{cc}^{+}\to\Xi_{c}^{\prime0}l^{+}\nu_{l}$  & $1.03\times10^{-14}$  & $2.83\times10^{-3}$  & $0.87$ & $\Omega_{bb}^{-}\to\Xi_{b}^{\prime0}l^{-}\bar{\nu}_{l}$  & $4.87\times10^{-17}$  & $5.92\times10^{-5}$  & $1.28$\tabularnewline
$\Omega_{cc}^{+}\to\Omega_{c}^{0}l^{+}\nu_{l}$  & $2.80\times10^{-13}$  & $7.67\times10^{-2}$  & $0.94$ & $\Omega_{bb}^{-}\to\Omega_{bc}^{0}l^{-}\bar{\nu}_{l}$  & $5.24\times10^{-14}$  & $6.37\times10^{-2}$  & $1.64$\tabularnewline
 &  &  &  & $\Omega_{bb}^{-}\to\Omega_{bc}^{\prime0}l^{-}\bar{\nu}_{l}$  & $2.55\times10^{-14}$  & $3.11\times10^{-2}$  & $0.89$\tabularnewline
\hline
$\Xi_{bc}^{+}\to\Lambda_{b}^{0}l^{+}\nu_{l}$  & $4.62\times10^{-15}$  & $1.71\times10^{-3}$  & $2.13$  & $\Xi_{bc}^{+}\to\Sigma_{c}^{++}l^{-}\bar{\nu}_{l}$  & $8.00\times10^{-17}$  & $2.97\times10^{-5}$  & $1.13$\tabularnewline
$\Xi_{bc}^{+}\to\Sigma_{b}^{0}l^{+}\nu_{l}$  & $5.54\times10^{-15}$  & $2.06\times10^{-3}$  & $0.79$  & $\Xi_{bc}^{+}\to\Xi_{cc}^{++}l^{-}\bar{\nu}_{l}$  & $4.26\times10^{-14}$  & $1.58\times10^{-2}$  & $2.21$\tabularnewline
$\Xi_{bc}^{+}\to\Xi_{b}^{0}l^{+}\nu_{l}$  & $4.89\times10^{-14}$  & $1.81\times10^{-2}$  & $2.70$  & $\Xi_{bc}^{0}\to\Lambda_{c}^{+}l^{-}\bar{\nu}_{l}$  & $1.76\times10^{-17}$  & $2.48\times10^{-6}$  & $6.24$\tabularnewline
$\Xi_{bc}^{+}\to\Xi_{b}^{\prime0}l^{+}\nu_{l}$  & $6.73\times10^{-14}$  & $2.50\times10^{-2}$  & $0.89$  & $\Xi_{bc}^{0}\to\Sigma_{c}^{+}l^{-}\bar{\nu}_{l}$  & $4.00\times10^{-17}$  & $5.65\times10^{-6}$  & $1.13$\tabularnewline
$\Xi_{bc}^{0}\to\Sigma_{b}^{-}l^{+}\nu_{l}$  & $1.10\times10^{-14}$  & $1.55\times10^{-3}$  & $0.79$  & $\Xi_{bc}^{0}\to\Xi_{cc}^{+}l^{-}\bar{\nu}_{l}$  & $4.26\times10^{-14}$  & $6.01\times10^{-3}$  & $2.21$\tabularnewline
$\Xi_{bc}^{0}\to\Xi_{b}^{-}l^{+}\nu_{l}$  & $4.85\times10^{-14}$  & $6.85\times10^{-3}$  & $2.71$  & $\Omega_{bc}^{0}\to\Xi_{c}^{+}l^{-}\bar{\nu}_{l}$  & $1.40\times10^{-17}$  & $4.69\times10^{-6}$  & $6.21$\tabularnewline
$\Xi_{bc}^{0}\to\Xi_{b}^{\prime-}l^{+}\nu_{l}$  & $6.73\times10^{-14}$  & $9.51\times10^{-3}$  & $0.89$  & $\Omega_{bc}^{0}\to\Xi_{c}^{\prime+}l^{-}\bar{\nu}_{l}$  & $3.27\times10^{-17}$  & $1.09\times10^{-5}$  & $1.16$\tabularnewline
$\Omega_{bc}^{0}\to\Xi_{b}^{-}l^{+}\nu_{l}$  & $2.93\times10^{-15}$  & $9.81\times10^{-4}$  & $2.73$  & $\Omega_{bc}^{0}\to\Omega_{cc}^{+}l^{-}\bar{\nu}_{l}$  & $4.11\times10^{-14}$  & $1.37\times10^{-2}$  & $2.15$\tabularnewline
$\Omega_{bc}^{0}\to\Xi_{b}^{\prime-}l^{+}\nu_{l}$  & $3.96\times10^{-15}$  & $1.33\times10^{-3}$  & $0.90$  &  &  &  & \tabularnewline
$\Omega_{bc}^{0}\to\Omega_{b}^{-}l^{+}\nu_{l}$  & $1.01\times10^{-13}$  & $3.36\times10^{-2}$  & $1.03$  &  &  &  & \tabularnewline
\hline
$\Xi_{bc}^{\prime+}\to\Lambda_{b}^{0}l^{+}\nu_{l}$  & $6.24\times10^{-15}$  & $2.31\times10^{-3}$  & $0.74$  & $\Xi_{bc}^{\prime+}\to\Sigma_{c}^{++}l^{-}\bar{\nu}_{l}$  & $3.31\times10^{-17}$  & $1.23\times10^{-5}$  & $5.75$\tabularnewline
$\Xi_{bc}^{\prime+}\to\Sigma_{b}^{0}l^{+}\nu_{l}$  & $2.02\times10^{-15}$  & $7.50\times10^{-4}$  & $3.75$  & $\Xi_{bc}^{\prime+}\to\Xi_{cc}^{++}l^{-}\bar{\nu}_{l}$  & $1.86\times10^{-14}$  & $6.90\times10^{-3}$  & $0.95$\tabularnewline
$\Xi_{bc}^{\prime+}\to\Xi_{b}^{0}l^{+}\nu_{l}$  & $5.91\times10^{-14}$  & $2.19\times10^{-2}$  & $0.88$  & $\Xi_{bc}^{\prime0}\to\Lambda_{c}^{+}l^{-}\bar{\nu}_{l}$  & $1.38\times10^{-17}$  & $1.95\times10^{-6}$  & $1.21$\tabularnewline
$\Xi_{bc}^{\prime+}\to\Xi_{b}^{\prime0}l^{+}\nu_{l}$  & $2.65\times10^{-14}$  & $9.83\times10^{-3}$  & $4.33$  & $\Xi_{bc}^{\prime0}\to\Sigma_{c}^{+}l^{-}\bar{\nu}_{l}$  & $1.65\times10^{-17}$  & $2.34\times10^{-6}$  & $5.76$\tabularnewline
$\Xi_{bc}^{\prime0}\to\Sigma_{b}^{-}l^{+}\nu_{l}$  & $4.01\times10^{-15}$  & $5.67\times10^{-4}$  & $3.78$  & $\Xi_{bc}^{\prime0}\to\Xi_{cc}^{+}l^{-}\bar{\nu}_{l}$  & $1.86\times10^{-14}$  & $2.63\times10^{-3}$  & $0.95$\tabularnewline
$\Xi_{bc}^{\prime0}\to\Xi_{b}^{-}l^{+}\nu_{l}$  & $5.84\times10^{-14}$  & $8.26\times10^{-3}$  & $0.88$  & $\Omega_{bc}^{\prime0}\to\Xi_{c}^{+}l^{-}\bar{\nu}_{l}$  & $1.14\times10^{-17}$  & $3.81\times10^{-6}$  & $1.27$\tabularnewline
$\Xi_{bc}^{\prime0}\to\Xi_{b}^{\prime-}l^{+}\nu_{l}$  & $2.65\times10^{-14}$  & $3.75\times10^{-3}$  & $4.33$  & $\Omega_{bc}^{\prime0}\to\Xi_{c}^{\prime+}l^{-}\bar{\nu}_{l}$  & $1.35\times10^{-17}$  & $4.52\times10^{-6}$  & $5.85$\tabularnewline
$\Omega_{bc}^{\prime0}\to\Xi_{b}^{-}l^{+}\nu_{l}$  & $3.38\times10^{-15}$  & $1.13\times10^{-3}$  & $0.92$  & $\Omega_{bc}^{\prime0}\to\Omega_{cc}^{+}l^{-}\bar{\nu}_{l}$  & $1.85\times10^{-14}$  & $6.18\times10^{-3}$  & $0.95$\tabularnewline
$\Omega_{bc}^{\prime0}\to\Xi_{b}^{\prime-}l^{+}\nu_{l}$  & $1.62\times10^{-15}$  & $5.42\times10^{-4}$  & $4.25$  &  &  &  & \tabularnewline
$\Omega_{bc}^{\prime0}\to\Omega_{b}^{-}l^{+}\nu_{l}$  & $4.40\times10^{-14}$  & $1.47\times10^{-2}$  & $4.76$  &  &  &  & \tabularnewline
\hline\hline
\end{tabular}%
\end{table}
\begin{table}
\caption{The decay widths, branching ratios and $\Gamma_{L}/\Gamma_{T}$s for the  transition ${1}/{2}\to{1}/{2}$ with FCNC.}\label{Tab:branchingratio22}
\begin{tabular}{l|c|c|c|l|c|c|c}
\hline \hline
channels  & $\Gamma/\text{~GeV}$  & ${\cal B}$  & $\Gamma_{L}/\Gamma_{T}$ & channels & $\Gamma/\text{~GeV}$  & ${\cal B}$  & $\Gamma_{L}/\Gamma_{T}$\tabularnewline
\hline
$\Xi_{bb}^{0}\to\Lambda_{b}^{0}e^{+}e^{-}$  & $4.15\times10^{-21}$  & $2.33\times10^{-9}$  & $5.28$ & $\Xi_{bb}^{0}\to\Xi_{b}^{0}e^{+}e^{-}$  & $1.62\times10^{-19}$  & $9.13\times10^{-8}$  & $4.70$\tabularnewline
$\Xi_{bb}^{0}\to\Sigma_{b}^{0}e^{+}e^{-}$  & $1.05\times10^{-20}$  & $5.91\times10^{-9}$  & $0.90$ & $\Xi_{bb}^{0}\to\Xi_{b}^{\prime0}e^{+}e^{-}$  & $4.32\times10^{-19}$  & $2.43\times10^{-7}$  & $0.85$\tabularnewline
$\Xi_{bb}^{-}\to\Sigma_{b}^{-}e^{+}e^{-}$  & $2.10\times10^{-20}$  & $1.18\times10^{-8}$  & $0.90$ & $\Xi_{bb}^{-}\to\Xi_{b}^{-}e^{+}e^{-}$  & $1.62\times10^{-19}$  & $9.12\times10^{-8}$  & $4.69$\tabularnewline
$\Omega_{bb}^{-}\to\Xi_{b}^{-}e^{+}e^{-}$  & $3.79\times10^{-21}$  & $4.61\times10^{-9}$  & $5.24$ & $\Xi_{bb}^{-}\to\Xi_{b}^{\prime-}e^{+}e^{-}$  & $4.32\times10^{-19}$  & $2.43\times10^{-7}$  & $0.85$\tabularnewline
$\Omega_{bb}^{-}\to\Xi_{b}^{\prime-}e^{+}e^{-}$  & $9.71\times10^{-21}$  & $1.18\times10^{-8}$  & $0.90$ & $\Omega_{bb}^{-}\to\Omega_{b}^{-}e^{+}e^{-}$  & $8.05\times10^{-19}$  & $9.79\times10^{-7}$  & $0.85$\tabularnewline
$\Xi_{bb}^{0}\to\Lambda_{b}^{0}\mu^{+}\mu^{-}$  & $3.98\times10^{-21}$  & $2.24\times10^{-9}$  & $6.88$ & $\Xi_{bb}^{0}\to\Xi_{b}^{0}\mu^{+}\mu^{-}$  & $1.56\times10^{-19}$  & $8.75\times10^{-8}$  & $5.99$\tabularnewline
$\Xi_{bb}^{0}\to\Sigma_{b}^{0}\mu^{+}\mu^{-}$  & $8.69\times10^{-21}$  & $4.89\times10^{-9}$  & $1.33$ & $\Xi_{bb}^{0}\to\Xi_{b}^{\prime0}\mu^{+}\mu^{-}$  & $3.61\times10^{-19}$  & $2.03\times10^{-7}$  & $1.20$\tabularnewline
$\Xi_{bb}^{-}\to\Sigma_{b}^{-}\mu^{+}\mu^{-}$  & $1.74\times10^{-20}$  & $9.77\times10^{-9}$  & $1.33$ & $\Xi_{bb}^{-}\to\Xi_{b}^{-}\mu^{+}\mu^{-}$  & $1.56\times10^{-19}$  & $8.75\times10^{-8}$  & $5.99$\tabularnewline
$\Omega_{bb}^{-}\to\Xi_{b}^{-}\mu^{+}\mu^{-}$  & $3.63\times10^{-21}$  & $4.41\times10^{-9}$  & $6.90$ & $\Xi_{bb}^{-}\to\Xi_{b}^{\prime-}\mu^{+}\mu^{-}$  & $3.61\times10^{-19}$  & $2.03\times10^{-7}$  & $1.20$\tabularnewline
$\Omega_{bb}^{-}\to\Xi_{b}^{\prime-}\mu^{+}\mu^{-}$  & $7.98\times10^{-21}$  & $9.71\times10^{-9}$  & $1.35$ & $\Omega_{bb}^{-}\to\Omega_{b}^{-}\mu^{+}\mu^{-}$  & $6.70\times10^{-19}$  & $8.14\times10^{-7}$  & $1.21$\tabularnewline
$\Xi_{bb}^{0}\to\Lambda_{b}^{0}\tau^{+}\tau^{-}$  & $1.51\times10^{-22}$  & $8.49\times10^{-11}$  & $5.83$ & $\Xi_{bb}^{0}\to\Xi_{b}^{0}\tau^{+}\tau^{-}$  & $6.68\times10^{-21}$  & $3.76\times10^{-9}$  & $5.71$\tabularnewline
$\Xi_{bb}^{0}\to\Sigma_{b}^{0}\tau^{+}\tau^{-}$  & $3.39\times10^{-22}$  & $1.91\times10^{-10}$  & $1.16$ & $\Xi_{bb}^{0}\to\Xi_{b}^{\prime0}\tau^{+}\tau^{-}$  & $1.54\times10^{-20}$  & $8.65\times10^{-9}$  & $1.05$\tabularnewline
$\Xi_{bb}^{-}\to\Sigma_{b}^{-}\tau^{+}\tau^{-}$  & $6.76\times10^{-22}$  & $3.80\times10^{-10}$  & $1.16$ & $\Xi_{bb}^{-}\to\Xi_{b}^{-}\tau^{+}\tau^{-}$  & $6.65\times10^{-21}$  & $3.74\times10^{-9}$  & $5.69$\tabularnewline
$\Omega_{bb}^{-}\to\Xi_{b}^{-}\tau^{+}\tau^{-}$  & $1.22\times10^{-22}$  & $1.49\times10^{-10}$  & $5.52$ & $\Xi_{bb}^{-}\to\Xi_{b}^{\prime-}\tau^{+}\tau^{-}$  & $1.54\times10^{-20}$  & $8.65\times10^{-9}$  & $1.05$\tabularnewline
$\Omega_{bb}^{-}\to\Xi_{b}^{\prime-}\tau^{+}\tau^{-}$  & $2.96\times10^{-22}$  & $3.60\times10^{-10}$  & $1.17$ & $\Omega_{bb}^{-}\to\Omega_{b}^{-}\tau^{+}\tau^{-}$  & $2.78\times10^{-20}$  & $3.37\times10^{-8}$  & $1.08$\tabularnewline
\hline
$\Xi_{bc}^{+}\to\Lambda_{c}^{+}e^{+}e^{-}$  & $3.71\times10^{-21}$  & $1.37\times10^{-9}$  & $5.29$ & $\Xi_{bc}^{+}\to\Xi_{c}^{+}e^{+}e^{-}$  & $1.19\times10^{-19}$  & $4.43\times10^{-8}$  & $4.90$\tabularnewline
$\Xi_{bc}^{+}\to\Sigma_{c}^{+}e^{+}e^{-}$  & $9.04\times10^{-21}$  & $3.35\times10^{-9}$  & $0.81$ & $\Xi_{bc}^{+}\to\Xi_{c}^{\prime+}e^{+}e^{-}$  & $2.97\times10^{-19}$  & $1.10\times10^{-7}$  & $0.77$\tabularnewline
$\Xi_{bc}^{0}\to\Sigma_{c}^{0}e^{+}e^{-}$  & $1.81\times10^{-20}$  & $2.56\times10^{-9}$  & $0.81$ & $\Xi_{bc}^{0}\to\Xi_{c}^{0}e^{+}e^{-}$  & $1.19\times10^{-19}$  & $1.69\times10^{-8}$  & $4.90$\tabularnewline
$\Omega_{bc}^{0}\to\Xi_{c}^{0}e^{+}e^{-}$  & $3.03\times10^{-21}$  & $1.01\times10^{-9}$  & $5.14$ & $\Xi_{bc}^{0}\to\Xi_{c}^{\prime0}e^{+}e^{-}$  & $2.97\times10^{-19}$  & $4.20\times10^{-8}$  & $0.77$\tabularnewline
$\Omega_{bc}^{0}\to\Xi_{c}^{\prime0}e^{+}e^{-}$  & $7.69\times10^{-21}$  & $2.57\times10^{-9}$  & $0.82$ & $\Omega_{bc}^{0}\to\Omega_{c}^{0}e^{+}e^{-}$  & $5.17\times10^{-19}$  & $1.73\times10^{-7}$  & $0.78$\tabularnewline
$\Xi_{bc}^{+}\to\Lambda_{c}^{+}\mu^{+}\mu^{-}$  & $3.54\times10^{-21}$  & $1.31\times10^{-9}$  & $7.20$ & $\Xi_{bc}^{+}\to\Xi_{c}^{+}\mu^{+}\mu^{-}$  & $1.13\times10^{-19}$  & $4.18\times10^{-8}$  & $7.17$\tabularnewline
$\Xi_{bc}^{+}\to\Sigma_{c}^{+}\mu^{+}\mu^{-}$  & $7.66\times10^{-21}$  & $2.84\times10^{-9}$  & $1.11$ & $\Xi_{bc}^{+}\to\Xi_{c}^{\prime+}\mu^{+}\mu^{-}$  & $2.41\times10^{-19}$  & $8.93\times10^{-8}$  & $1.15$\tabularnewline
$\Xi_{bc}^{0}\to\Sigma_{c}^{0}\mu^{+}\mu^{-}$  & $1.53\times10^{-20}$  & $2.17\times10^{-9}$  & $1.11$ & $\Xi_{bc}^{0}\to\Xi_{c}^{0}\mu^{+}\mu^{-}$  & $1.13\times10^{-19}$  & $1.59\times10^{-8}$  & $7.16$\tabularnewline
$\Omega_{bc}^{0}\to\Xi_{c}^{0}\mu^{+}\mu^{-}$  & $2.89\times10^{-21}$  & $9.68\times10^{-10}$  & $6.95$ & $\Xi_{bc}^{0}\to\Xi_{c}^{\prime0}\mu^{+}\mu^{-}$  & $2.41\times10^{-19}$  & $3.41\times10^{-8}$  & $1.15$\tabularnewline
$\Omega_{bc}^{0}\to\Xi_{c}^{\prime0}\mu^{+}\mu^{-}$  & $6.52\times10^{-21}$  & $2.18\times10^{-9}$  & $1.13$ & $\Omega_{bc}^{0}\to\Omega_{c}^{0}\mu^{+}\mu^{-}$  & $4.19\times10^{-19}$  & $1.40\times10^{-7}$  & $1.17$\tabularnewline
$\Xi_{bc}^{+}\to\Lambda_{c}^{+}\tau^{+}\tau^{-}$  & $3.28\times10^{-22}$  & $1.22\times10^{-10}$  & $12.5$ & $\Xi_{bc}^{+}\to\Xi_{c}^{+}\tau^{+}\tau^{-}$  & $8.64\times10^{-21}$  & $3.21\times10^{-9}$  & $11.9$\tabularnewline
$\Xi_{bc}^{+}\to\Sigma_{c}^{+}\tau^{+}\tau^{-}$  & $6.92\times10^{-22}$  & $2.57\times10^{-10}$  & $1.68$ & $\Xi_{bc}^{+}\to\Xi_{c}^{\prime+}\tau^{+}\tau^{-}$  & $1.73\times10^{-20}$  & $6.41\times10^{-9}$  & $1.72$\tabularnewline
$\Xi_{bc}^{0}\to\Sigma_{c}^{0}\tau^{+}\tau^{-}$  & $1.39\times10^{-21}$  & $1.96\times10^{-10}$  & $1.67$ & $\Xi_{bc}^{0}\to\Xi_{c}^{0}\tau^{+}\tau^{-}$  & $8.60\times10^{-21}$  & $1.22\times10^{-9}$  & $11.8$\tabularnewline
$\Omega_{bc}^{0}\to\Xi_{c}^{0}\tau^{+}\tau^{-}$  & $2.12\times10^{-22}$  & $7.09\times10^{-11}$  & $9.20$ & $\Xi_{bc}^{0}\to\Xi_{c}^{\prime0}\tau^{+}\tau^{-}$  & $1.73\times10^{-20}$  & $2.44\times10^{-9}$  & $1.71$\tabularnewline
$\Omega_{bc}^{0}\to\Xi_{c}^{\prime0}\tau^{+}\tau^{-}$  & $5.17\times10^{-22}$  & $1.73\times10^{-10}$  & $1.55$ & $\Omega_{bc}^{0}\to\Omega_{c}^{0}\tau^{+}\tau^{-}$  & $2.62\times10^{-20}$  & $8.77\times10^{-9}$  & $1.60$\tabularnewline
\hline
$\Xi_{bc}^{\prime+}\to\Lambda_{c}^{+}e^{+}e^{-}$  & $3.23\times10^{-21}$  & $1.20\times10^{-9}$  & $0.84$ & $\Xi_{bc}^{\prime+}\to\Xi_{c}^{+}e^{+}e^{-}$  & $1.08\times10^{-19}$  & $4.02\times10^{-8}$  & $0.82$\tabularnewline
$\Xi_{bc}^{\prime+}\to\Sigma_{c}^{+}e^{+}e^{-}$  & $3.50\times10^{-21}$  & $1.30\times10^{-9}$  & $4.76$ & $\Xi_{bc}^{\prime+}\to\Xi_{c}^{\prime+}e^{+}e^{-}$  & $1.15\times10^{-19}$  & $4.25\times10^{-8}$  & $4.60$\tabularnewline
$\Xi_{bc}^{\prime0}\to\Sigma_{c}^{0}e^{+}e^{-}$  & $7.01\times10^{-21}$  & $9.90\times10^{-10}$  & $4.76$ & $\Xi_{bc}^{\prime0}\to\Xi_{c}^{0}e^{+}e^{-}$  & $1.08\times10^{-19}$  & $1.53\times10^{-8}$  & $0.82$\tabularnewline
$\Omega_{bc}^{\prime0}\to\Xi_{c}^{0}e^{+}e^{-}$  & $2.78\times10^{-21}$  & $9.30\times10^{-10}$  & $0.87$ & $\Xi_{bc}^{\prime0}\to\Xi_{c}^{\prime0}e^{+}e^{-}$  & $1.15\times10^{-19}$  & $1.62\times10^{-8}$  & $4.59$\tabularnewline
$\Omega_{bc}^{\prime0}\to\Xi_{c}^{\prime0}e^{+}e^{-}$  & $2.93\times10^{-21}$  & $9.80\times10^{-10}$  & $4.75$ & $\Omega_{bc}^{\prime0}\to\Omega_{c}^{0}e^{+}e^{-}$  & $1.97\times10^{-19}$  & $6.58\times10^{-8}$  & $4.60$\tabularnewline
$\Xi_{bc}^{\prime+}\to\Lambda_{c}^{+}\mu^{+}\mu^{-}$  & $2.70\times10^{-21}$  & $1.00\times10^{-9}$  & $1.20$ & $\Xi_{bc}^{\prime+}\to\Xi_{c}^{+}\mu^{+}\mu^{-}$  & $8.72\times10^{-20}$  & $3.23\times10^{-8}$  & $1.26$\tabularnewline
$\Xi_{bc}^{\prime+}\to\Sigma_{c}^{+}\mu^{+}\mu^{-}$  & $3.34\times10^{-21}$  & $1.24\times10^{-9}$  & $6.40$ & $\Xi_{bc}^{\prime+}\to\Xi_{c}^{\prime+}\mu^{+}\mu^{-}$  & $1.08\times10^{-19}$  & $4.00\times10^{-8}$  & $6.68$\tabularnewline
$\Xi_{bc}^{\prime0}\to\Sigma_{c}^{0}\mu^{+}\mu^{-}$  & $6.68\times10^{-21}$  & $9.44\times10^{-10}$  & $6.40$ & $\Xi_{bc}^{\prime0}\to\Xi_{c}^{0}\mu^{+}\mu^{-}$  & $8.72\times10^{-20}$  & $1.23\times10^{-8}$  & $1.26$\tabularnewline
$\Omega_{bc}^{\prime0}\to\Xi_{c}^{0}\mu^{+}\mu^{-}$  & $2.32\times10^{-21}$  & $7.77\times10^{-10}$  & $1.25$ & $\Xi_{bc}^{\prime0}\to\Xi_{c}^{\prime0}\mu^{+}\mu^{-}$  & $1.08\times10^{-19}$  & $1.52\times10^{-8}$  & $6.67$\tabularnewline
$\Omega_{bc}^{\prime0}\to\Xi_{c}^{\prime0}\mu^{+}\mu^{-}$  & $2.79\times10^{-21}$  & $9.34\times10^{-10}$  & $6.36$ & $\Omega_{bc}^{\prime0}\to\Omega_{c}^{0}\mu^{+}\mu^{-}$  & $1.85\times10^{-19}$  & $6.19\times10^{-8}$  & $6.68$\tabularnewline
$\Xi_{bc}^{\prime+}\to\Lambda_{c}^{+}\tau^{+}\tau^{-}$  & $1.60\times10^{-22}$  & $5.93\times10^{-11}$  & $0.90$ & $\Xi_{bc}^{\prime+}\to\Xi_{c}^{+}\tau^{+}\tau^{-}$  & $4.26\times10^{-21}$  & $1.58\times10^{-9}$  & $0.91$\tabularnewline
$\Xi_{bc}^{\prime+}\to\Sigma_{c}^{+}\tau^{+}\tau^{-}$  & $2.70\times10^{-22}$  & $1.00\times10^{-10}$  & $8.06$ & $\Xi_{bc}^{\prime+}\to\Xi_{c}^{\prime+}\tau^{+}\tau^{-}$  & $7.27\times10^{-21}$  & $2.70\times10^{-9}$  & $8.91$\tabularnewline
$\Xi_{bc}^{\prime0}\to\Sigma_{c}^{0}\tau^{+}\tau^{-}$  & $5.40\times10^{-22}$  & $7.63\times10^{-11}$  & $8.04$ & $\Xi_{bc}^{\prime0}\to\Xi_{c}^{0}\tau^{+}\tau^{-}$  & $4.26\times10^{-21}$  & $6.02\times10^{-10}$  & $0.91$\tabularnewline
$\Omega_{bc}^{\prime0}\to\Xi_{c}^{0}\tau^{+}\tau^{-}$  & $1.27\times10^{-22}$  & $4.24\times10^{-11}$  & $0.86$ & $\Xi_{bc}^{\prime0}\to\Xi_{c}^{\prime0}\tau^{+}\tau^{-}$  & $7.25\times10^{-21}$  & $1.02\times10^{-9}$  & $8.86$\tabularnewline
$\Omega_{bc}^{\prime0}\to\Xi_{c}^{\prime0}\tau^{+}\tau^{-}$  & $1.86\times10^{-22}$  & $6.21\times10^{-11}$  & $6.86$ & $\Omega_{bc}^{\prime0}\to\Omega_{c}^{0}\tau^{+}\tau^{-}$  & $1.02\times10^{-20}$  & $3.41\times10^{-9}$  & $7.60$\tabularnewline
\hline \hline
\end{tabular}%
\end{table}
\begin{table}
\caption{The decay widths, branching ratios and $\Gamma_{L}/\Gamma_{T}$s for the  transition ${1}/{2}\to{3}/{2}$ with the charge current.}
\label{Tab:branchingv23}
\begin{tabular}{l|c|c|c|l|c|c|c}
\hline\hline
channels  & $\Gamma/\text{~GeV}$  & ${\cal B}$  & $\Gamma_{L}/\Gamma_{T}$  & channels  & $\Gamma/\text{~GeV}$  & ${\cal B}$  & $\Gamma_{L}/\Gamma_{T}$\tabularnewline
\hline
$\Xi_{cc}^{++}\to\Sigma_{c}^{*+}l^{+}\nu_{l}$  & $1.43\times10^{-15}$  & $5.55\times10^{-4}$  & $0.92$ & $\Xi_{bb}^{0}\to\Sigma_{b}^{*+}l^{-}\bar{\nu}_{l}$  & $2.33\times10^{-17}$  & $1.31\times10^{-5}$  & $0.94$\tabularnewline
$\Xi_{cc}^{+}\to\Sigma_{c}^{*0}l^{+}\nu_{l}$  & $2.85\times10^{-15}$  & $1.95\times10^{-4}$  & $0.92$ & $\Xi_{bb}^{-}\to\Sigma_{b}^{*0}l^{-}\bar{\nu}_{l}$  & $1.16\times10^{-17}$  & $6.52\times10^{-6}$  & $0.94$\tabularnewline
$\Omega_{cc}^{+}\to\Xi_{c}^{\prime*0}l^{+}\nu_{l}$  & $1.35\times10^{-15}$  & $3.69\times10^{-4}$  & $0.93$ & $\Omega_{bb}^{-}\to\Xi_{b}^{\prime*0}l^{-}\bar{\nu}_{l}$  & $1.11\times10^{-17}$  & $1.35\times10^{-5}$  & $0.97$\tabularnewline
$\Xi_{cc}^{++}\to\Xi_{c}^{\prime*+}l^{+}\nu_{l}$  & $1.74\times10^{-14}$  & $6.76\times10^{-3}$  & $1.08$ & $\Xi_{bb}^{0}\to\Xi_{bc}^{*+}l^{-}\bar{\nu}_{l}$  & $3.68\times10^{-15}$  & $2.07\times10^{-3}$  & $0.42$\tabularnewline
$\Xi_{cc}^{+}\to\Xi_{c}^{\prime*0}l^{+}\nu_{l}$  & $1.74\times10^{-14}$  & $1.19\times10^{-3}$  & $1.08$ & $\Xi_{bb}^{-}\to\Xi_{bc}^{*0}l^{-}\bar{\nu}_{l}$  & $3.68\times10^{-15}$  & $2.07\times10^{-3}$  & $0.42$\tabularnewline
$\Omega_{cc}^{+}\to\Omega_{c}^{*0}l^{+}\nu_{l}$  & $3.45\times10^{-14}$  & $9.45\times10^{-3}$  & $1.07$ & $\Omega_{bb}^{-}\to\Omega_{bc}^{*0}l^{-}\bar{\nu}_{l}$  & $4.57\times10^{-15}$  & $5.56\times10^{-3}$  & $0.45$\tabularnewline
\hline
$\Xi_{bc}^{+}\to\Sigma_{b}^{*0}l^{+}\nu_{l}$  & $1.16\times10^{-15}$  & $4.31\times10^{-4}$  & $0.69$ & $\Xi_{bc}^{+}\to\Sigma_{c}^{*++}l^{-}\bar{\nu}_{l}$  & $3.55\times10^{-17}$  & $1.31\times10^{-5}$  & $0.89$\tabularnewline
$\Xi_{bc}^{0}\to\Sigma_{b}^{*-}l^{+}\nu_{l}$  & $2.29\times10^{-15}$  & $3.24\times10^{-4}$  & $0.69$ & $\Xi_{bc}^{0}\to\Sigma_{c}^{*+}l^{-}\bar{\nu}_{l}$  & $1.77\times10^{-17}$  & $2.51\times10^{-6}$  & $0.89$\tabularnewline
$\Omega_{bc}^{0}\to\Xi_{b}^{\prime*-}l^{+}\nu_{l}$  & $7.38\times10^{-16}$  & $2.47\times10^{-4}$  & $0.81$ & $\Omega_{bc}^{0}\to\Xi_{c}^{\prime*+}l^{-}\bar{\nu}_{l}$  & $1.37\times10^{-17}$  & $4.59\times10^{-6}$  & $0.95$\tabularnewline
$\Xi_{bc}^{+}\to\Xi_{b}^{\prime*0}l^{+}\nu_{l}$  & $1.36\times10^{-14}$  & $5.04\times10^{-3}$  & $0.78$ & $\Xi_{bc}^{+}\to\Xi_{cc}^{*++}l^{-}\bar{\nu}_{l}$  & $1.06\times10^{-14}$  & $3.92\times10^{-3}$  & $1.46$\tabularnewline
$\Xi_{bc}^{0}\to\Xi_{b}^{\prime*-}l^{+}\nu_{l}$  & $1.30\times10^{-14}$  & $1.84\times10^{-3}$  & $0.79$ & $\Xi_{bc}^{0}\to\Xi_{cc}^{*+}l^{-}\bar{\nu}_{l}$  & $1.06\times10^{-14}$  & $1.49\times10^{-3}$  & $1.46$\tabularnewline
$\Omega_{bc}^{0}\to\Omega_{b}^{*-}l^{+}\nu_{l}$  & $1.50\times10^{-14}$  & $5.03\times10^{-3}$  & $1.00$ & $\Omega_{bc}^{0}\to\Omega_{cc}^{*+}l^{-}\bar{\nu}_{l}$  & $7.31\times10^{-15}$  & $2.44\times10^{-3}$  & $1.21$\tabularnewline
\hline
$\Xi_{bc}^{\prime+}\to\Sigma_{b}^{*0}l^{+}\nu_{l}$  & $3.48\times10^{-15}$  & $1.29\times10^{-3}$  & $0.69$ & $\Xi_{bc}^{\prime+}\to\Sigma_{c}^{*++}l^{-}\bar{\nu}_{l}$  & $1.06\times10^{-16}$  & $3.94\times10^{-5}$  & $0.89$\tabularnewline
$\Xi_{bc}^{\prime0}\to\Sigma_{b}^{*-}l^{+}\nu_{l}$  & $6.87\times10^{-15}$  & $9.71\times10^{-4}$  & $0.69$ & $\Xi_{bc}^{\prime0}\to\Sigma_{c}^{*+}l^{-}\bar{\nu}_{l}$  & $5.32\times10^{-17}$  & $7.52\times10^{-6}$  & $0.89$\tabularnewline
$\Omega_{bc}^{\prime0}\to\Xi_{b}^{\prime*-}l^{+}\nu_{l}$  & $2.21\times10^{-15}$  & $7.40\times10^{-4}$  & $0.81$ & $\Omega_{bc}^{\prime0}\to\Xi_{c}^{\prime*+}l^{-}\bar{\nu}_{l}$  & $4.12\times10^{-17}$  & $1.38\times10^{-5}$  & $0.95$\tabularnewline
$\Xi_{bc}^{\prime+}\to\Xi_{b}^{\prime*0}l^{+}\nu_{l}$  & $4.08\times10^{-14}$  & $1.51\times10^{-2}$  & $0.78$ & $\Xi_{bc}^{\prime+}\to\Xi_{cc}^{*++}l^{-}\bar{\nu}_{l}$  & $3.17\times10^{-14}$  & $1.17\times10^{-2}$  & $1.46$\tabularnewline
$\Xi_{bc}^{\prime0}\to\Xi_{b}^{\prime*-}l^{+}\nu_{l}$  & $3.90\times10^{-14}$  & $5.51\times10^{-3}$  & $0.79$ & $\Xi_{bc}^{\prime0}\to\Xi_{cc}^{*+}l^{-}\bar{\nu}_{l}$  & $3.17\times10^{-14}$  & $4.48\times10^{-3}$  & $1.46$\tabularnewline
$\Omega_{bc}^{\prime0}\to\Omega_{b}^{*-}l^{+}\nu_{l}$  & $4.51\times10^{-14}$  & $1.51\times10^{-2}$  & $1.00$ & $\Omega_{bc}^{\prime0}\to\Omega_{cc}^{*+}l^{-}\bar{\nu}_{l}$  & $2.19\times10^{-14}$  & $7.33\times10^{-3}$  & $1.21$\tabularnewline
\hline\hline
\end{tabular}
\end{table}

\begin{table}
\caption{The decay widths, branching ratios and $\Gamma_{L}/\Gamma_{T}$s for
the transition ${1}/{2}\to{3}/{2}$ with FCNC.}
\label{Tab:branching23fcnc} %
\begin{tabular}{l|c|c|c|l|c|c|c}
\hline \hline
channels  & $\Gamma/\text{~GeV}$  & ${\cal B}$  & $\Gamma_{L}/\Gamma_{T}$ & channels  & $\Gamma/\text{~GeV}$  & ${\cal B}$  & $\Gamma_{L}/\Gamma_{T}$\tabularnewline
\hline
$\Xi_{bb}^{0}\to\Sigma_{b}^{*0}e^{+}e^{-}$  & $3.27\times10^{-21}$  & $1.84\times10^{-9}$  & $0.80$ & $\Xi_{bb}^{0}\to\Xi_{b}^{\prime*0}e^{+}e^{-}$  & $1.45\times10^{-19}$  & $8.15\times10^{-8}$  & $0.75$\tabularnewline
$\Xi_{bb}^{-}\to\Sigma_{b}^{*-}e^{+}e^{-}$  & $6.52\times10^{-21}$  & $3.67\times10^{-9}$  & $0.80$ & $\Xi_{bb}^{-}\to\Xi_{b}^{\prime*-}e^{+}e^{-}$  & $1.43\times10^{-19}$  & $8.05\times10^{-8}$  & $0.74$\tabularnewline
$\Omega_{bb}^{-}\to\Xi_{b}^{\prime*-}e^{+}e^{-}$  & $3.06\times10^{-21}$  & $3.72\times10^{-9}$  & $0.81$ & $\Omega_{bb}^{-}\to\Omega_{b}^{*-}e^{+}e^{-}$  & $2.71\times10^{-19}$  & $3.30\times10^{-7}$  & $0.74$\tabularnewline
$\Xi_{bb}^{0}\to\Sigma_{b}^{*0}\mu^{+}\mu^{-}$  & $2.56\times10^{-21}$  & $1.44\times10^{-9}$  & $1.35$ & $\Xi_{bb}^{0}\to\Xi_{b}^{\prime*0}\mu^{+}\mu^{-}$  & $1.19\times10^{-19}$  & $6.69\times10^{-8}$  & $1.12$\tabularnewline
$\Xi_{bb}^{-}\to\Sigma_{b}^{*-}\mu^{+}\mu^{-}$  & $5.11\times10^{-21}$  & $2.87\times10^{-9}$  & $1.35$ & $\Xi_{bb}^{-}\to\Xi_{b}^{\prime*-}\mu^{+}\mu^{-}$  & $1.17\times10^{-19}$  & $6.58\times10^{-8}$  & $1.12$\tabularnewline
$\Omega_{bb}^{-}\to\Xi_{b}^{\prime*-}\mu^{+}\mu^{-}$  & $2.40\times10^{-21}$  & $2.92\times10^{-9}$  & $1.36$ & $\Omega_{bb}^{-}\to\Omega_{b}^{*-}\mu^{+}\mu^{-}$  & $2.22\times10^{-19}$  & $2.70\times10^{-7}$  & $1.12$\tabularnewline
$\Xi_{bb}^{0}\to\Sigma_{b}^{*0}\tau^{+}\tau^{-}$  & $1.27\times10^{-22}$  & $7.13\times10^{-11}$  & $1.76$ & $\Xi_{bb}^{0}\to\Xi_{b}^{\prime*0}\tau^{+}\tau^{-}$  & $8.18\times10^{-21}$  & $4.60\times10^{-9}$  & $1.87$\tabularnewline
$\Xi_{bb}^{-}\to\Sigma_{b}^{*-}\tau^{+}\tau^{-}$  & $2.52\times10^{-22}$  & $1.42\times10^{-10}$  & $1.76$ & $\Xi_{bb}^{-}\to\Xi_{b}^{\prime*-}\tau^{+}\tau^{-}$  & $7.96\times10^{-21}$  & $4.48\times10^{-9}$  & $1.88$\tabularnewline
$\Omega_{bb}^{-}\to\Xi_{b}^{\prime*-}\tau^{+}\tau^{-}$  & $1.14\times10^{-22}$  & $1.39\times10^{-10}$  & $1.75$ & $\Omega_{bb}^{-}\to\Omega_{b}^{*-}\tau^{+}\tau^{-}$  & $1.45\times10^{-20}$  & $1.76\times10^{-8}$  & $1.85$\tabularnewline
\hline
$\Xi_{bc}^{+}\to\Sigma_{c}^{*+}e^{+}e^{-}$  & $3.51\times10^{-21}$  & $1.30\times10^{-9}$  & $0.71$ & $\Xi_{bc}^{+}\to\Xi_{c}^{\prime*+}e^{+}e^{-}$  & $1.24\times10^{-19}$  & $4.61\times10^{-8}$  & $0.66$\tabularnewline
$\Xi_{bc}^{0}\to\Sigma_{c}^{*0}e^{+}e^{-}$  & $7.02\times10^{-21}$  & $9.92\times10^{-10}$  & $0.71$ & $\Xi_{bc}^{0}\to\Xi_{c}^{\prime*0}e^{+}e^{-}$  & $1.24\times10^{-19}$  & $1.76\times10^{-8}$  & $0.66$\tabularnewline
$\Omega_{bc}^{0}\to\Xi_{c}^{\prime*0}e^{+}e^{-}$  & $2.77\times10^{-21}$  & $9.26\times10^{-10}$  & $0.75$ & $\Omega_{bc}^{0}\to\Omega_{c}^{*0}e^{+}e^{-}$  & $2.07\times10^{-19}$  & $6.92\times10^{-8}$  & $0.68$\tabularnewline
$\Xi_{bc}^{+}\to\Sigma_{c}^{*+}\mu^{+}\mu^{-}$  & $3.09\times10^{-21}$  & $1.15\times10^{-9}$  & $0.91$ & $\Xi_{bc}^{+}\to\Xi_{c}^{\prime*+}\mu^{+}\mu^{-}$  & $1.07\times10^{-19}$  & $3.98\times10^{-8}$  & $0.86$\tabularnewline
$\Xi_{bc}^{0}\to\Sigma_{c}^{*0}\mu^{+}\mu^{-}$  & $6.18\times10^{-21}$  & $8.74\times10^{-10}$  & $0.91$ & $\Xi_{bc}^{0}\to\Xi_{c}^{\prime*0}\mu^{+}\mu^{-}$  & $1.07\times10^{-19}$  & $1.52\times10^{-8}$  & $0.86$\tabularnewline
$\Omega_{bc}^{0}\to\Xi_{c}^{\prime*0}\mu^{+}\mu^{-}$  & $2.41\times10^{-21}$  & $8.07\times10^{-10}$  & $0.98$ & $\Omega_{bc}^{0}\to\Omega_{c}^{*0}\mu^{+}\mu^{-}$  & $1.77\times10^{-19}$  & $5.91\times10^{-8}$  & $0.92$\tabularnewline
$\Xi_{bc}^{+}\to\Sigma_{c}^{*+}\tau^{+}\tau^{-}$  & $4.19\times10^{-22}$  & $1.55\times10^{-10}$  & $1.36$ & $\Xi_{bc}^{+}\to\Xi_{c}^{\prime*+}\tau^{+}\tau^{-}$  & $1.33\times10^{-20}$  & $4.91\times10^{-9}$  & $1.37$\tabularnewline
$\Xi_{bc}^{0}\to\Sigma_{c}^{*0}\tau^{+}\tau^{-}$  & $8.38\times10^{-22}$  & $1.18\times10^{-10}$  & $1.36$ & $\Xi_{bc}^{0}\to\Xi_{c}^{\prime*0}\tau^{+}\tau^{-}$  & $1.33\times10^{-20}$  & $1.87\times10^{-9}$  & $1.37$\tabularnewline
$\Omega_{bc}^{0}\to\Xi_{c}^{\prime*0}\tau^{+}\tau^{-}$  & $2.55\times10^{-22}$  & $8.53\times10^{-11}$  & $1.37$ & $\Omega_{bc}^{0}\to\Omega_{c}^{*0}\tau^{+}\tau^{-}$  & $1.72\times10^{-20}$  & $5.77\times10^{-9}$  & $1.37$\tabularnewline
\hline
$\Xi_{bc}^{\prime+}\to\Sigma_{c}^{*+}e^{+}e^{-}$  & $1.05\times10^{-20}$  & $3.90\times10^{-9}$  & $0.71$ & $\Xi_{bc}^{\prime+}\to\Xi_{c}^{\prime*+}e^{+}e^{-}$  & $3.73\times10^{-19}$  & $1.38\times10^{-7}$  & $0.66$\tabularnewline
$\Xi_{bc}^{\prime0}\to\Sigma_{c}^{*0}e^{+}e^{-}$  & $2.11\times10^{-20}$  & $2.98\times10^{-9}$  & $0.71$ & $\Xi_{bc}^{\prime0}\to\Xi_{c}^{\prime*0}e^{+}e^{-}$  & $3.73\times10^{-19}$  & $5.27\times10^{-8}$  & $0.66$\tabularnewline
$\Omega_{bc}^{\prime0}\to\Xi_{c}^{\prime*0}e^{+}e^{-}$  & $8.31\times10^{-21}$  & $2.78\times10^{-9}$  & $0.75$ & $\Omega_{bc}^{\prime0}\to\Omega_{c}^{*0}e^{+}e^{-}$  & $6.21\times10^{-19}$  & $2.08\times10^{-7}$  & $0.68$\tabularnewline
$\Xi_{bc}^{\prime+}\to\Sigma_{c}^{*+}\mu^{+}\mu^{-}$  & $9.27\times10^{-21}$  & $3.44\times10^{-9}$  & $0.91$ & $\Xi_{bc}^{\prime+}\to\Xi_{c}^{\prime*+}\mu^{+}\mu^{-}$  & $3.22\times10^{-19}$  & $1.19\times10^{-7}$  & $0.86$\tabularnewline
$\Xi_{bc}^{\prime0}\to\Sigma_{c}^{*0}\mu^{+}\mu^{-}$  & $1.85\times10^{-20}$  & $2.62\times10^{-9}$  & $0.91$ & $\Xi_{bc}^{\prime0}\to\Xi_{c}^{\prime*0}\mu^{+}\mu^{-}$  & $3.22\times10^{-19}$  & $4.55\times10^{-8}$  & $0.86$\tabularnewline
$\Omega_{bc}^{\prime0}\to\Xi_{c}^{\prime*0}\mu^{+}\mu^{-}$  & $7.24\times10^{-21}$  & $2.42\times10^{-9}$  & $0.98$ & $\Omega_{bc}^{\prime0}\to\Omega_{c}^{*0}\mu^{+}\mu^{-}$  & $5.31\times10^{-19}$  & $1.77\times10^{-7}$  & $0.92$\tabularnewline
$\Xi_{bc}^{\prime+}\to\Sigma_{c}^{*+}\tau^{+}\tau^{-}$  & $1.26\times10^{-21}$  & $4.66\times10^{-10}$  & $1.36$ & $\Xi_{bc}^{\prime+}\to\Xi_{c}^{\prime*+}\tau^{+}\tau^{-}$  & $3.98\times10^{-20}$  & $1.47\times10^{-8}$  & $1.37$\tabularnewline
$\Xi_{bc}^{\prime0}\to\Sigma_{c}^{*0}\tau^{+}\tau^{-}$  & $2.51\times10^{-21}$  & $3.55\times10^{-10}$  & $1.36$ & $\Xi_{bc}^{\prime0}\to\Xi_{c}^{\prime*0}\tau^{+}\tau^{-}$  & $3.98\times10^{-20}$  & $5.62\times10^{-9}$  & $1.37$\tabularnewline
$\Omega_{bc}^{\prime0}\to\Xi_{c}^{\prime*0}\tau^{+}\tau^{-}$  & $7.65\times10^{-22}$  & $2.56\times10^{-10}$  & $1.37$ & $\Omega_{bc}^{\prime0}\to\Omega_{c}^{*0}\tau^{+}\tau^{-}$  & $5.17\times10^{-20}$  & $1.73\times10^{-8}$  & $1.37$\tabularnewline
\hline \hline
\end{tabular}
\end{table}

Some comments on the results for phenomenological observables are given as follows.
\begin{itemize}
\item It can be seen in Tabs.~\ref{Tab:branchingv22}-\ref{Tab:branching23fcnc} that the decay widths for the four cases have the following hierarchical difference.
\begin{eqnarray}
  &&\Gamma{\rm(the~transition~1/2\to 1/2~with~charged~current)}>\Gamma{\rm(the~transition~1/2\to 3/2~with~charged~current)}\nonumber\\
  &&>\Gamma{\rm(the~transition~1/2\to 1/2~with~FCNC)}>\Gamma{\rm(the~transition~1/2\to 3/2~with~FCNC)}.\nonumber
\end{eqnarray}
In the transition $1/2\to 1/2$ and $1/2\to 3/2$ with FCNC cases, the decay widths are very close to each other for $l=e/\mu$ cases, while it is about one order of magnitude smaller for $l=\tau$ case. This can be attributed to the much smaller phase space for $l=\tau$ case.
\item A reasonable modification with momentum-space wave function $\Psi^{SS_{z}}$ in the case of an axial-vector diquark involved is performed in this work in Eqs.~(\ref{eq:momentum_wave_function_1/2}) and (\ref{eq:momentum_wave_function_1/2gamma}).
While, in Refs.~\cite{Wang:2017mqp,Xing:2018lre,Zhao:2018mrg}, the momentum-space wave function $\Psi^{SS_{z}}$ in the case of an axial-vector diquark involved is defined as
\begin{eqnarray}	
& \Psi^{SS_{z}}(\tilde{p}_{1},\tilde{p}_{2},\lambda_{1},\lambda_{2})=\frac{A}{\sqrt{2(p_{1}\cdot\bar{P}+m_{1}M_{0})}}\bar{u}(p_{1},\lambda_{1})\Gamma u(\bar{P},S_{z})\phi(x,k_{\perp}), \label{eq:momentum_wave_function_1/2zhao}\\
&\Gamma  =-\frac{1}{\sqrt{3}}\gamma_{5}\left(\slashed\epsilon^{*}(p_{2},\lambda_{2})\right) \quad {\rm with}\quad A=\sqrt{\frac{3(m_{1}M_{0}+p_{1}\cdot\bar{P})}{3m_{1}M_{0}+p_{1}\cdot\bar{P}+2(p_{1}\cdot p_{2})(p_{2}\cdot\bar{P})/m_{2}^{2}}}.\label{eq:momentum_wave_function_1/2gammazhao}
\end{eqnarray}
In Ref.~\cite{Wang:2017mqp} the extraction approach is different from the one used in this work and in Refs.~\cite{Xing:2018lre,Zhao:2018mrg}.
In order to find out the impact on the form factors and decay widths of these two factors: the extraction method of the form factors and the baryon wave function related to the axial vector diquark, we list numerical results of theses form factors of the three decay channels and the corresponding partial decay widths in Tab.~\ref{difference}. The corresponding numerical results in Refs.~\cite{Wang:2017mqp,Xing:2018lre,Zhao:2018mrg} are also given in Tab.~\ref{difference}.

\begin{table}
\caption{The comparison of form factors and decay widths $\Gamma$ between this work (This) and Zhao's work~\cite{Wang:2017mqp,Xing:2018lre,Zhao:2018mrg}, and "SE" means ``the same extraction method of the form factors as this work".}
\label{difference}
\begin{center}
\begin{tabular}{c|c|c|c|c|c|c|c|c|c|c|c|c|c}
\hline\hline
channel & $f_{1,S}$ & $f_{2,S}$ & $f_{3,S}$ & $g_{1,S}$ & $g_{2,S}$ & $g_{3,S}$ & $f_{1,A}$ & $f_{2,A}$ & $f_{3,A}$ & $g_{1,A}$ & $g_{2,A}$ & $g_{3,A}$&$\Gamma/{\rm GeV}$\tabularnewline
\hline
$\Xi_{cc}^{++}\to\Lambda_{c}^{+}l^{+}\nu_{l}$[This] & 0.495 & -0.621 & 0.832 & 0.332 & 1.004 & -2.957 & 0.489 & 0.290 & 0.648 & -0.111 & -0.325 & 0.943&$7.97\times10^{-15}$\tabularnewline
\hline
$\Xi_{cc}^{++}\to\Lambda_{c}^{+}l^{+}\nu_{l}$~[SE]& 0.495 & -0.621 & 0.832 & 0.332 & 1.004 & -2.957 & 0.479 & 0.268 &0.650& -0.111 & -0.307 &1.702&$7.96\times10^{-15}$\tabularnewline
\hline
$\Xi_{cc}^{++}\to\Lambda_{c}^{+}l^{+}\nu_{l}$~\cite{Wang:2017mqp}& 0.653 & -0.738 &  & 0.533 & -0.053 &  & 0.637 & 0.725 &  & -0.167 & -0.028 & &$1.05\times10^{-14}$ \tabularnewline
\hline\hline
channel & $f_{1,A}$ & $f_{2,A}$ & $f_{3,A}$ & $g_{1,A}$ & $g_{2,A}$ & $g_{3,A}$ & $f_{1,A}^{T}$ & $f_{2,A}^{T}$ & $g_{1,A}^{T}$ & $g_{2,A}^{T}$ &  & &$\Gamma/{\rm GeV}$\tabularnewline
\hline
$\Xi_{bb}^{0}\to\Xi_{b}^{0}e^{+}e^{-}$ [This]& 0.140 & 0.123 & -0.066 & -0.041 & -0.017 & 0.130&$0.134$ & -0.061 &$-0.054$ & -0.042 &  & &$1.62\times10^{-19}$  \tabularnewline
\hline
$\Xi_{bb}^{0}\to\Xi_{b}^{0}e^{+}e^{-}$~\cite{Xing:2018lre}& 0.138 & 0.132 & -0.068 & -0.030 & -0.055 & 0.261 &$0.134$& -0.066 & $0.032$ & -0.049  &  & &$1.98\times10^{-19}$\tabularnewline
\hline\hline
channel & $f_{1,A}$ & $f_{2,A}$ & $f_{3,A}$ & $f_{4,A}$ & $g_{1,A}$ & $g_{2,A}$ & $g_{3,A}$ & $g_{4,A}$ &  &  &  & &$\Gamma/{\rm GeV}$\tabularnewline
\hline
$\Xi_{cc}^{++}\to\Sigma_{c}^{*+}l^{+}\nu_{l}$[This] & -0.979 & -0.645 & 0.047 & -1.969 & -5.792 & -3.602 & 0.947 & 0.393 &  &  &  & &$1.43\times10^{-15}$ \tabularnewline
\hline
$\Xi_{cc}^{++}\to\Sigma_{c}^{*+}l^{+}\nu_{l}$~\cite{Zhao:2018mrg}& -1.121 & 1.845 & -1.703 & -1.827 & -8.292 & -5.262 & 0.942 & 0.295 &  &  &  & &$1.26\times10^{-15}$ \tabularnewline
\hline\hline
\end{tabular}
\end{center}
\end{table}
Firstly, comparing each first two lines for $\Xi_{cc}^{++}\to\Lambda_{c}^{+}l^{+}\nu_{l}$, $\Xi_{bb}^{0}\to\Xi_{b}^{0}e^{+}e^{-}$ and $\Xi_{cc}^{++}\to\Sigma_{c}^{*+}l^{+}\nu_{l}$, we could find that partial decay width differences coming from the different wave function with axial-vector diquark are small, but there are some differences among the form factors. Secondly, comparing the second line and third line of the channel $\Xi_{cc}^{++}\to\Lambda_{c}^{+}l^{+}\nu_{l}$, we could find the extracting approach of these form factors will bring in some effect in the form factors and decay widths; So the effect in the form factors and decay widths brought in by the extraction approach is much larger than that of the definition of the wave function $\Psi^{SS_{z}}$.
\item Since there exist uncertainties in the lifetimes of the parent baryons, there may be some small fluctuations in the results for branching ratios. Form Tab.~\ref{Tab:branchingratio22}, we may find that
     \begin{eqnarray}
     &&{\cal B}(\Xi^{++}_{cc}\to \Xi^{\prime +}_{c}l^{+}\nu_{l})=5.57\times10^{-2},\quad
    {\cal B}(\Xi^{++}_{cc}\to \Xi^{+}_{c}l^{+}\nu_{l})=3.40\times10^{-2},\\
     &&{\cal B}(\Omega^{+}_{cc}\to \Omega^{0}_{c}l^{+}\nu_{l})=7.67\times10^{-2},\quad
     {\cal B}(\Xi^{+}_{bc}\to \Xi^{\prime 0}_{b}l^{+}\nu_{l})=2.50\times10^{-2}.
     \end{eqnarray}
    These channels may be firstly examined at experimental facilities like   LHC or BelleII.
\item Take the four processes $\Xi^{++}_{cc}\to\Lambda_{c}^{+}l^{+}\nu_{l}$, $\Xi^{++}_{cc}\to\Sigma_{c}^{*+}l^{+}\nu_{l}$, $\Xi^{0}_{bb}\to\Xi_{b}^{0}e^{+}e^{-}$ and $\Xi^{0}_{bb}\to\Xi_{b}^{\prime*0}e^{+}e^{-}$ as examples. The uncertainties for the partial decay widths caused by the model parameters and the single pole assumption for $c\to d,s$ channels are listed as
      \begin{eqnarray}
      \Gamma(\Xi^{++}_{cc}\to\Lambda_{c}^{+}l^{+}\nu_{l})&=&(7.97\pm0.65\pm1.28\pm1.55\pm1.65)\times10^{-15}~{\rm GeV},\nonumber\\
      \Gamma(\Xi^{++}_{cc}\to\Sigma_{c}^{*+}l^{+}\nu_{l})&=&(1.43\pm0.23\pm0.29\pm0.29\pm0.16)\times10^{-15}~{\rm GeV},
      \label{eq:vverrors}
      \end{eqnarray}
where these errors come from $\beta_{i}$, $\beta_{f}$, $m_{\rm di}$ and $m_{\rm pole}$ respectively;
\begin{eqnarray}
      \Gamma(\Xi^{0}_{bb}\to\Xi_{b}^{0}e^{+}e^{-})&=&(1.62\pm0.69\pm0.96\pm0.17)\times10^{-19}~{\rm GeV},\nonumber\\
      \Gamma(\Xi^{0}_{bb}\to\Xi_{b}^{\prime*0}e^{+}e^{-})&=&(1.45\pm0.19\pm0.70\pm0.43)\times10^{-19}~{\rm GeV},
      \label{eq:fcncerrors}
\end{eqnarray}
where these errors come from $\beta_{i}$, $\beta_{f}$, $m_{\rm di}$, respectively.
Taking $\Xi_{cc}^{++}\to\Lambda^{+}_{c}$ as an example, the error estimates for the form factors can be found in Tab.~\ref{Tab:errorsformfractors}.
\begin{table}
\caption{Error estimates for the form factors, taking $\Xi_{cc}^{++}\to\Lambda^{+}_{c}$ as an example. The first number is the central value, and following 3 errors come from $\beta_{i}=\beta_{\Xi_{cc}^{++}}$, $\beta_{f}=\beta_{\Lambda^{+}_{c}}$ and $m_{di}=m_{(cu)}$, respectively. These parameters are all varied by $10\%$.}
\label{Tab:errorsformfractors}
\begin{tabular}{c|c|c|c}
\hline\hline
$F$ & $F(0)$ & $F$ & $F(0)$\tabularnewline
\hline
$f_{1,S}^{\Xi_{cc}^{++}\to\Lambda_{c}^{+}}$ & $0.495\pm0.020\pm0.034\pm0.042$ & $f_{1,A}^{\Xi_{cc}^{++}\to\Lambda_{c}^{+}}$ & $0.489\pm0.019\pm0.034\pm0.042$\tabularnewline
\hline
$f_{2,S}^{\Xi_{cc}^{++}\to\Lambda_{c}^{+}}$ & $-0.621\pm0.119\pm0.065\pm0.227$ & $f_{2,A}^{\Xi_{cc}^{++}\to\Lambda_{c}^{+}}$ & $0.290\pm0.074\pm0.080\pm0.199$\tabularnewline
\hline
$f_{3,S}^{\Xi_{cc}^{++}\to\Lambda_{c}^{+}}$ & $0.832\pm0.130\pm0.165\pm0.202$ & $f_{3,A}^{\Xi_{cc}^{++}\to\Lambda_{c}^{+}}$ & $0.648\pm0.122\pm0.170\pm0.194$\tabularnewline
\hline
$g_{1,S}^{\Xi_{cc}^{++}\to\Lambda_{c}^{+}}$ & $0.332\pm0.020\pm0.004\pm0.086$ & $g_{1,A}^{\Xi_{cc}^{++}\to\Lambda_{c}^{+}}$ & $-0.111\pm0.007\pm0.001\pm0.003$\tabularnewline
\hline
$g_{2,S}^{\Xi_{cc}^{++}\to\Lambda_{c}^{+}}$ & $1.004\pm0.059\pm0.199\pm0.170$ & $g_{2,A}^{\Xi_{cc}^{++}\to\Lambda_{c}^{+}}$ & $-0.325\pm0.021\pm0.065\pm0.058$\tabularnewline
\hline
$g_{3,S}^{\Xi_{cc}^{++}\to\Lambda_{c}^{+}}$ & $-2.957\pm0.973\pm0.804\pm0.731$ & $g_{3,A}^{\Xi_{cc}^{++}\to\Lambda_{c}^{+}}$ & $0.943\pm0.330\pm0.264\pm0.247$\tabularnewline
\hline\hline
\end{tabular}
\end{table}
It can be seen from Eqs.~(\ref{eq:vverrors}-\ref{eq:fcncerrors}) and Tab.~\ref{Tab:errorsformfractors} that, the uncertainties caused by these parameters may be sizable.
\item The ratios $\Gamma_{L}/\Gamma_{T}$s have the following rule:
\begin{eqnarray}
&&c\to d:~\Gamma_{L}/\Gamma_{T}(\Xi_{cc}^{++}\to\Sigma_{c}^{+}l^{+}\nu_{l})=
\Gamma_{L}/\Gamma_{T}(\Xi_{cc}^{+}\to\Sigma_{c}^{0}l^{+}\nu_{l})=
\Gamma_{L}/\Gamma_{T}(\Omega_{cc}^{+}\to\Xi_{c}^{\prime0}l^{+}\nu_{l}),\label{eq:cdcc}\\
&&c\to s:~\Gamma_{L}/\Gamma_{T}(\Xi_{cc}^{++}\to\Xi_{c}^{\prime+}l^{+}\nu_{l})=
\Gamma_{L}/\Gamma_{T}(\Xi_{cc}^{+}\to\Xi_{c}^{\prime0}l^{+}\nu_{l})=
\Gamma_{L}/\Gamma_{T}(\Omega_{cc}^{+}\to\Omega_{c}^{0}l^{+}\nu_{l}),\label{eq:cscc}
\end{eqnarray}
for these decay channels in Eq.~(\ref{eq:cdcc}) have the same decay in quark level and the final single heavy baryons all in the sextets which leads to the same overlapping factors in the SU(3) sysmetry. In other decay channels the similar relations exist.
 For the transition $1/2\to3/2$, we have the following relations:
 \begin{eqnarray}
 &&\Gamma_{L}/\Gamma_{T}(B_{bc}^{\prime}\to B_{b}^{*}l^{+}\nu_{l})=\Gamma_{L}/\Gamma_{T}(B_{bc}\to B_{b}^{*}l^{+}\nu_{l}), \label{eq:cdsbc}\\
  &&\Gamma_{L}/\Gamma_{T}(B_{bc}^{\prime}\to B_{c}^{*}l^{-}\bar\nu_{l})=\Gamma_{L}/\Gamma_{T}(B_{bc}\to B_{c}^{*}l^{-}\bar\nu_{l}),\label{eq:bucbc}\\
 &&\Gamma_{L}/\Gamma_{T}(B_{bc}^{\prime}\to B_{c}^{*}l^{+}l^{-})=\Gamma_{L}/\Gamma_{T}(B_{bc}\to B_{c}^{*}l^{+}l^{-}),\label{eq:bdsbc}
 \end{eqnarray}
 for these decay channels have the same decay in quark level and the spin of final single heavy baryons are all $3/2$, only with the axial-vector diquark spectator which have the same overlapping factors and form factors. As shown in Figs.~\ref{fig:decaywidthbbtob22}-\ref{fig:decaywidthbbtobfcnc23} and Tabs.~\ref{Tab:branchingv22}-~\ref{Tab:branching23fcnc}, these decay channels with same $\Gamma_{L}/\Gamma_{T}$ have the similar plots of the dependence of $d\Gamma_{L}/dq^2$ and $d\Gamma_{T}/dq^2$ on $q^2$.
 \item From Figs.~\ref{fig:decaywidthbbtobFCNC22} and \ref{fig:decaywidthbbtobfcnc23}, it can be found that, at small $q^2$, there are some divergence of the $d\Gamma_{L}/dq^2$ and $d\Gamma_{T}/dq^2$ for the cases
 ${\cal B}_{bb}\to{\cal B}_{b}^{(*)} e^{+}e^{-}/ \mu^{+}\mu^{-}$ and ${\cal B}_{bb}\to{\cal B}_{bc}^{(*)} l^{-}\bar{\nu}_{l}$, because $\frac{1}{\sqrt{q^2}}$ is included in their helicity amplitudes shown with Eqs.~(\ref{eq:helicty23v}), (\ref{eq:hv22}), (\ref{eq:helicty22fcnc}),
  (\ref{eq:hv232}), (\ref{eq:helicty23fcnc}).
 \end{itemize}

\subsection{SU(3) symmetry for semileptonic decays}
\renewcommand{\thetable}{D\arabic{table}}
Recently, an analysis of weak decays of doubly-heavy baryons based on flavor symmetry is available in Ref.~\cite{Wang:2017azm,Shi:2017dto}. In the SU(3) symmetry limit,  there exist the a number of relations
among these semileptonic decay widths, which we are going to examine in the following.

\begin{itemize}
	
\item For $c\to d,s$ process, we have
\begin{align*}
&\Gamma(\Xi_{cc}^{++}\to\Lambda_{c}^{+}l^{+}\nu_{l}) =\Gamma(\Omega_{cc}^{+}\to\Xi_{c}^{0}l^{+}\nu_{l}),\quad
\Gamma(\Xi_{cc}^{++}\to\Xi_{c}^{+}l^{+}\nu_{l})  =\Gamma(\Xi_{cc}^{+}\to\Xi_{c}^{0}l^{+}\nu_{l}),\\
&\Gamma(\Xi_{cc}^{+}\to\Sigma_{c}^{(*)0}l^{+}\nu_{l})=2\Gamma(\Xi_{cc}^{++}\to\Sigma_{c}^{(*)+}l^{+}\nu_{l})  =2\Gamma(\Omega_{cc}^{+}\to\Xi_{c}^{\prime(*)0}l^{+}\nu_{l}),\\
&\Gamma(\Omega_{cc}^{+}\to\Omega_{c}^{(*)0}l^{+}\nu_{l}) =2\Gamma(\Xi_{cc}^{++}\to\Xi_{c}^{\prime(*)+}l^{+}\nu_{l}) =2\Gamma(\Xi_{cc}^{+}\to\Xi_{c}^{\prime(*)0}l^{+}\nu_{l}),\\
&\Gamma(\Xi_{bc}^{(\prime)+}\to\Lambda_{b}^{0}l^{+}\nu_{l}) =\Gamma(\Omega_{bc}^{(\prime)0}\to\Xi_{b}^{-}l^{+}\nu_{l}),\quad
\Gamma(\Xi_{bc}^{(\prime)+}\to\Xi_{b}^{0}l^{+}\nu_{l})  =2\Gamma(\Xi_{bc}^{(\prime)0}\to\Xi_{b}^{-}l^{+}\nu_{l}),\\
&2\Gamma(\Xi_{bc}^{(\prime)+}\to\Sigma_{b}^{(*)0}l^{+}\nu_{l})  =\Gamma(\Xi_{bc}^{(\prime)0}\to\Sigma_{b}^{(*)-}l^{+}\nu_{l})=2\Gamma(\Omega_{bc}^{(\prime)0}\to\Xi_{b}^{\prime(*)-}l^{+}\nu_{l}),\\
&2\Gamma(\Xi_{bc}^{(\prime)+}\to\Xi_{b}^{\prime(*)0}l^{+}\nu_{l}) =2\Gamma(\Xi_{bc}^{(\prime)0}\to\Xi_{b}^{\prime(*)-}l^{+}\nu_{l})=\Gamma(\Omega_{bc}^{(\prime)0}\to\Omega_{b}^{(*)-}l^{+}\nu_{l}).
\end{align*}
\item For $b\to u,c$ process, we have
\begin{align*}
&\Gamma(\Xi_{bb}^{-}\to\Lambda_{b}^{0}l^{-}\bar{\nu}_{l})  =\Gamma(\Omega_{bb}^{-}\to\Xi_{b}^{0}l^{-}\bar{\nu}_{l}),\\
&\Gamma(\Xi_{bb}^{0}\to\Xi_{bc}^{(*)+}l^{-}\bar{\nu}_{l})  =\Gamma(\Xi_{bb}^{-}\to\Xi_{bc}^{(*)0}l^{-}\bar{\nu}_{l})=\Gamma(\Omega_{bb}^{-}\to\Omega_{bc}^{(*)0}l^{-}\bar{\nu}_{l}),\\
&\Gamma(\Xi_{bb}^{0}\to\Sigma_{b}^{(*)+}l^{-}\bar{\nu}_{l})  =2\Gamma(\Xi_{bb}^{-}\to\Sigma_{b}^{(*)0}l^{-}\bar{\nu}_{l})=2\Gamma(\Omega_{bb}^{-}\to\Xi_{b}^{\prime(*)0}l^{-}\bar{\nu}_{l}),\\
&\Gamma(\Xi_{bc}^{(\prime)+}\to\Xi_{cc}^{(*)++}l^{-}\bar{\nu}_{l})  =\Gamma(\Xi_{bc}^{(\prime)0}\to\Xi_{cc}^{(*)+}l^{-}\bar{\nu}_{l})=\Gamma(\Omega_{bc}^{(\prime)0}\to\Omega_{cc}^{(*)+}l^{-}\bar{\nu}_{l}),\\
&\Gamma(\Xi_{bc}^{(\prime)+}\to\Sigma_{c}^{(*)++}l^{-}\bar{\nu}_{l}) =2\Gamma(\Xi_{bc}^{(\prime)0}\to\Sigma_{c}^{(*)+}l^{-}\bar{\nu}_{l})=2\Gamma(\Omega_{bc}^{(\prime)0}\to\Xi_{c}^{\prime(*)+}l^{-}\bar{\nu}_{l}).
\end{align*}

\end{itemize}

According to the flavor SU(3) symmetry, there exist the following
relations among these FCNC process. These relations can be readily
derived using the overlapping factors given in Tab. \ref{Tab:overlapping_factors_22}.
\begin{itemize}
\item For $b\to d$ process, we have
	\begin{eqnarray}
	&  & \Gamma(\Xi_{bb}^{0}\to\Lambda_{b}^{0}l^{+}l^{-})=\Gamma(\Omega_{bb}^{-}\to\Xi_{b}^{-}l^{+}l^{-}),\nonumber \\
	&  & 2\Gamma(\Xi_{bb}^{0}\to\Sigma_{b}^{(*)0}l^{+}l^{-})=\Gamma(\Xi_{bb}^{-}\to\Sigma_{b}^{(*)-}l^{+}l^{-})
=2\Gamma(\Omega_{bb}^{-}\to\Xi_{b}^{\prime(*)0}l^{+}l^{-}),\nonumber\\
	&  & \Gamma(\Xi_{bc}^{(\prime)+}\to\Lambda_{c}^{+}l^{+}l^{-})=\Gamma(\Omega_{bc}^{(\prime)0}\to\Xi_{c}^{0}l^{+}l^{-}),\nonumber \\
	&  & 2\Gamma(\Xi_{bc}^{(\prime)+}\to\Sigma_{c}^{(*)+}l^{+}l^{-})=\Gamma(\Xi_{bc}^{(\prime)0}\to\Sigma_{c}^{(*)0}l^{+}l^{-})
=2\Gamma(\Omega_{bc}^{(\prime)0}\to\Xi_{c}^{\prime(*)0}l^{+}l^{-}).\nonumber
	\end{eqnarray}

	\item For $b\to s$ process, we have
	\begin{eqnarray}
	&  & \Gamma(\Xi_{bb}^{0}\to\Xi_{b}^{0}l^{+}l^{-})=\Gamma(\Xi_{bb}^{-}\to\Xi_{b}^{-}l^{+}l^{-}),\nonumber \\
	&  & 2\Gamma(\Xi_{bb}^{0}\to\Xi_{b}^{\prime(*)0}l^{+}l^{-})=2\Gamma(\Xi_{bb}^{-}\to\Xi_{b}^{\prime(*)-}l^{+}l^{-})
=\Gamma(\Omega_{bb}^{-}\to\Omega_{b}^{(*)-}l^{+}l^{-}),\nonumber \\
	&  & \Gamma(\Xi_{bc}^{(\prime)+}\to\Xi_{c}^{+}l^{+}l^{-})=\Gamma(\Xi_{bc}^{(\prime)0}\to\Xi_{c}^{0}l^{+}l^{-}),\nonumber \\
	&  & 2\Gamma(\Xi_{bc}^{(\prime)+}\to\Xi_{c}^{\prime(*)+}l^{+}l^{-})=2\Gamma(\Xi_{bc}^{(\prime)0}\to\Xi_{c}^{\prime(*)0}l^{+}l^{-})
=\Gamma(\Omega_{bc}^{(\prime)0}\to\Omega_{c}^{(*)0}l^{+}l^{-}).\nonumber
	\end{eqnarray}
	
\end{itemize}	
Comparing the above equations predicted by SU(3) symmetry with the corresponding results in this work, we have the following remarks:
\begin{itemize}
	\item  most of our numerical results are respected very well with the SU(3) symmetry relations, except for the following ones
	\begin{align}
	\Gamma(\Xi_{cc}^{++}\to\Lambda_{c}^{+}l^{+}\nu_{l}) &= \Gamma(\Omega_{cc}^{+}\to\Xi_{c}^{0}l^{+}\nu_{l}),\quad \Gamma(\Xi_{bc}^{+}\to\Lambda_{b}^{0}l^{+}\nu_{l})  = \Gamma(\Omega_{bc}^{0}\to\Xi_{b}^{-}l^{+}\nu_{l}),\nonumber\\ \Gamma(\Xi_{bc}^{+}\to\Sigma_{b}^{(*)0}l^{+}\nu_{l}) &= \Gamma(\Omega_{bc}^{0}\to\Xi_{b}^{\prime(*)-}l^{+}\nu_{l}),\quad \Gamma(\Xi_{bc}^{+}\to\Xi_{b}^{\prime(*)0}l^{+}\nu_{l}) = \frac{1}{2}\Gamma(\Omega_{bc}^{0}\to\Omega_{b}^{(*)-}l^{+}\nu_{l}),\nonumber\\ \Gamma(\Xi_{bc}^{0}\to\Sigma_{c}^{(*)+}l^{-}\bar{\nu}_{l}) &= \Gamma(\Omega_{bc}^{0}\to\Xi_{c}^{\prime(*)+}l^{-}\bar{\nu}_{l}).\label{eq:su3_breaking}
\end{align}
These five relations  are broken considerably: larger  than 20\% but still less than 50\% using the definition of $({\rm Max}[\Gamma_{{\rm LHS}},\Gamma_{{\rm RHS}}]-{\rm Min}[\Gamma_{{\rm LHS}},\Gamma_{{\rm RHS}}])/{\rm Max}[\Gamma_{{\rm LHS}},\Gamma_{{\rm RHS}}]$. Since the mass difference between the $u$ and $d$ quark has been neglected in this work, the isospin symmetry is well respected. But since the strange quark is much heavier,  the SU(3) relations for the channels involving   $u,d$ quark and $s$ quark can be significantly broken. All relations given in  Eq.~(\ref{eq:su3_breaking})  are of this type.

\item The first 4 relations  in Eq.~(\ref{eq:su3_breaking}) involve the $c$ quark decay but  the last one involves the $b$ quark decay. It indicates that the $c$ quark decay modes tend to break SU(3) symmetry easily. This can be understood since  the phase space of the $c$ quark decay is smaller, and thus the decay amplitude  is more  sensitive to the mass of the initial and final baryons.
\item SU(3) symmetry breaking is larger for the $Qs$ diquark involved case than that for the $Qu/Qd$ diquark involved case with $Q=b,c$. SU(3) symmetry breaking is larger for the $cq$ diquark involved case than that for the $bq$ diquark involved case with $q=u,d,s$.
\item SU(3) symmetry breaking is smaller for $l=e/\mu$ cases than that for $l=\tau$ case. This can be attributed to the much smaller phase space for $l=\tau$ case. Smaller phase space is more sensitive to the variation of the masses of baryons in the initial and final states.
	\end{itemize}

\section{Conclusion}

In this paper, we have  presented  a
systematic investigation of  transition form factors of   doubly heavy baryon  decays in the light front approach. Our main results for the form factors with the $q^2$ distributions are collected in Tabs.~\ref{Tab:ff_ccc}-\ref{Tab:fcnc32_bd}. The present analysis is the sequel and update of the  previous works on weak decays of doubly heavy baryons, Ref.~\cite{Wang:2017mqp,Zhao:2018mrg,Xing:2018lre}. It improves upon the  previous work by
\begin{itemize}
\item adding predictions for the FCNC process with the spin-1/2 to spin-3/2 transition;
\item using a new extraction of the Lorentz structure for the form factors;
\item updating  the wave function for spin-$1/2$ baryons states with an  axial-vector diquark ;
\item updating  the wave function for spin-$3/2$ baryons states;
\item presenting  a new derivation of  the overlapping factors using flavor  SU(3) symmetry approach;
\item presenting   the momentum distribution   of form factors.
\end{itemize}
Using these form factors, we have also preformed the calculation of phenomenological observables of these corresponding semileptonic weak decays of doubly heavy baryons with the results  shown in Tabs.~\ref{Tab:branchingv22}-\ref{Tab:branching23fcnc}. The flavor  SU(3) symmetry and sources of symmetry breaking are also discussed in great details.
We find that,
\begin{itemize}
\item most  branching ratios for spin-1/2 to spin-1/2  with $c\to d,s$ processes
are at the order  $10^{-3}\sim10^{-2}$,  which might be  examined at experimental facilities at LHC or Belle-II;
\item the different extraction approaches could give   sizable differences to form factors;
\item the uncertainties of form factors and decay widths caused by model parameters are sizable;
\item the ratios $\Gamma_{L}/\Gamma_{T}$s have the rules shown in Eqs.~(\ref{eq:cdcc})-(\ref{eq:bdsbc}), for these decay channels have the same decay in quark level and the same overlapping factors and form factors;
\item most of our results are comparable to the theoretical results in Refs.~\cite{Wang:2017mqp,Zhao:2018mrg,Xing:2018lre};
\item since the mass difference between the $u$ and $d$ quark has been neglected and the strange quark is much heavier, the SU(3) relations shown in Eq.~(\ref{eq:su3_breaking}) for the channels involving $u,d$ quark and $s$ quark can be  broken;

\item the SU(3) symmetry breaking is sizable in the charmed
baryon decays, while for the bottomed case the SU(3) symmetry breaking is small.
\end{itemize}

This work completes the study of form factors in the traditional light-front quark model with the quark-diquark constituent viewpoint.   We hope our phenomenology predictions for these semi-leptonic decays could be tested by LHCb and other experiments in the future.

\section*{Acknowledgements}
We thank Prof. Wei Wang,  Fu-Sheng Yu and Zhen-Xing Zhao for fruitful discussions.
This work is supported in part by National
Natural Science Foundation of China under Grants No.
11735010, 11911530088, and 11765012,  Natural Science Foundation of Shanghai under Grants
No.~15DZ2272100, and by Key Laboratory for Particle
Physics, Astrophysics and Cosmology, Ministry of Education.
\appendix

\section{Wave functions in initial and final states}
\label{app:wave_functions}
	
\subsection{Wave functions in the standard flavor-spin basis}

The wave functions in the flavor space can also be found in Ref.~\cite{Chau:1995gk}.
The wave functions of the doubly heavy baryons in the standard flavor-spin basis are given as follows.

For ${\cal B}_{QQq}$ ($\Xi_{cc}^{++,+}$, $\Omega_{cc}^{+}$, $\Xi_{bb}^{0,-}$ and $\Omega_{bb}^{-}$), their flavor-spin functions are given as
\begin{equation}
|{\cal B}_{QQq},\uparrow\rangle=(QQq)\left(\frac{1}{\sqrt{6}}(\uparrow\downarrow\uparrow+\downarrow\uparrow\uparrow-2\uparrow\uparrow\downarrow)\right),
\end{equation}
with $q=u,d,s$ and $Q=c,b$.

For the baryons with two different heavy quarks ${\cal B}_{Q_{1}Q_{2}q}$ ( $\Xi_{bc}^{+,0}$ and $\Omega_{bc}^{0}$), their flavor-spin functions can be given as
\begin{equation}
|{\cal B}_{Q_{1}Q_{2}q},\uparrow\rangle=\Big(\frac{1}{\sqrt{2}}(Q_{1}Q_{2}+Q_{2}Q_{1})q\Big)\left(\frac{1}{\sqrt{6}}(\uparrow\downarrow\uparrow+\downarrow\uparrow\uparrow-2\uparrow\uparrow\downarrow)\right),
\end{equation}
where the two different heavy quarks are symmetry.
While for the two different heavy quarks in baryons are anti-symmetry, the flavor-spin functions of these baryons ${\cal B}^{\prime}_{Q_{1}Q_{2}q}$ ($\Xi_{bc}^{\prime+}$, $\Xi_{bc}^{\prime0}$ and $\Omega_{bc}^{\prime0}$) are
\begin{equation}
|{\cal B}_{Q_{1}Q_{2}q},\uparrow\rangle=\Big(\frac{1}{\sqrt{2}}(Q_{1}Q_{2}-Q_{2}Q_{1})q\Big)\left(\frac{1}{\sqrt{2}}(\uparrow\downarrow\uparrow-\downarrow\uparrow\uparrow)\right),
\end{equation}
where $q=u,d,s$, $Q_{1}=b$ and $Q_{2}=c$.

Then the flavor-spin wave functions of the singly heavy baryons in the final states
are given with the following functions.
	
The flavor-spin wave functions of ${\cal B}_{Qqq}^{\boldsymbol{6}}$ ($\Sigma_{c}^{++,0}$, $\Omega_{c}^{0}$, $\Sigma_{b}^{+,-}$ and $\Omega_{b}^{-}$) are
\begin{equation}
|{\cal B}_{Qqq}^{\boldsymbol{6}},\uparrow\rangle=(qqQ)\left(\frac{1}{\sqrt{6}}(\uparrow\downarrow\uparrow+\downarrow\uparrow\uparrow-2\uparrow\uparrow\downarrow)\right),\label{eq:wave_cqq}
\end{equation}
where $q=u,d,s$ and $Q=c,b$.
	
When the two light quarks in the baryons ${\cal B}_{Qq_{1}q_{2}}^{\boldsymbol{6}}$ ($\Sigma_{c}^{+}$, $\Xi_{c}^{\prime+,\prime0}$, $\Sigma_{b}^{0}$ and $\Xi_{b}^{\prime0,\prime-}$) are different and symmetry, their flavor-spin wave functions are
\begin{equation}
|{\cal B}_{Qq_{1}q_{2}}^{\boldsymbol{6}},\uparrow\rangle=\left(\frac{1}{\sqrt{2}}(q_{1}q_{2}+q_{2}q_{1})Q\right)\left(\frac{1}{\sqrt{6}}(\uparrow\downarrow\uparrow+\downarrow\uparrow\uparrow-2\uparrow\uparrow\downarrow)\right),\label{eq:wave_6cq1q2}
\end{equation}
with $(q_{1},q_{2})=(u,d),(u,s),(d,s)$ and $Q=c,b$.
While for the two different light quarks in baryons are anti-symmetry, the flavor-spin functions of these baryons
${\cal B}_{Qq_{1}q_{2}}^{\bar{\boldsymbol{3}}}$ ($\Lambda_{c}^{+}$, $\Xi_{c}^{+,0}$, $\Lambda_{b}^{0}$ and $\Xi_{b}^{0,-}$) are
\begin{equation}
|{\cal B}_{Qq_{1}q_{2}}^{\bar{\boldsymbol{3}}},\uparrow\rangle=\left(\frac{1}{\sqrt{2}}(q_{1}q_{2}-q_{2}q_{1})Q\right)\left(\frac{1}{\sqrt{2}}(\uparrow\downarrow\uparrow-\downarrow\uparrow\uparrow)\right),\label{eq:wave_3cq1q2}
\end{equation}
with $(q_{1},q_{2})=(u,d),(u,s),(d,s)$ and $Q=c,b$.
\subsection{Wave functions in the diquark basis}
	
As we know, the coupling of the two angular momenta $j_{1}=1$ and $j_{2}=\frac{1}{2}$ is
\begin{equation}
|J=\frac{1}{2},M=\frac{1}{2}\rangle=\sqrt{\frac{2}{3}}|m_{1}=1,m_{2}=-\frac{1}{2}\rangle-\sqrt{\frac{1}{3}}|m_{1}=0,m_{2}=\frac{1}{2}\rangle.
\end{equation}
Then the baryon state with an axial-vector diquark could be defined as follows,
\begin{equation}
|q_{1}(q_{2}q_{3})_{A},\uparrow\rangle\equiv\sqrt{\frac{2}{3}}q_{1}\downarrow(q_{2}q_{3})_{11}-\sqrt{\frac{1}{3}}q_{1}\uparrow(q_{2}q_{3})_{10},
\end{equation}
with $(q_{2}q_{3})_{11}=(q_{2}q_{3})(\uparrow\uparrow)$ and $(q_{2}q_{3})_{10}=(q_{2}q_{3})\left(\frac{1}{\sqrt{2}}(\uparrow\downarrow+\downarrow\uparrow)\right)$.
For the baryon state with a scalar diquark, we have the following definition
\begin{equation}
|q_{1}(q_{2}q_{3})_{S},\uparrow\rangle\equiv q_{1}\uparrow(q_{2}q_{3})_{S},
\end{equation}
with $(q_{2}q_{3})_{S}=(q_{2}q_{3})_{00}=(q_{2}q_{3})\left(\frac{1}{\sqrt{2}}(\uparrow\downarrow-\downarrow\uparrow)\right)$.
	
Using the above definitions, the following two equations can be proved
\begin{eqnarray}
q_{1}q_{2}q_{3}\left(\frac{1}{\sqrt{2}}(\uparrow\downarrow\uparrow-\downarrow\uparrow\uparrow)\right) & = & -\frac{1}{2}|q_{1}(q_{2}q_{3})_{S},\uparrow\rangle-\frac{\sqrt{3}}{2}|q_{1}(q_{2}q_{3})_{A},\uparrow\rangle,\\
q_{1}q_{2}q_{3}\left(\frac{1}{\sqrt{6}}(\uparrow\downarrow\uparrow+\downarrow\uparrow\uparrow-2\uparrow\uparrow\downarrow)\right) & = & -\frac{\sqrt{3}}{2}|q_{1}(q_{2}q_{3})_{S},\uparrow\rangle+\frac{1}{2}|q_{1}(q_{2}q_{3})_{A},\uparrow\rangle.
\end{eqnarray}
Then in the diquark basis, the flavor-spin wave functions of the initial and final baryon states can be derived  with the help of above expressions and are shown in Subsec.\ref{sec:flavorspin}.

	
\section{Helicity amplitude}\label{appendix:helicity}
The helicity amplitude of the decays in this work are derived from the effective Hamilton of semi-leptonic decays.
Here we take the process $c\to s l^{+}\nu_{l}$ as an example to derive the helicity amplitude. Firstly, the effective Hamilton of the process $c\to s l^{+}\nu_{l}$ is
\begin{eqnarray}
  {\cal H}_{eff}(c\to sl^{+}\nu_{l}) &=& \frac{G_F}{\sqrt{2}}V_{cs}^{*}[\bar{s}\gamma^{\mu}(1-\gamma_5)c]
  [\bar{\nu}_{l}\gamma_{\mu}(1-\gamma_5)l].
\end{eqnarray}
Secondly, the amplitude of the decay $\Xi_{cc}^{++}\to \Xi_{c}^{+} l^{+}\nu_{l}$ can be written as
\begin{eqnarray}
  i{\cal M}(2\pi)^4\delta^4(P-P^{\prime}-q) &=&
  -i\int d^4x\langle\Xi_{c}^{+}l^{+}\nu_{l}|{\cal H}_{eff}|\Xi_{cc}^{++}\rangle
  \nonumber\\
  &=&-i\int d^4x \frac{G_F}{\sqrt{2}}V_{cs}^{*}
  \langle\Xi_{c}^{+}|\bar{s}\gamma^{\mu}(1-\gamma_5)c|\Xi_{cc}^{++}\rangle
  \langle l^{+}\nu_{l}|\bar{\nu}_{l}\gamma_{\mu}(1-\gamma_5)l|0\rangle
  \nonumber\\
  &=&-i\int d^4x\frac{G_F}{\sqrt{2}}V_{cs}^{*}
  \langle\Xi_{c}^{+}|\bar{s}\gamma^{\mu}(1-\gamma_5)c|\Xi_{cc}^{++}\rangle g_{\mu\nu}
  \langle l^{+}\nu_{l}|\bar{\nu}_{l}\gamma^{\nu}(1-\gamma_5)l|0\rangle
  \nonumber\\
   &=&-i\int d^4x\frac{G_F}{\sqrt{2}}V_{cs}^{*}\Big[
  \langle\Xi_{c}^{+}|\bar{s}\gamma^{\mu}(1-\gamma_5)c|\Xi_{cc}^{++}\rangle \varepsilon^*_{W\mu}(t) \times
  \langle l^{+}\nu_{l}|\bar{\nu}_{l}\gamma^{\nu}(1-\gamma_5)l|0\rangle\varepsilon_{W\nu}(t)
  \nonumber\\
  &&-
  \sum_{\lambda_W=0,\pm}\langle\Xi_{c}^{+}|\bar{s}\gamma^{\mu}(1-\gamma_5)c|\Xi_{cc}^{++}\rangle\varepsilon^*_{W,\mu}(\lambda_W)
  \times \langle l^{+}\nu_{l}|\bar{\nu}_{l}\gamma^{\nu}(1-\gamma_5)l|0\rangle\varepsilon_{W,\nu}(\lambda_W)\Big].\label{eq:helicityfulu}
\end{eqnarray}
Here $W$ can be regarded as a virtual propagation, and we use $g_{\mu\nu}=\varepsilon^*_{\mu}(t)\varepsilon_{\nu}(t)
-\sum_{\lambda}\varepsilon^*_{\mu}(\lambda)\varepsilon_{\nu}(\lambda)$
in deriving the last line the above equations.

Then the the hadronic helicity amplitude can be defined as
\begin{eqnarray}
  & &\langle\Xi_{c}^{+}(P^{\prime},\lambda^{\prime})|\bar{s}\gamma^{\mu}(1-\gamma_5)c|\Xi_{cc}^{++}(P,\lambda)\rangle\varepsilon^*_{W,\mu}(\lambda_W) \nonumber\\
  &=&\langle\Xi_{c}^{+}(P^{\prime},\lambda^{\prime})|\bar{s}\gamma^{\mu}c|\Xi_{cc}^{++}(P,\lambda)\rangle\varepsilon^*_{W,\mu}(\lambda_W)-
  \langle\Xi_{c}^{+}(P^{\prime},\lambda^{\prime})|\bar{s}\gamma^{\mu}\gamma_5c|\Xi_{cc}^{++}(P,\lambda)\rangle\varepsilon^*_{W,\mu}(\lambda_W) \nonumber\\
  &=& H^V_{\lambda',\lambda_W}-H^A_{\lambda',\lambda_W}.\label{eq:helicityfulu1}
\end{eqnarray}
 While the leptonic part amplitude can be calculated by the follow equation,
 \begin{eqnarray}
  \langle l^{+}\nu_{l}|\bar{\nu}_{l}\gamma^{\nu}(1-\gamma_5)l|0\rangle\varepsilon_{W,\nu}(\lambda_W)
  = \bar{\nu}_{l}\slashed\varepsilon_{W}(\lambda_W)(1-\gamma_5)l.
\end{eqnarray}
In this work, the dynamics involved in the the hadronic amplitude are all in the rest frame of the initial states ${\cal B}_{i}$,
\begin{gather*}
{\cal B}_{i}: P^{\mu}=(M,0,0,0),\quad{\cal B}_{f}: P'^{\mu}=(E',0,0,-|\vec{P}^{\prime}|),\quad W: q^{\mu}=(E_W,0,0,|\vec{P}^{\prime}|),\\
\varepsilon^{\mu}_W(\pm1) =\frac{1}{\sqrt{2}}(0,\mp1,-i,0),\quad\varepsilon^{\mu}_W(0)=\frac{1}{\sqrt{q^2}}(|\vec{P}^{\prime}|,0,0,E_W),\quad\varepsilon^{\mu}_W(t) = \frac{1}{\sqrt{q^2}}(E_W,0,0,|\vec{P}^{\prime}|).
\end{gather*}
While the dynamics involved in the the leptonic part are all in the rest frame of the virtual vector particle $W$. The spinor expressions involved in this work are given as follows:
\begin{gather*}
u(P,\lambda=\frac{1}{2})= \sqrt{E+M}
\begin{pmatrix} \phi^1  \\ \frac{|\vec p|}{E+M}\phi^1
\end{pmatrix},\quad\phi^1=
\begin{pmatrix} \cos\frac{\theta}{2}e^{-i\frac{\phi}{2}} \\ e^{i\frac{\phi}{2}}\sin\frac{\theta}{2}
\end{pmatrix},
\\
u(P,\lambda=-\frac{1}{2})= \sqrt{E+M}
\begin{pmatrix} \phi^2  \\ -\frac{|\vec p|}{E+M}\phi^2
\end{pmatrix},\quad
\phi^2=
\begin{pmatrix} -\sin\frac{\theta}{2}e^{-i\frac{\phi}{2}} \\
 \cos\frac{\theta}{2}e^{i\frac{\phi}{2}}
\end{pmatrix},
\end{gather*}
here $\lambda$ denotes the helicity of the spinor and ($\theta$,$\phi$) are the direction of the momentum of the initial and final particles. In this work, we take the direction of the momentum of virtual vector particle $W$ as positive with $(\theta,\phi)=(0,0)$ and the direction of the momentum of the final baryons as negative direction with $(\theta,\phi)=(\pi,\pi)$, which are shown in Fig.\ref{fig:dynamic}.

The spinor expressions of anti-particles in the leptonic part are
\begin{gather*}
\nu(p,\lambda=\frac{1}{2})=\sqrt{E+M}
\begin{pmatrix} \frac{|\vec p|}{E+M}\phi^2 \\ -\phi^2
\end{pmatrix},\quad
\nu(p,\lambda=-\frac{1}{2})=\sqrt{E+M}
\begin{pmatrix} \frac{|\vec p|}{E+M}\phi^1 \\ \phi^1
\end{pmatrix}
\end{gather*}
In this work, the direction of the momentum of $l^+$ is $(\theta_l,\phi_{l})$,
 and the direction of the momentum of $\nu$ is $(\pi-\theta_l,\phi_{l}+\pi)$.

 The hadronic helicity amplitude of the transition ${\cal B}_{QQ^{\prime}}(P,S=1/2)\to {\cal B}_{Q^{\prime}}^{*}(P^{\prime},S^{\prime}=3/2)$ can be get in a similar way with ${\cal B}_{QQ^{\prime}}(P,S=1/2)\to {\cal B}_{Q^{\prime}}^{*}(P^{\prime},S^{\prime}=1/2)$ shown by Eqs.~(\ref{eq:helicityfulu})-(\ref{eq:helicityfulu1}). But the spinor for the final baryon ${\cal B}_{Q^{\prime}}^{*}(P^{\prime},S^{\prime}=3/2)$ is a vectorial spinor $u_\alpha(P^{\prime},S^{\prime}=3/2)$ which is the coupling of one spinor $u(P^{\prime},s_{1}=1/2)$ and a polarization vector $\varepsilon(P^{\prime},s_{2}=1)$ and the detail coupling formula are given as follows,
\begin{align}
u_\alpha(P^{\prime},S^{\prime}=3/2,\lambda=\pm3/2)=&u(P^{\prime},\lambda_{1}=\pm1/2)\varepsilon_\alpha(P^{\prime} ,\lambda_2 =\pm1),\notag\\
u_\alpha(P^{\prime},S^{\prime}=3/2,\lambda=\pm1/2)=&\frac{1}{\sqrt{3}}u(P^{\prime},\lambda_{1}=\mp1/2)
\varepsilon_\alpha(P^{\prime} ,\lambda_2 =\pm1)+\sqrt{\frac{2}{3}}
u(P^{\prime},\lambda_{1}=\pm1/2)\varepsilon_\alpha(P^{\prime} ,\lambda_2 =0),\notag
\end{align}
with
\begin{align}
\varepsilon_\alpha(P^{\prime} ,\lambda_2 =\pm1)=\frac{1}{\sqrt{2}}(0,\pm1,-i,0)&,\quad\varepsilon_\alpha(P^{\prime},
\lambda_2 =0)=\frac{1}{M^{\prime}}(|\vec P^{\prime}|,0,0,E^{\prime}),\\
u(P^{\prime},\lambda_{1}=+1/2)= \sqrt{E^{\prime}+M^{\prime}}
\begin{pmatrix} 0  \\ i \\0\\i\frac{|\vec P^{\prime}|}{E^{\prime}+M^{\prime}}
\end{pmatrix}&,\quad
\text{and}\quad
u(\vec p,\lambda=-\frac{1}{2})= \sqrt{E^{\prime}+M^{\prime}}
\begin{pmatrix} i \\ 0 \\-i\frac{|\vec P^{\prime}|}{E^{\prime}+M^{\prime}}\\0
\end{pmatrix}.
\end{align}

\section{Overlapping factors in flavor SU(3) symmetry}\label{su3approach}
The overlapping factors can also be calculated using the flavor SU(3) symmetry. The doubly heavy baryons triplets are   given as
\begin{equation}
[{\cal B}_{cc}]_{i}=
\begin{pmatrix}
\Xi_{cc}^{++}&
 \Xi_{cc}^{+}&
 \Omega_{cc}^{+}
\end{pmatrix},\quad
[{\cal B}_{bc}^{(\prime)}]_{i}=
\begin{pmatrix}
\Xi_{bc}^{(\prime)+}&
 \Xi_{bc}^{(\prime)0}&
 \Omega_{bc}^{(\prime)0}
\end{pmatrix},\quad
[{\cal B}_{bb}]_{i}= \begin{pmatrix}
\Xi_{bb}^{0}&
 \Xi_{bb}^{-}&
 \Omega_{bb}^{-}
\end{pmatrix},
\end{equation}
and the singly heavy baryons anti-triplets are
\begin{equation}
[{\cal B}_{c}]^{[ij]}_{\bar 3}=
\begin{pmatrix}
0&\Lambda_{c}^{+}&\Xi_{c}^{+} \\
 -\Lambda_{c}^{+}&0&\Xi_{c}^{0}\\
 -\Xi_{c}^{+}&-\Xi_{c}^{0}&0
\end{pmatrix},
\quad[{\cal B}_{b}]^{[ij]}_{\bar 3}=
\begin{pmatrix}
 0&\Lambda_{b}^{0}&\Xi_{b}^{0} \\
 -\Lambda_{b}^{0}&0&\Xi_{b}^{-}\\
 -\Xi_{b}^{0}&-\Xi_{b}^{-}&0
\end{pmatrix},
\end{equation}
the singly heavy baryons sextets are
\begin{equation}
[{\cal B}_{c}]^{\{ij\}}_{6}=
\begin{pmatrix}
\Sigma_{c}^{++}&\frac{1}{\sqrt{2}}\Sigma_{c}^{+}&\frac{1}{\sqrt{2}}\Xi_{c}^{\prime+} \\
 \frac{1}{\sqrt{2}}\Sigma_{c}^{+}&\Sigma_{c}^{0}&\frac{1}{\sqrt{2}}\Xi_{c}^{\prime0}\\
 \frac{1}{\sqrt{2}}\Xi_{c}^{\prime+}&\frac{1}{\sqrt{2}}\Xi_{c}^{\prime0}&\Omega_{c}^{0}
\end{pmatrix},
\quad[{\cal B}_{b}]^{\{ij\}}_{6}=
\begin{pmatrix}
\Sigma_{b}^{+}&\frac{1}{\sqrt{2}}\Sigma_{b}^{0}&\frac{1}{\sqrt{2}}\Xi_{b}^{\prime0} \\
 \frac{1}{\sqrt{2}}\Sigma_{b}^{0}&\Sigma_{b}^{-}&\frac{1}{\sqrt{2}}\Xi_{b}^{\prime-}\\
 \frac{1}{\sqrt{2}}\Xi_{b}^{\prime0}&\frac{1}{\sqrt{2}}\Xi_{b}^{\prime-}&\Omega_{b}^{-}
\end{pmatrix}.
\end{equation}
For ${\cal B}_{cc}\to {\cal B}_{c}$ transitions with $c\to u,d,s$, the SU(3) amplitude could be written as:
\begin{eqnarray}
C_{S}&=&[{\cal B}_{cc}]_{i}[{\cal B}_{c}]^{[ij]}_{\bar 3}O_{j}^{1}+[{\cal B}_{cc}]_{i}[{\cal B}_{c}]^{\{ij\}}_{6}O_{j}^{2},
\quad j=u,d,s,\nonumber\\
&&\quad\text{with}\quad O_{u,d,s}^{1}=\begin{pmatrix}
\frac{\sqrt{6}}{4}\\0\\ 0
\end{pmatrix},\begin{pmatrix}
0\\\frac{\sqrt{6}}{4}\\ 0
\end{pmatrix},\begin{pmatrix}
0\\0\\ \frac{\sqrt{6}}{4}
\end{pmatrix},\quad\text{and}\quad O_{u,d,s}^{2}=\begin{pmatrix}
-\frac{3}{2}\\0\\ 0
\end{pmatrix},\begin{pmatrix}
0\\-\frac{3}{2}\\ 0
\end{pmatrix},\begin{pmatrix}
0\\0\\-\frac{3}{2}
\end{pmatrix},\\
C_{A}&=&[{\cal B}_{cc}]_{i}[{\cal B}_{c}]^{[ij]}_{\bar 3}O_{j}^{1}+[{\cal B}_{cc}]_{i}[{\cal B}_{c}]^{\{ij\}}_{6}O_{j}^{2},
\quad j=u,d,s,\nonumber\\
&&\quad\text{with}\quad O_{u,d,s}^{1}=\begin{pmatrix}
\frac{\sqrt{6}}{4}\\0\\ 0
\end{pmatrix},\begin{pmatrix}
0\\\frac{\sqrt{6}}{4}\\ 0
\end{pmatrix},\begin{pmatrix}
0\\0\\ \frac{\sqrt{6}}{4}
\end{pmatrix},\quad\text{and}\quad
O_{u,d,s}^{2}=
\begin{pmatrix}
\frac{1}{2}\\0\\ 0
\end{pmatrix},
\begin{pmatrix}
0\\\frac{1}{2}\\ 0
\end{pmatrix},
\begin{pmatrix}
0\\0\\ \frac{1}{2}
\end{pmatrix}.
\end{eqnarray}
For ${\cal B}_{bc}\to {\cal B}_{b}$ transitions with $c\to u,d,s$, it is:
\begin{eqnarray}
C_{S}&=&[{\cal B}_{bc}]_{i}[{\cal B}_{b}]^{[ij]}_{\bar 3}O_{j}^{1}+[{\cal B}_{bc}]_{i}[{\cal B}_{b}]^{\{ij\}}_{6}O_{j}^{2},
\quad j=u,d,s,\nonumber\\
&&\quad\text{with}\quad O_{u,d,s}^{1}=
\begin{pmatrix}
\frac{\sqrt{3}}{4}\\0\\ 0
\end{pmatrix},\begin{pmatrix}
0\\\frac{\sqrt{3}}{4}\\ 0
\end{pmatrix},\begin{pmatrix}
0\\0\\ \frac{\sqrt{3}}{4}
\end{pmatrix},
\quad\text{and}\quad O_{u,d,s}^{2}=
\begin{pmatrix}
-\frac{3\sqrt{2}}{4}\\0\\ 0
\end{pmatrix},\begin{pmatrix}
0\\-\frac{3\sqrt{2}}{4}\\ 0
\end{pmatrix},\begin{pmatrix}
0\\0\\ -\frac{3\sqrt{2}}{4}
\end{pmatrix},\\
C_{A}&=&[{\cal B}_{bc}]_{i}[{\cal B}_{b}]^{[ij]}_{\bar 3}O_{j}^{1}+[{\cal B}_{bc}]_{i}[{\cal B}_{b}]^{\{ij\}}_{6}O_{j}^{2},
\quad j=u,d,s,\nonumber\\
&&\quad\text{with}\quad O_{u,d,s}^{1}=
\begin{pmatrix}
\frac{\sqrt{3}}{4}\\0\\ 0
\end{pmatrix},
\begin{pmatrix}
0\\\frac{\sqrt{3}}{4}\\ 0
\end{pmatrix},
\begin{pmatrix}
0\\0\\ \frac{\sqrt{3}}{4}
\end{pmatrix},
\quad\text{and}\quad O_{u,d,s}^{2}=
\begin{pmatrix}
\frac{\sqrt{2}}{4}\\0\\ 0
\end{pmatrix},\begin{pmatrix}
0\\\frac{\sqrt{2}}{4}\\ 0
\end{pmatrix},\begin{pmatrix}
0\\0\\\frac{\sqrt{2}}{4}
\end{pmatrix}.
\end{eqnarray}	
For ${\cal B}_{bc}^{\prime}\to {\cal B}_{b}$ transitions with $c\to u,d,s$, the amplitude is
\begin{eqnarray}
C_{S}&=&[{\cal B}_{bc}^{\prime}]_{i}[{\cal B}_{b}]^{[ij]}_{\bar 3}O_{j}^{1}+[{\cal B}_{bc}^{\prime}]_{i}[{\cal B}_{b}]^{\{ij\}}_{6}O_{j}^{2},
\quad j=u,d,s,\nonumber\\
&&\quad\text{with}\quad O_{u,d,s}^{1}=
\begin{pmatrix}
-\frac{1}{4}\\0\\ 0
\end{pmatrix},\begin{pmatrix}
0\\-\frac{1}{4}\\ 0
\end{pmatrix},\begin{pmatrix}
0\\0\\ -\frac{1}{4}
\end{pmatrix},
\quad\text{and}\quad O_{u,d,s}^{2}=
\begin{pmatrix}
\frac{\sqrt{6}}{4}\\0\\ 0
\end{pmatrix},\begin{pmatrix}
0\\\frac{\sqrt{6}}{4}\\ 0
\end{pmatrix},\begin{pmatrix}
0\\0\\ \frac{\sqrt{6}}{4}
\end{pmatrix},\\
C_{A}&=&[{\cal B}_{bc}^{\prime}]_{i}[{\cal B}_{b}]^{[ij]}_{\bar 3}O_{j}^{1}+[{\cal B}_{bc}^{\prime}]_{i}[{\cal B}_{b}]^{\{ij\}}_{6}O_{j}^{2},
\quad j=u,d,s,\nonumber\\
&&\quad\text{with}\quad O_{u,d,s}^{1}=
\begin{pmatrix}
\frac{3}{4}\\0\\ 0
\end{pmatrix},
\begin{pmatrix}
0\\\frac{3}{4}\\ 0
\end{pmatrix},
\begin{pmatrix}
0\\0\\ \frac{3}{4}
\end{pmatrix},
\quad\text{and}\quad O_{u,d,s}^{2}=
\begin{pmatrix}
\frac{\sqrt{6}}{4}\\0\\ 0
\end{pmatrix},\begin{pmatrix}
0\\\frac{\sqrt{6}}{4}\\ 0
\end{pmatrix},\begin{pmatrix}
0\\0\\\frac{\sqrt{6}}{4}
\end{pmatrix}.
\end{eqnarray}	
For ${\cal B}_{bb}\to {\cal B}_{b}$ transitions with $b\to u,d,s$, it is
\begin{eqnarray}
C_{S}&=&[{\cal B}_{bb}]_{i}[{\cal B}_{b}]^{[ij]}_{\bar 3}O_{j}^{1}+[{\cal B}_{bb}]_{i}[{\cal B}_{b}]^{\{ij\}}_{6}O_{j}^{2},
\quad j=u,d,s,\nonumber\\
&&\quad\text{with}
\quad O_{u,d,s}^{1}=
\begin{pmatrix}
\frac{\sqrt{6}}{4}\\0\\ 0
\end{pmatrix},
\begin{pmatrix}
0\\\frac{\sqrt{6}}{4}\\ 0
\end{pmatrix},
\begin{pmatrix}
0\\0\\ \frac{\sqrt{6}}{4}
\end{pmatrix},
\quad\text{and}\quad O_{u,d,s}^{2}=
\begin{pmatrix}
-\frac{3}{2}\\0\\ 0
\end{pmatrix},
\begin{pmatrix}
0\\-\frac{3}{2}\\ 0
\end{pmatrix},
\begin{pmatrix}
0\\0\\ -\frac{3}{2}
\end{pmatrix},\\
C_{A}&=&[{\cal B}_{bb}]_{i}[{\cal B}_{b}]^{[ij]}_{\bar 3}O_{j}^{1}+[{\cal B}_{bb}]_{i}[{\cal B}_{b}]^{\{ij\}}_{6}O_{j}^{2},
\quad j=u,d,s,\nonumber\\
&&\quad\text{with}\quad O_{u,d,s}^{1}=
\begin{pmatrix}
\frac{\sqrt{6}}{4}\\0\\ 0
\end{pmatrix},
\begin{pmatrix}
0\\\frac{\sqrt{6}}{4}\\ 0
\end{pmatrix},\begin{pmatrix}
0\\0\\ \frac{\sqrt{6}}{4}
\end{pmatrix},
\quad\text{and}\quad O_{u,d,s}^{2}=
\begin{pmatrix}
\frac{1}{2}\\0\\ 0
\end{pmatrix},
\begin{pmatrix}
0\\\frac{1}{2}\\ 0
\end{pmatrix},
\begin{pmatrix}
0\\0\\ \frac{1}{2}
\end{pmatrix}.
\end{eqnarray}	
The  ${\cal B}_{bc}\to {\cal B}_{c}$ transitions with $b\to u,d,s$ have the results
\begin{eqnarray}
C_{S}&=&[{\cal B}_{bc}]_{i}[{\cal B}_{c}]^{[ij]}_{\bar 3}O_{j}^{1}+[{\cal B}_{bc}]_{i}[{\cal B}_{c}]^{\{ij\}}_{6}O_{j}^{2},
\quad j=u,d,s,\nonumber\\
&&\quad\text{with}\quad O_{u,d,s}^{1}=
\begin{pmatrix}
\frac{\sqrt{3}}{4}\\0\\ 0
\end{pmatrix},
\begin{pmatrix}
0\\\frac{\sqrt{3}}{4}\\ 0
\end{pmatrix},
\begin{pmatrix}
0\\0\\ \frac{\sqrt{3}}{4}
\end{pmatrix},\quad\text{and}\quad O_{u,d,s}^{2}=
\begin{pmatrix}
-\frac{3\sqrt{2}}{4}\\0\\ 0
\end{pmatrix},
\begin{pmatrix}
0\\-\frac{3\sqrt{2}}{4}\\ 0
\end{pmatrix},
\begin{pmatrix}
0\\0\\ -\frac{3\sqrt{2}}{4}
\end{pmatrix},\\
C_{A}&=&[{\cal B}_{bc}]_{i}[{\cal B}_{c}]^{[ij]}_{\bar 3}O_{j}^{1}+[{\cal B}_{bc}]_{i}[{\cal B}_{c}]^{\{ij\}}_{6}O_{j}^{2},
\quad j=u,d,s,\nonumber\\
&&\quad\text{with}\quad O_{u,d,s}^{1}=\begin{pmatrix}
\frac{\sqrt{3}}{4}\\0\\ 0
\end{pmatrix},
\begin{pmatrix}
0\\\frac{\sqrt{3}}{4}\\ 0
\end{pmatrix},
\begin{pmatrix}
0\\0\\ \frac{\sqrt{3}}{4}
\end{pmatrix},\quad\text{and}\quad O_{u,d,s}^{2}=\begin{pmatrix}
\frac{\sqrt{2}}{4}\\0\\ 0
\end{pmatrix},
\begin{pmatrix}
0\\\frac{\sqrt{2}}{4}\\ 0
\end{pmatrix},
\begin{pmatrix}
0\\0\\\frac{\sqrt{2}}{4}
\end{pmatrix}.
\end{eqnarray}	
For ${\cal B}_{bc}^{\prime}\to {\cal B}_{c}$ transitions with $b\to u,d,s$, the amplitude is:
\begin{eqnarray}
C_{S}&=&[{\cal B}_{bc}^{\prime}]_{i}[{\cal B}_{c}]^{[ij]}_{\bar 3}O_{j}^{1}+[{\cal B}_{bc}^{\prime}]_{i}[{\cal B}_{c}]^{\{ij\}}_{6}O_{j}^{2},
\quad j=u,d,s,\nonumber\\
&&\quad\text{with}\quad O_{u,d,s}^{1}=
\begin{pmatrix}
\frac{1}{4}\\0\\ 0
\end{pmatrix},
\begin{pmatrix}
0\\\frac{1}{4}\\ 0
\end{pmatrix},
\begin{pmatrix}
0\\0\\ \frac{1}{4}
\end{pmatrix},\quad\text{and}\quad O_{u,d,s}^{2}=
\begin{pmatrix}
-\frac{\sqrt{6}}{4}\\0\\ 0
\end{pmatrix},
\begin{pmatrix}
0\\-\frac{\sqrt{6}}{4}\\ 0
\end{pmatrix},
\begin{pmatrix}
0\\0\\ -\frac{\sqrt{6}}{4}
\end{pmatrix},\\
C_{A}&=&[{\cal B}_{bc}^{\prime}]_{i}[{\cal B}_{c}]^{[ij]}_{\bar 3}O_{j}^{1}+[{\cal B}_{bc}^{\prime}]_{i}[{\cal B}_{c}]^{\{ij\}}_{6}O_{j}^{2},
\quad j=u,d,s,\nonumber\\
&&\quad\text{with}\quad O_{u,d,s}^{1}=\begin{pmatrix}
-\frac{3}{4}\\0\\ 0
\end{pmatrix},
\begin{pmatrix}
0\\-\frac{3}{4}\\ 0
\end{pmatrix},
\begin{pmatrix}
0\\0\\ -\frac{3}{4}
\end{pmatrix},\quad\text{and}\quad O_{u,d,s}^{2}=\begin{pmatrix}
-\frac{\sqrt{6}}{4}\\0\\ 0
\end{pmatrix},
\begin{pmatrix}
0\\-\frac{\sqrt{6}}{4}\\ 0
\end{pmatrix},
\begin{pmatrix}
0\\0\\-\frac{\sqrt{6}}{4}
\end{pmatrix}.
\end{eqnarray}

\end{document}